\documentclass  [11pt,letterpaper]{article}
\pdfoutput=1
\usepackage     {jheppub}
\usepackage     {journals}
\usepackage     {verbatim}
\usepackage     [mathscr]{eucal}
\usepackage     {bm}
\bibpunct	{[}{]}{,}{n}{}{;}

\def\Nfour	{\mathcal{N}\,{=}\,4}

\def\Nc		{N_\text{c}}

\def\tr		{\text{tr}\,}
\def\half	{\tfrac{1}{2}}
\def\third	{\tfrac{1}{3}}
\def\fourth	{\tfrac{1}{4}}
\def\x		{\bm{x}}
\def\rh		{r_{\rm h}}
\def\rhbar	{\bar r_{\rm h}}
\def\dim	{\nu}
\def\ghat	{\hat g}
\def\Too	{T^{00}}
\def\Tor	{T^{0r}}
\def\Trr	{T^{rr}}
\def\stress	{s}
\def\grad	{\widetilde \nabla}
\def\Gam	{\widetilde \Gamma}
\def\suck	[#1]#2{\includegraphics[#1]{#2}}	

\def\tocsqueeze	{\vspace*{-3.7pt}}

\title          {Numerical solution of gravitational dynamics
		in asymptotically anti-de Sitter spacetimes}
\author[a]      {Paul~M.~Chesler,}
\author[b]	{Laurence~G.~Yaffe}
\affiliation[a]	{Department of Physics, MIT, Cambridge MA 02139, USA}
\affiliation[b] {Department of Physics, University of Washington,
                Seattle WA 98195, USA}
\emailAdd	{pchesler@mit.edu}
\emailAdd       {yaffe@phys.washington.edu}

\abstract
    {%
    A variety of gravitational dynamics problems in asymptotically
    anti-de Sitter (AdS) spacetime are amenable to efficient numerical solution
    using a common approach involving a null slicing of spacetime based
    on infalling geodesics, convenient exploitation of the residual
    diffeomorphism freedom, and use of spectral methods for discretizing
    and solving the resulting differential equations.
    Relevant issues and choices leading to this approach are
    discussed in detail.
    Three examples, motivated by applications to non-equilibrium
    dynamics in strongly coupled gauge theories,
    are discussed as instructive test cases.
    These are gravitational descriptions of homogeneous isotropization,
    collisions of planar shocks,
    and turbulent fluid flows in two spatial dimensions.
    }

\keywords	{general relativity, gauge-gravity correspondence, 
		 quark-gluon plasma}
\arxivnumber	{1309.1439}
\begin		{document}

\advance\textheight 55pt
\maketitle
\thispagestyle{empty}
\advance\textheight -55pt

\addtocontents	{toc}{\tocsqueeze}
\addtocontents	{toc}{\tocsqueeze}
\section	{Introduction}

The advent of gauge-gravity duality (or ``holography'') has revolutionized
the study of strongly interacting field theories.%
\footnote
    {%
    See, for example, refs.~\cite{%
    Maldacena:1997re,
    Aharony:1999ti,
    D'Hoker:2002aw, 
    Gubser:2009md}
    and references therein.
    }
In gauge theories with gravitational duals, holography maps problems
involving non-equilibrium quantum dynamics, in the limit of
strong coupling and large gauge group rank, into problems involving
classical gravitational dynamics in higher dimensions.
Consequently, holography provides unique opportunities to study
strongly-coupled non-equilibrium dynamics --- provided one can
actually solve the associated classical gravitational dynamics.

Gauge theories with known gravitational duals are generally
conformal field theories (CFTs)
or conformal theories deformed by relevant operators;
for such theories the dual gravitational description involves
a 5D spacetime with asymptotically anti-de Sitter (AdS) geometry.%
\footnote
    {%
    The Klebanov-Strassler cascading gauge theory \cite{Klebanov:2000hb}
    is an example of a more complicated theory whose dual geometry is not
    asymptotically anti-de Sitter.
    }
Much work to date has explored near-equilibrium phenomena via holography.
Examples include the study of
viscosity and related transport coefficients
\cite{%
Policastro:2001yc,
Kovtun:2004de,
Son:2006em,
Karch:2008fa},
more general aspects of dissipative hydrodynamics
\cite{%
Baier:2007ix,
Bhattacharyya:2008jc,
Hubeny:2011hd},
quasi-normal modes and near-equilibrium response
\cite{%
Starinets:2002br,
Kovtun:2005ev},
and dynamics of probes such as heavy quarks
\cite{%
Herzog:2006gh,
CasalderreySolana:2006rq,
Friess:2006fk,
Chesler:2007sv,
Chesler:2011nc}
or light quarks
\cite{Chesler:2008wd,Chesler:2008uy}.%
\footnote
    {%
    For additional prior work in this area, see the
    recent review \cite{CasalderreySolana:2011us},
    which is focused on applications to heavy ion collisions,
    and references therein.
    }

There has been much less work on far-from-equilibrium dynamics, as
this requires the solution of gravitational dynamics with non-trivial
initial conditions and (except in extremely special cases)
such solutions can only be found numerically.%
\footnote
    {%
    For a broad perspective on numerical relativity and applications
    to high energy physics, see ref.~\cite{Cardoso:2012qm}.
    }
Despite the difficulty, notable progress has been made on gravitational
initial value problems involving asymptotically AdS geometries.%
\footnote
    {%
    In the numerical relativity community, the phrase ``initial value problem''
    is sometimes viewed as referring, specifically, to dynamical evolution
    problems based on spacelike Cauchy surfaces.
    A ``characteristic'' formulation refers to evolution schemes based on
    null surfaces.
    We will treat the phrase ``initial value problem'' as encompassing
    dynamical evolution schemes with initial data given on either
    spacelike or null surfaces.
    }
Recent work applying holography to far-from-equilibrium dynamics
includes studies of
isotropization in spatially homogeneous systems
\cite{CY:isotropize,Heller:2012km,Heller:2013oxa},
boost-invariant expansion with transverse homogeneity
\cite{CY:boostinvar,Heller:2012je}
or with radial flow
\cite{vanderSchee:2013pia},
spherically symmetric thermalization
\cite{Bantilan:2012vu},
collisions of planar shocks
\cite{CY:shocks,Casalderrey-Solana:2013aba,Casalderrey-Solana:2013sxa},
and turbulence in 2D fluids
\cite{Adams:2013vsa}.
With the exception of the last example,
all these problems have a sufficient degree of symmetry that the 5D Einstein
equations reduce to either $1{+}1$ or $2{+}1$ dimensional partial differential
equations (PDEs).
An obvious goal of current and future work is the solution of
initial value problems involving lower degrees of symmetry.

In this paper, we discuss the computational challenges involved in
solving, numerically, asymptotically anti-de Sitter gravitational
initial value problems.
We describe in detail a particular approach which we have found to be
effective in a series of progressively more complex applications,
three of which will be examined as specific test cases:
homogeneous isotropization, planar shock wave collisions,
and turbulent 2D fluid flows.
Results on homogeneous isotropization have been previously reported
in ref.~\cite{CY:isotropize};
the degree of symmetry for this problem reduces the
5D Einstein equations to a set of coupled 1+1 dimensional PDEs.
Initial results for planar shock wave collisions were reported in
ref.~\cite{CY:shocks};
in this case the 5D Einstein equations reduce to 2+1 dimensional PDEs.
Studies of fluid flows in two spatial dimensions, using holography,
involve the solution of 4D Einstein equations \cite{Adams:2013vsa}.
With no simplifying symmetry restrictions,
this case requires solving coupled 3+1 dimensional PDEs.
We present results for each of these test cases and discuss
both the associated physics and
computational issues such as stability and accuracy.
The results presented in this paper extend and complement earlier work.
In particular, for colliding planar shock waves,
we show that it is possible to perform numerically stable,
accurate, calculations without adding any artificial background
energy density, as was done in refs.~%
\cite{CY:shocks,Casalderrey-Solana:2013aba,Casalderrey-Solana:2013sxa}.
We study stocks of different thicknesses,
as in ref.~\cite{Casalderrey-Solana:2013aba},
but integrate farther in time.
In agreement with ref.~\cite{Casalderrey-Solana:2013aba},
we find that collisions of relatively thin shocks are
\emph{not} well approximated as boost invariant.
However, we show that the resulting hydrodynamic flow
may be characterized as \emph{locally} boost invariant,
in a sense which we discuss in section \ref{sec:shocks}.

For simplicity of presentation,
most discussion in this paper is limited to problems involving solutions
to pure Einstein gravity which are asymptotic to the Poincar\'e patch of
anti-de Sitter space \cite{Maldacena:1997re}.
Many interesting extensions are only touched upon or left to future work.
These include generalizations of these methods to problems involving
non-flat boundary geometry (e.g., global AdS asymptotics,
or explicit time-dependent boundary geometries 
\cite{CY:isotropize,CY:boostinvar}),
additional compact dimensions
(e.g., dynamics of initial states in $\Nfour$ super-Yang-Mills (SYM)
theory which are not invariant under the full $SU(4)$ $R$-symmetry),
or additional dynamical fields (dilaton-gravity, Maxwell-Einstein, etc.).%
\footnote
    {%
    As 5D Einstein gravity is a consistent truncation of 10D IIB supergravity
    on AdS$_5 \times S^5$,
    all the 5D pure Einstein gravity solutions we discuss may be viewed
    as supergravity solutions which describe the dynamics of
    $\Nfour$ SYM states invariant under the $SU(4)_R$ symmetry.
    }

\addtocontents	{toc}{\tocsqueeze}
\section	{Setup and conventions}\label{sec:setup}

Gauge/gravity duality relates certain quantum field theories in $D$ spacetime
dimensions to gravitational physics in $D{+}1$ dimensions.
(As noted above, we are not considering problems in which
the dynamics of additional compact dimensions play any role
in the gravitational description.)
We consider quantum field theories (QFTs) in $D$ dimensional
flat Minkowski spacetime,
and hence will be interested in gravitational solutions
describing $D+1$ dimensional geometries with boundary, for which the
boundary geometry is $D$ dimensional Minkowski space.
Using Fefferman-Graham coordinates, the resulting asymptotically AdS
metric may be represented as
\cite{FeffermanGraham, Fefferman:2007rka, deHaro:2000xn, Skenderis:2002wp}%
\footnote
    {%
    Our metric signature convention is mostly plus.
    Uppercase Latin letters $M,N,\cdots = 0,\,\cdots,D$ will be used as $D+1$
    dimensional spacetime indices.
    Greek letters $\mu,\nu,\cdots = 0,\,\cdots,D{-}1$ are used as $D$
    dimensional spacetime indices in the dual quantum field theory,
    and lower case Latin letters $i,j,k,\cdots {} = 1,\,\cdots,D{-}1$
    are used for $D{-}1$ dimensional spatial indices.
    The usual Minkowski metric tensor
    $\eta = \|\eta_{\mu\nu}\| \equiv \mathrm{diag}(-1,+1,\cdots,+1)$.
    }
\begin{equation}
    ds^2
    =
    \frac {L^2}{\rho^2}
    \left[\,
	g_{\mu\nu}(x,\rho) \, dx^\mu \, dx^\nu + d\rho^2
    \right],
\label{eq:FeffGraham}
\end{equation}
where $\rho$ is a ``bulk'' radial coordinate such that the
spacetime boundary lies at $\rho = 0$, with $\{ x^\mu \}$
denoting the $D$ remaining ``boundary'' coordinates.
We use $L$ as the spacetime curvature scale;
it is related to the cosmological constant via
\begin{equation}
    \Lambda = -\half D(D{-}1) / L^2 \,.
\end{equation}
The metric functions $g_{\mu\nu}(x,\rho)$ have a near-boundary
asymptotic expansion in integer powers of $\rho$, with
the leading term equal to the desired Minkowski boundary metric
and subleading terms starting at order~$\rho^D$,%
\footnote
    {%
    If the boundary metric is not flat, then additional terms
    involving even powers of $\rho$ below order $\rho^D$ are present,
    as well as logarithmic terms starting with $\rho^D \ln \rho$
    when $D$ is even
    \cite{deHaro:2000xn}.
    \label{fn:nonflat}
    }
\begin{equation}
    g_{\mu\nu}(x,\rho) \sim \eta_{\mu\nu}
    + \sum_{n=D}^\infty \> g_{\mu\nu}^{(n)}(x)\, \rho^{n} \,.
\label{eq:FGasym}
\end{equation}

It will be convenient to use a rescaled stress-energy tensor
\begin{equation}
    \widehat T_{\mu\nu}(x) = \frac {T_{\mu\nu}(x)}{\kappa} \,,\qquad
    \kappa \equiv \frac {D L^{D-1}}{16\pi G_N} \,,
\label{eq:Thatdef}
\end{equation}
where $G_N$ is the $D{+}1$ dimensional Newton gravitational constant.
[For $D=4$, Newton's constant is related to the dual $SU(\Nc)$ SYM theory via
$G_N = \frac{\pi}{2} L^3/\Nc^2$, so $\kappa = \Nc^2/(2\pi^2)$.]
The coefficient of the first sub-leading term in the
near-boundary expansion (\ref{eq:FGasym})
determines the boundary stress-energy tensor, which
coincides with the expectation value of the (rescaled)
stress-energy tensor in the dual QFT,
\begin{equation}
    \label{eq:FGstress}
    \langle \widehat T_{\mu\nu}(x) \rangle
    = g_{\mu\nu}^{(D)} (x) \,.
\end{equation}
Einstein's equations imply boundary stress-energy conservation
and tracelessness
\cite{FeffermanGraham, deHaro:2000xn, Skenderis:2002wp},
\begin{equation}
    \nabla_\mu \, \langle \widehat T^{\mu\nu}(x) \rangle = 0 \,,
    \qquad
    \langle \widehat T^\mu_{\;\,\mu} \rangle = 0 \,.
\label{eq:conservation}
\end{equation}

Given a non-vanishing stress-energy expectation value,
one may define an associated velocity field $u(x)$ 
and (rescaled) proper energy density $\varepsilon(x)$ as the
timelike eigenvector and corresponding eigenvalue
of the stress-energy tensor,
\begin{equation}
    \langle \widehat T^\mu_{\;\,\nu}(x) \rangle \> u^\nu(x)
    =
    -\varepsilon(x) \, u^\mu(x) \,,
\label{eq:flowfield}
\end{equation}
(with normalization $u(x)^2 \equiv -1$),
provided $\langle \widehat T_{\mu\nu}\rangle$
(or $-\langle \widehat T_{\mu\nu}\rangle$)
satisfies the weak energy condition.%
\footnote
    {%
    If $\pm\widehat T^{\mu\nu}$ fail to satisfy the weak
    energy condition,
    then the matrix $\|\widehat T^\mu_{\;\,\nu}\|$ can have complex conjugate pairs
    of eigenvalues and no real time-like eigenvector.
    }
An observer moving with spacetime velocity $u(x)$ sees
an energy density equal to $\varepsilon(x)$ and vanishing energy flux.
For later convenience, let
$\widetilde u = u_\mu \, dx^\mu$ denote the one-form dual
to the vector field $u = u^\mu\, \partial_\mu$.


\addtocontents	{toc}{\tocsqueeze}
\section	{Computational strategy}

A basic issue affecting any numerical relativity
calculation is the choice of how to deal with the diffeomorphism
invariance of general relativity.
This lies at the heart of how one converts Einstein's equations
into a well-posed initial value problem.

One general approach is to choose an ansatz for the metric,
whose form will greatly restrict the remaining diffeomorphism freedom.
The Fefferman-Graham form (\ref{eq:FeffGraham}) is one such possibility.
The ansatz must allow an arbitrary metric,
consistent with the symmetries of the physical problem under consideration,
to be transformed into the chosen form by a suitable change of coordinates.
Even when this is possible in any local region, a given ansatz for the metric
may fail to provide good coordinates covering the entire domain of interest.
This is a known problem with the Fefferman-Graham form (\ref{eq:FeffGraham}).
Although convenient and useful for analyzing near-boundary behavior,
in solutions describing gravitational infall and horizon equilibration,
the Fefferman-Graham metric develops coordinate singularities in the
bulk and fails to remain regular across the future event horizon
\cite{Heller:2007qt,Heller:2008mb}.
Consequently, despite its utility for other purposes,
the Fefferman-Graham ansatz is not a good choice for numerical
initial value problems.

A different approach, avoiding the need to commit to some specific global form
of coordinates, is provided by the ADM formalism in which Cauchy surfaces
are arbitrary spacelike slices of the geometry, and some chosen lapse function
and shift vector field relate the coordinates on neighboring spacelike slices
foliating the geometry \cite{Arnowitt:1959ah,Arnowitt:1962hi}.
This approach has been widely used in numerical relativity calculations
in asymptotically Minkowski space
\cite{Lehner:2001wq,Baumgarte:2002jm,BaumgarteShapiro}.
However, employing this approach has some practical downsides.
Implementing this method (particularly when combined with adaptive
mesh refinement) is complex.
One must formulate a scheme for dynamically choosing lapse and shift
vectors, or make some a-priori choice, in a manner which, one hopes,
will allow the foliation to remain regular throughout the spacetime
region of interest.
Achieving a numerically stable scheme can be problematic
\cite{Lehner:2001wq,Alcubierre:1999rt,Yo:2002bm,Knapp:2002fm,Calabrese:2005ft}.

\subsection	{Metric ansatz}

We have chosen to employ the first approach involving a metric ansatz,
one which is specifically tailored to gravitational infall problems.
The metric ansatz is a generalization of traditional ingoing
Eddington-Finkelstein coordinates for black holes.%
\footnote
    {%
    Previous work \cite{Bhattacharyya:2008jc,Heller:2008mb}
    studying late time behavior of
    solutions which approach stationary black brane
    solutions convincingly demonstrates the virtues of using
    generalized Eddington-Finkelstein coordinates for this class
    of asymptotically AdS gravitational infall problems.
    }
It is based on a null slicing of spacetime constructed
from infalling null geodesics,
and will lead to a characteristic formulation of gravitational dynamics.%
\footnote
    {%
    For useful prior discussions of characteristic formulations of relativity,
    see ref.~\cite{Winicour:1998tz} and references therein.
    }
The general form of the metric is
\begin{equation}
    ds^2 =
    \frac{r^2}{L^2} \> g_{\mu\nu}(x,r) \, dx^\mu \, dx^\nu
    -2 \, w_\mu(x) \, dx^\mu \, dr
    \,,
\label{eq:ansatz1}
\end{equation}
where $r$ is a non-inverted bulk radial coordinate
(so the spacetime boundary lies at $r = \infty$),
and $\{ x^\mu \}$ denote the $D$ remaining boundary coordinates.%
\footnote
    {%
    The inverse metric
    $
	G^{MN} =
	\left(
	    \begin{array}{cc}
		(L/r)^2 \, (g^{\mu\nu} {-} w^\mu w^\nu/w^2)
		& -w^\mu/w^2
	    \\
		-w^\nu/w^2  & -(r/L)^2/w^2
	    \end{array}
	\right),
    $
    with $w^\mu(x,r) \equiv g^{\mu\nu}(x,r) \, w_\nu(x)$.
    }
The boundary one-form $\widetilde w = w_\mu \, dx^\mu$ appearing in the
second term is independent of the radial coordinate $r$.
This one-form is assumed to be timelike and,
without loss of generality, may be taken to satisfy $\widetilde w^{\,2} = -1$
(using the boundary metric discussed below).
A more explicit representation of the metric $g_{\mu\nu}$ which
describes the geometry on fixed-$r$ slices will be introduced in
section \ref{sec:Einstein}.

From the ansatz (\ref{eq:ansatz1}), one immediately sees that
lines along which $r$ varies while the other coordinates
are held fixed are null curves.
One may easily check that these curves are
infalling null geodesics
for which $r$ is an affine parameter.
Therefore, the vector $\partial_r$ is a directional derivative along
infalling null geodesics.
At the boundary ($r = \infty$), an observer whose $D$-velocity components
equal $w^\mu$ would describe these geodesics as representing
trajectories of comoving objects at rest in his frame;
their tangent vectors are normal to the $D{-}1$ spatial basis
vectors in the observer's frame.
In our coordinates, these geodesics remain purely radial throughout the
bulk geometry.

The form of the metric ansatz (\ref{eq:ansatz1}) is preserved by two
types of residual diffeomorphisms:
arbitrary $D$-dimensional diffeomorphisms (independent of $r$),
\begin{equation}
    x^\mu \to \bar x^\mu \equiv f^\mu(x) \,,
\label{eq:residualdiffeo}
\end{equation}
and
arbitrary shifts in the radial coordinate
(depending on $x$),
\begin{equation}
    r \to \bar r \equiv r + \delta\lambda(x) \,.
\label{eq:rshift}
\end{equation}
The diffeomorphism freedom (\ref{eq:residualdiffeo}) may be used to
transform the boundary one-form $\widetilde w$ into a standard form such as
\begin{equation}
    w_\mu(x) = -\delta^{\,0}_\mu \,.
\label{eq:stdu}
\end{equation}
This simple choice will be used in the examples presented
in section \ref{sec:examples}.
Alternatively, one could choose to require that the boundary one-form
$\widetilde w$ coincide with the flow field $\widetilde u$ which will
(eventually) be extracted from the boundary stress-energy tensor
via eq.~(\ref{eq:flowfield}).
Circumstances in which this may be desirable will be 
discussed in the next subsection.

\subsection	{Boundary metric and asymptotic behavior}

We are interested in solutions to Einstein's equations for which the
boundary geometry is flat Minkowski space.
Using the ansatz (\ref{eq:ansatz1}),
such solutions may be expanded, asymptotically,
in inverse powers of $r$,
\begin{equation}
    g_{\mu\nu}(x,r) \sim h_{\mu\nu}(x)
    + \sum_{n=1}^\infty \> g_{\mu\nu}^{(n)}(x) \, r^{-n} \,.
\label{eq:EFexpansion}
\end{equation}
The leading term $h_{\mu\nu}(x)$ is the $D$ dimensional boundary metric.
This equals the $r \to \infty$ limit of the
induced metric obtained by restricting the $D{+}1$ dimensional metric
(\ref{eq:ansatz1}) to $r = \mathrm{const.}$ slices,
$
    \left. ds^2 \right|_{r = \mathrm{const.}}
    =
    (r^2/L^2) \, g_{\mu\nu}(x,r) \, dx^\mu dx^\nu
$,
after rescaling to remove the overall $r^2/L^2$ factor.
The order-$D$ coefficient $g_{\mu\nu}^{(D)}$
in expansion (\ref{eq:EFexpansion})
cannot be determined solely by a
near-boundary analysis;
the value of this coefficient
(which depends on the solution throughout the bulk)
determines the boundary stress-energy tensor in
a manner similar to the Fefferman-Graham case.
With $w^\alpha \equiv h^{\alpha\beta} \, w_\beta$, one finds
\begin{equation}
    \langle \widehat T_{\mu\nu} \rangle
    =
    g_{\mu\nu}^{(D)}
    +
    D^{-1} \, w^\alpha g_{\alpha\beta}^{(D)} w^\beta \> h_{\mu\nu} 
    \,.
\label{eq:Tmunu}
\end{equation}

The boundary metric $h_{\mu\nu}$
may be chosen to equal the standard Minkowski metric,
\begin{equation}
    h_{\mu\nu}(x) = \eta_{\mu\nu} \,.
\label{eq:leadingasymp}
\end{equation}
But demanding a flat boundary geometry
does not obligate one to use Cartesian Minkowski space coordinates.
Use of the boundary metric (\ref{eq:leadingasymp}) represents
a further, arbitrary choice of coordinates on the boundary geometry.
Alternatively, one may choose to describe Minkowski space using some set of
coordinates $\{ x^\mu \}$ which are non-trivially (and non-linearly) related
to a set of Cartesian Minkowski coordinates $\{ y^\alpha \}$, so that
\begin{equation}
    h_{\mu\nu}(x) =
    \frac {\partial y^\alpha(x)}{\partial x^\mu} \,
    \frac {\partial y^\beta(x)}{\partial x^\nu} \>
    \eta_{\alpha\beta} \,.
\label{eq:bdrymetric}
\end{equation}

For some problems, the standard choice (\ref{eq:leadingasymp})
of Minkowski boundary metric is sufficient. 
This will be the case for the specific examples presented in subsequent
sections.
For other problems, exploiting the freedom of using non-Cartesian boundary
coordinates, with corresponding boundary metric (\ref{eq:bdrymetric}),
is helpful.
This is true, for example, in problems involving
cylindrical or spherical symmetry in the dual field theory, 
where it is natural to use boundary coordinates adapted to
that symmetry.

We believe that exploiting the freedom to choose non-Cartesian boundary
coordinates will also be helpful in problems involving highly relativistic
fluid flow with large gradients of flow velocity.
For such problems, it will undoubtedly be preferable to choose
the congruence of radial null geodesics underlying the 
ansatz (\ref{eq:ansatz1}) to involve geodesics describing infalling
matter which is at rest (or nearly at rest) in the
{\em local fluid rest frame} ---
not at rest with respect to some globally defined inertial Lorentz frame
which is necessarily divorced from any local physics of interest.
This implies that one would like to use the $D$-dimensional
diffeomorphism freedom (\ref{eq:residualdiffeo}) to set the
boundary one-form $\widetilde w$ appearing in the ansatz (\ref{eq:ansatz1})
equal to the flow field $\widetilde u$, as suggested earlier.%
\footnote
    {%
    Although some other choice will be needed in spacetime regions where
    the stress-energy tensor fails to satisfy the weak energy condition
    and the fluid flow field $\widetilde u$ is ill-defined.
    }
However, simplifying features in the
form of the resulting Einstein equations (discussed next)
are easier to exploit, numerically, if one uses
the diffeomorphism freedom (\ref{eq:residualdiffeo})
to transform the choice
\begin{equation}
    \widetilde w(x) = \widetilde u(x) \,,\qquad
    h_{\mu\nu}(x) = \eta_{\mu\nu} \,,
\end{equation}
to a formally equivalent description where the boundary one-form
$\widetilde w$ has the standard form (\ref{eq:stdu}) and the complexity of the
actual fluid flow is isolated in non-trivial boundary coordinates,
\begin{equation}
    w_\mu(x) = -\delta^{\,0}_\mu \,,\qquad
    h_{\mu\nu}(x) = 
    \tfrac {\partial y^\alpha(x)}{\partial x^\mu} \,
    \tfrac {\partial y^\beta(x)}{\partial x^\nu} \>
    \eta_{\alpha\beta} \,,
\end{equation}
for some choice of $y^\alpha(x)$.
This amounts to changing from an Eulerian to a Lagrangian description
of fluid mechanics.
The desired diffeomorphism is one for which $y^\alpha(x)$,
for fixed values of the spatial coordinates $\x \equiv \{ x^i \}$
and varying $x^0$,
gives the worldline of a fluid cell labeled by $\x$.
If one chooses $x^0$ to coincide with proper time along this worldline,
then the required diffeomorphism is one for which
\begin{equation}
    w^\alpha(x) = \frac {\partial y^\alpha(x)}{\partial x^0} \,.
\end{equation}

\subsection	{Horizons and IR cutoffs}

The radial direction in AdS is related, via holography,
to the energy scale in the dual conformal field theory.
Dynamics arbitrarily deep in the bulk correspond to arbitrarily
low energy processes in the quantum field theory.
With bounded resources, any numerical calculation 
can only be accurate over a finite dynamic range.
So it is inevitable that some form of high energy (UV)
and low energy (IR) cutoff will be necessary in any numerical calculation.
An effective UV cutoff is imposed by the discretization used when
solving differential equations; this will be discussed below.
Here, we focus on IR issues.

The Poincar\'e horizon of AdS$_{D+1}$ is the locus of events beyond
which no signal can reach any boundary observer.
Any infalling perturbation will distort the geometry and hence perturb
the Poincar\'e horizon.  
A perturbation with uniform (boundary) energy density can deform
the geometry to AdS-Schwarzschild (AdS-BH) form
\cite{Witten:1998zw},
describing a black brane embedded in asymptotically anti-de Sitter space.
This geometry has a non-compact planar event horizon, with an associated
temperature which is related to the radial position
of the horizon.
In more general cases of gravitational infall in asymptotically AdS space,
one should expect a time-dependent geometry which, at least at late times,
will resemble the AdS-BH solution in a local ``tubewise'' fashion
\cite{Bhattacharyya:2008jc}.

The essential point is that a non-compact event horizon, with the topology
of a plane, may be regarded as an effective IR cutoff.
From a holographic perspective, the energy scale of this IR cutoff
is set by the local temperature of the horizon.
Events beyond this horizon cannot affect any physics extracted by a
boundary observer.

In a numerical calculation of the evolving geometry,
one is free to excise the portion of spacetime beyond such an event horizon.
However, the location of the event horizon cannot be determined
without knowing the entire future spacetime geometry (because gravitational
infall arbitrarily far in the future can change which null congruence
is picked out as the event horizon).
Of more practical utility is the identification of an apparent horizon,
or outermost marginally trapped surface
which, if it exists, will lie inside the true event horizon.%
\footnote
    {%
    For more discussion of event horizons and apparent horizons
    see, for example, refs.~\cite{Wald:1984rg,Booth:2005qc}.
    Apparent horizons depend on the foliation of spacetime.
    We are exclusively concerned with apparent horizons on our
    $t = \mathrm{const.}$ null slices of the geometry.
    }

We will require initial data such that, at some initial time $t_0$,
there exists an apparent horizon at some radial position $r = \rh(t_0,\x)$.
And we will require that this horizon smoothly evolve into an apparent
horizon located at radial position $r = \rh(t,\x)$ on subsequent time slices.
Hence, we are assuming that there exists an apparent horizon
which, on every time slice, has a planar topology and whose radial position
is a smooth function of $\x$ and $t$.
The location of this apparent horizon will function as an IR cutoff
and will be the boundary of our computational domain.
The modification of initial data needed to create or adjust the location of
such an apparent horizon is simple: it corresponds, in the dual
field theory, to adding a small background energy density.
Explicit examples will be discussed in the context of our test cases below.

We will find that some of the fields in our metric ansatz grow,
in a power-law fashion, as one moves deeper into the bulk.
This can lead to increasingly large problems with numerical loss of precision,
which will be discussed in more detail below.
Such precision loss can be ameliorated by increasing the IR cutoff, or
in other words, choosing initial data which leads to larger values of the
apparent horizon radius.

The bottom line is that excising the geometry inside the horizon is not
only allowable, it is necessary to avoid numerical problems.
The location of the apparent horizon may be tuned by suitably adjusting
what, in the dual field theory, is a small background energy density.

This is an appropriate point at which to discuss the limits of applicability
of our methods.
We require that the metric ansatz (\ref{eq:ansatz}) provide good coordinates
throughout the region of spacetime between the boundary and an
apparent horizon at some radial position $r = \rh(t,\x)$.
This could potentially fail if:
(\emph{a}) some coordinate singularity develops in the spacetime
region outside the apparent horizon, or
(\emph{b}) an apparent horizon of the assumed form does not exist.

\begin{figure}
    \hfil
    \suck[scale=0.45]{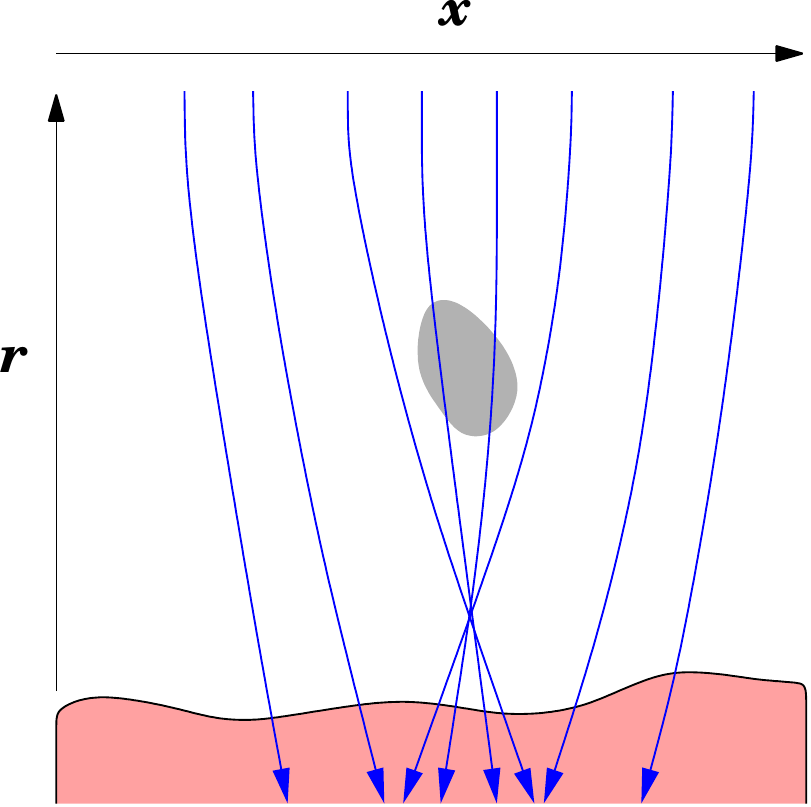}
    \hfil
    \hfil
    \hfil
    \suck[scale=0.45]{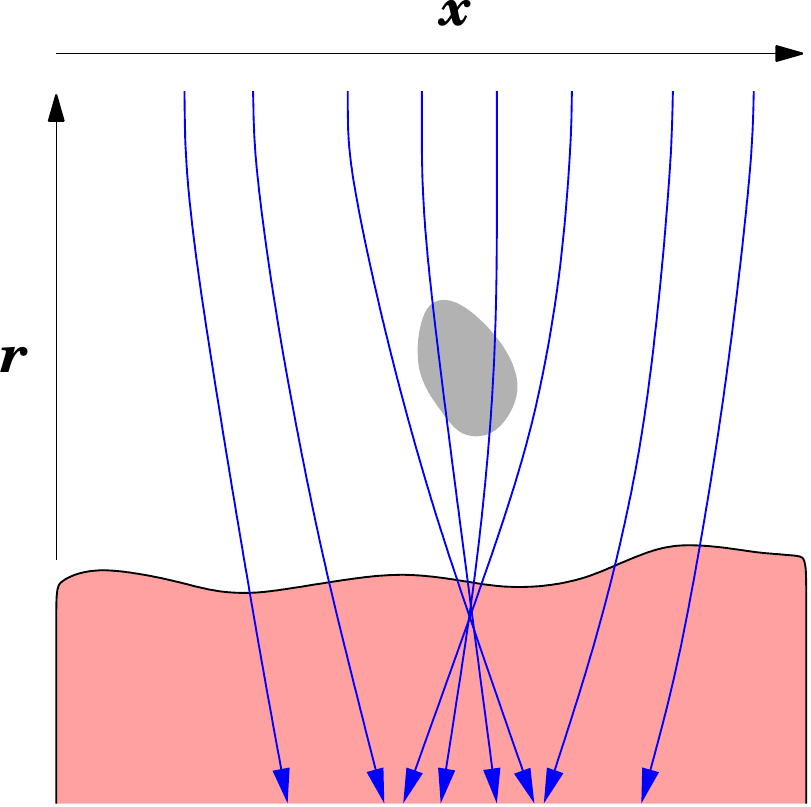}
    \hfil
\caption
    {%
    Focusing of null infalling radial geodesics and consequent formation
    of caustics.
    Only the radial direction and one spatial direction are shown.
    The grey shaded ``blob'' represents some perturbation in the geometry
    causing focusing of infalling geodesics.
    The shaded area at the bottom of each figure represents events
    behind the apparent horizon.
    Left panel: caustic formation outside the apparent horizon.
    Right panel: caustic hidden behind apparent horizon.
    }
\label{fig:caustics}
\end{figure}

Since our coordinates are directly tied to the congruence
of infalling radial null geodesics, possibility (\emph{a}) 
would mean some event is not uniquely identified by our coordinates
$(x,r)$, which label a particular infalling
radial geodesic (originating at point $x$ on the boundary),
together with an affine position $r$ along this geodesic.
This is precisely what happens when there is focusing of the geodesic
congruence, leading to intersections between differing geodesics. 
The boundary of the region where such intersections occur defines
a \emph{caustic}.
As illustrated schematically in fig.~\ref{fig:caustics}, a localized
perturbation will typically lead to geodesic focusing and
consequent formation of caustics.
Our method assumes that any such caustics lie outside the computational domain;
in other words, they must be hidden behind the apparent horizon.

\begin{figure}
    \suck[scale=0.40]{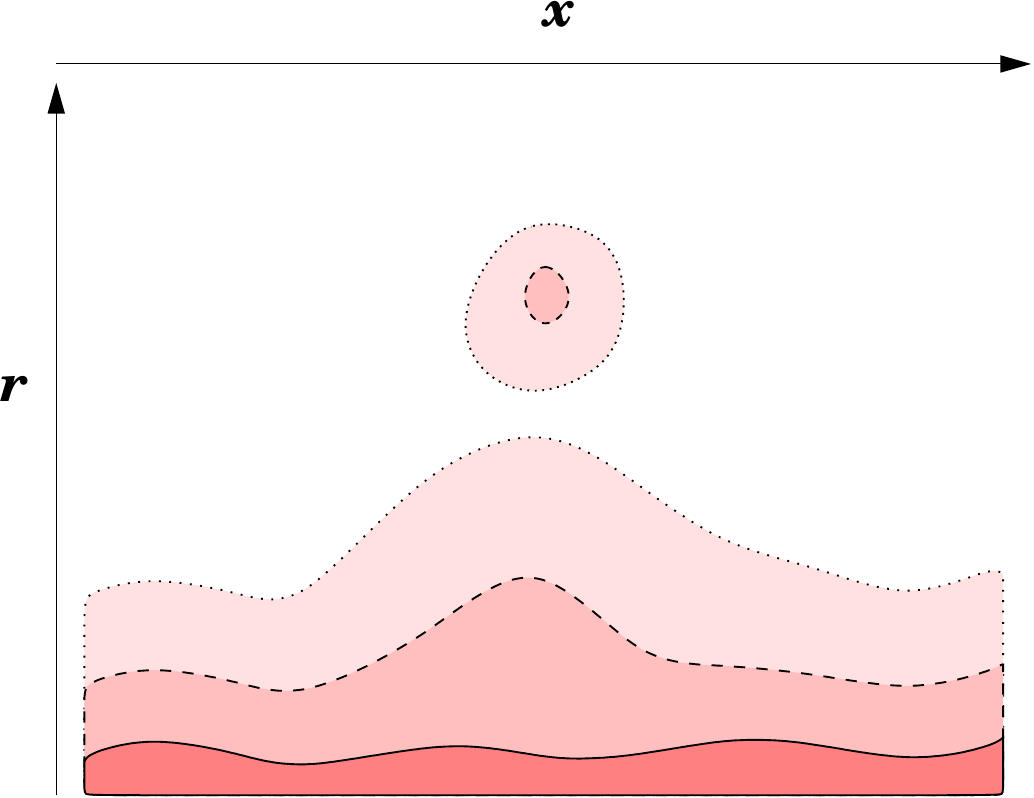} \hfill
    \suck[scale=0.40]{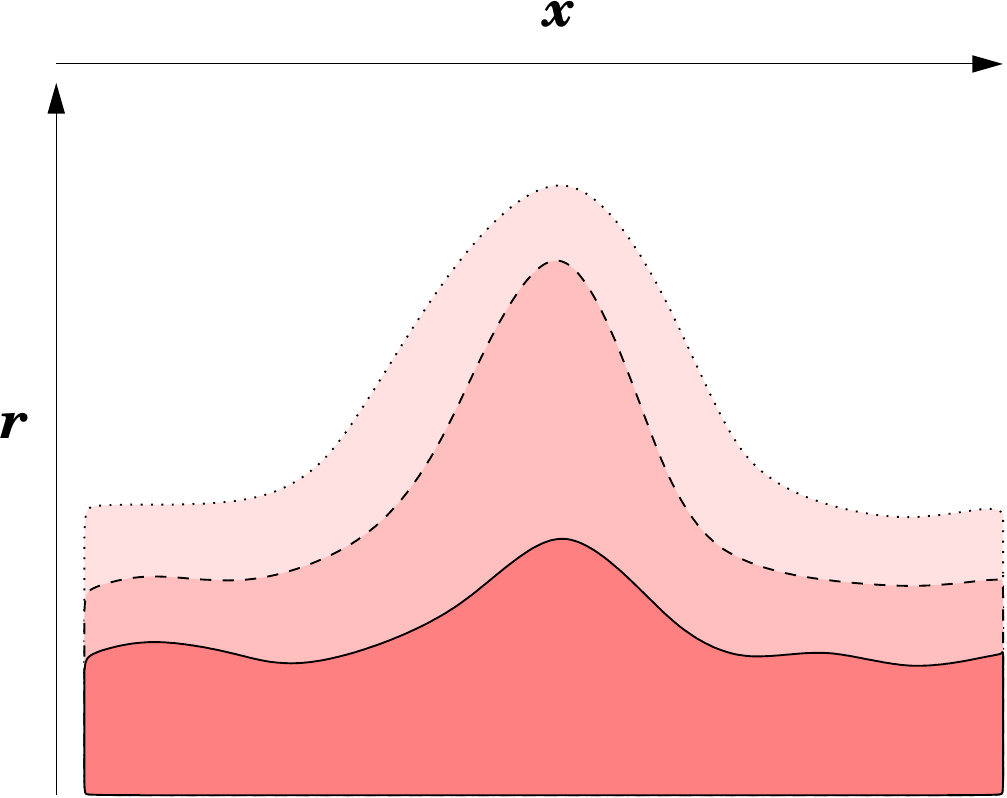}
\caption
    {%
    Possible forms of apparent horizon evolution induced by gravitational
    infall.
    Only the radial and one spatial direction are shown.
    The solid, dashed, and dotted lines,
    bounding progressively lighter shaded regions,
    show the position of the apparent horizon at three times
    $t_0$, $t_1$, and $t_2$, respectively, with $t_0 < t_1 < t_2$.
    Right panel: planar horizon topology at all times,
    to which our methods apply.
    Left panel: non-planar horizon topology (at times $t_1$ and $t_2$),
    requiring different computational methods.
    }
\label{fig:horizons}
\end{figure}

Possibility (\emph{b}), or non-existence of a planar topology apparent horizon,
can occur if the apparent horizon changes form discontinuously.
For example, gravitational infall could lead to the formation of a compact
trapped surface which is disconnected from a non-compact apparent horizon
lying deeper in the bulk.
This is illustrated schematically in fig.~\ref{fig:horizons}.
Of course, the formation of such a compact apparent horizon will likely
also lead to focusing and caustic formation in nearby 
infalling geodesics, so these two failure modes are interrelated.

In either case, the applicability of our methods should be restored if
the value of the IR cutoff is increased, i.e., if the position of the
non-compact planar topology horizon is pushed outward by increasing
the size of the background energy density in the dual theory,
as illustrated in the right panels of figs.~\ref{fig:caustics}
and \ref{fig:horizons}.
Consequently, for some problems, one should expect there to be a limit
on the maximum scale separation achievable between the IR cutoff
and the physics of interest.

Despite this limitation, we find that a large range of
interesting problems are amenable to solution using our methods.
In fact, we have yet to encounter difficulties with caustic formation
or horizon topology change in any gravitational infall problem
we have studied.
The underlying issue is one of relative scales.
The above described pathologies are likely to occur if one is studying
situations with variations in the geometry (or bulk sources) which are
spatially localized on a scale which is small compared to the gravitational
infall time associated with the apparent horizon.
In the dual field theory, this corresponds to states with spatial
structure on scales which are small compared to the length scale
(or inverse temperature) $\tau$ set by the energy density.
In a strongly coupled theory,
one expects fine spatial structure on scales small compared to $\tau$
to be washed out on the microscopic time scale $\tau$,
with negligible influence on the later evolution.
So, in practice, sources or initial conditions of most interest are those
with spatial size large compared to $\tau$.
Caustics generated by such sources should be hidden by an apparent
horizon whose infall time (or inverse local temperature)
is set by the microscopic scale $\tau$.

\subsection	{Einstein's equations}
\label		{sec:Einstein}

Turning Einstein's equations into a computable time-evolution scheme
necessitates a separation of these equations into those which
specify the dynamical evolution of the geometry,
and those which impose constraints on acceptable initial data
(or boundary data).
To exhibit key aspects of the explicit equations which emerge when
our metric ansatz (\ref{eq:ansatz1}) is inserted into
Einstein's equations,
\begin{equation}
    R^{MN} -\half R \, g^{MN} 
    + \Lambda \, g^{MN} = 0 \,,
\label{eq:Einstein}
\end{equation}
we use the diffeomorphism invariance (\ref{eq:residualdiffeo})
of the ansatz to specialize the boundary one-form $\widetilde w$
to the standard choice (\ref{eq:stdu}),
and rename the metric components in the ansatz,
\begin{align}
    \frac{r^2}{L^2} \, g_{00}(X) &\equiv -2A(X) \,,\quad
    \frac{r^2}{L^2} \, g_{0i}(X) \equiv -F_i(X) \,,\quad
    \frac{r^2}{L^2} \, g_{ij}(X) \equiv G_{ij}(X) \,,
\label{eq:names}
\end{align}
so that the line element (\ref{eq:ansatz1}) becomes%
\footnote
    {%
    We have redefined $A$ by a factor of two,
    and flipped the sign of $F_i$,
    relative to definitions in our earlier works
    \cite{CY:isotropize,CY:boostinvar,CY:shocks}.
    This change simplifies the forms (\ref{eq:AFtrans})--(\ref{eq:di})
    for the radial shift transformations and associated
    radially covariant derivatives presented below.
    Our metric ansatz (\ref{eq:ansatz}) is closely related
    to the null Bondi-Sachs form
    (see, for example, ref.~\cite{Winicour:1998tz}).
    The key difference is that our radial coordinate $r$ is
    an affine parameter along infalling geodesics, whereas
    the Bondi-Sachs metric uses a non-affine radial coordinate $\bar r$,
    chosen to make the determinant of the spatial metric
    a prescribed function of $\bar r$.
    }
\begin{equation}
    ds^2 =
    G_{ij}(X) \, dx^i \, dx^j
    + 2\, dt \left[ dr - A(X) \, dt - F_i(X) \, dx^i \right] .
\label{eq:ansatz}
\end{equation}
Here, $X = (x,r) \equiv (t,\x,r)$ denotes event coordinates in which
$t \equiv x^0$ is a null time coordinate,
$r$ remains the AdS radial coordinate, and
$\x \equiv \{ x_i \}$
denotes the remaining $D{-}1$ spatial coordinates.%
\footnote
    {%
    Using the symbol $v$ instead of $t$
    for the null time coordinate would be traditional,
    as this is customary in discussions of black hole
    geometries using Eddington-Finklestein (or Kruskal) coordinates.
    We choose not to do so, but readers should keep in mind that
    $t = \mathrm{const.}$ surfaces are null, not spacelike.
    Near the boundary our coordinates ($t,r$) are related to Fefferman-Graham
    coordinates ($x_{\rm FG}^0,\rho_{\rm FG}^{\phantom{|}}$)
    via $r = 1/\rho^{\phantom{|}}_{\rm FG}$ and
    $t = x_{\rm FG}^0 - \rho^{\phantom{|}}_{\rm FG}$.
    }
For later convenience, let
\begin{equation}
    \dim \equiv D{-}1
\label{eq:nudef}
\end{equation}
denote the
spatial dimensionality of the boundary theory.

Spatial ($\dim$ dimensional) diffeomorphisms are a
residual invariance of the form (\ref{eq:ansatz}) of the metric,
and transform the metric functions $G_{ij}$, $F_i$, and $A$
in the usual manner
(as components of a covariant tensor, one-form,
and scalar field, respectively).
As mentioned earlier,
arbitrary radial shifts,
\begin{equation}
    r \to \bar r = r + \delta\lambda(x) \,,
\label{eq:radialshift}
\end{equation}
also leave the form of the metric invariant.
Metric functions transform as
\begin{subequations}\label{eq:AFtrans}
\begin{align}
    A(x,r) &\to \bar A(x,\bar r) \equiv
    A(x,\bar r {-}\delta\lambda) + \,\partial_t\, \delta\lambda(x) \,,
\\
    F_i(x,r) &\to \bar F_i(x,\bar r) \equiv
    F_i(x,\bar r {-}\delta\lambda) + \partial_i\, \delta\lambda(x) \,,
\end{align}
\end{subequations}
while
$
    G_{ij}(x,r) \to \bar G_{ij}(x,\bar r) \equiv
    G_{ij}(x,\bar r {-}\delta\lambda)
$.
From the transformations (\ref{eq:AFtrans}), 
it is apparent that $A$ and $F_i$ function as the temporal
and spatial components of a ``radial shift'' gauge field.

In light of the spatial diffeomorphism invariance of the metric ansatz,
it must be possible to write explicit forms of the
resulting Einstein equations in a manner which is manifestly covariant
under spatial diffeomorphisms.
In addition, it is possible, and quite helpful, to write expressions
in a form which also makes invariance under the radial shift symmetry
manifest.
To do so, we introduce derivatives which transform covariantly
under \emph{both} radial shifts and spatial diffeomorphisms.
For the temporal derivative, this is accomplished by defining 
\begin{equation}
    d_+ \equiv \partial_t + A(X) \, \partial_r \,.
\label{eq:d+}
\end{equation}
As noted earlier, $\partial_r$ is a directional derivative
along {\em ingoing} radial null geodesics.
The $d_+$ derivative is the corresponding directional
derivative along the {\em outgoing} null geodesic
which passes through some event $X$ in the radial direction.

The analogous definition for spatial derivatives, acting
on (spatial) scalar functions, is
\begin{equation}
    d_i \equiv \partial_i + F_i(X) \, \partial_r \,.
\label{eq:di}
\end{equation}
Geometrically, these are derivatives along spacelike directions
which are orthogonal (at the event $X$) to the plane spanned
by tangents to ingoing and outgoing radial null geodesics.
In the derivatives (\ref{eq:d+}) and (\ref{eq:di}),
$A$ and $F_i$ act like gauge field components,
with $\partial_r$ the associated ``charge'' operator.
When acting on spatial tensor fields, one must augment
the derivative (\ref{eq:di}) with an affine connection,
which we denote by $\Gam^i{}_{jk}$,
to build a derivative which is also covariant under
spatial diffeomorphisms.
The required connection
is the usual Christoffel connection
associated with the spatial metric $G_{ij}$ except that,
to maintain radial shift invariance,
the spatial derivatives appearing in the definition of
the connection must be replaced by $d_i$ derivatives.
Hence,
\begin{subequations}\label{eq:Gammatilde}
\begin{align}
    \Gam^i_{jk}
    &\equiv
    \half \, G^{il}
    \left( d_k \, G_{lj} + d_j \, G_{lk} - d_l \, G_{jk} \right)
\\
    &=
    \half \, G^{il}
	\left( G_{lj,k} + G_{lk,j} - G_{jk,l}
	+ G'_{lj}\, F_k + G'_{lk}\, F_j - G'_{jk}\, F_l
    \right).
\end{align}
\end{subequations}
Here, and henceforth, we use primes to denote radial differentiation.

We denote by $\grad$
the resulting spatial and radially covariant derivative.
When displaying indices, we use a vertical bar ($|$),
instead of the usual semicolon, to indicate 
this modified covariant derivative.
So, for example, if $v$ is a spatial vector and
$\omega$ a spatial one-form, then
\begin{subequations}\label{eq:modderiv}
\begin{align}
    v^i{}_{|k}
    &\equiv ({\grad v})^i{}_k
    = d_k(v^i) + \Gam^i{}_{jk} \> v^j
    = v^i{}_{,k} + v'^{\,i} \, F_k + \Gam^i{}_{jk} \> v^j
    \,,\quad
\\
    \omega_{i|k}
    &\equiv ({\grad \omega})_{ik}
    = d_k(\omega_i) - \Gam^j{}_{ik} \> \omega_j
    = \omega_{i,k} + \omega'_i \, F_k - \Gam^j{}_{ik} \> \omega_j
    \,.
\end{align}
\end{subequations}
The modified covariant derivative is both metric compatible,
$G_{ij|k} = 0$, and torsion free, $\Gam^i{}_{jk} = \Gam^i{}_{kj}$.
Associated with our modified spatial covariant derivative is
a modified spatial Riemann curvature tensor,
$\widetilde R_{ijkl}$,
defined by the usual formula, but with our modified derivatives
replacing the usual derivatives.%
\footnote
    {%
    Explicitly,
    $
	\widetilde R^i{}_{jkl}
	\equiv
	d_k \Gam^i{}_{jl}
	-d_l \Gam^i{}_{jk}
	+ \Gam^i{}_{mk} \Gam^m{}_{jl}
	- \Gam^i{}_{ml} \Gam^m{}_{jk}
    $.
    The modified spatial Ricci tensor and scalar are given by
    the usual contractions,
    $\widetilde R_{jk} \equiv \widetilde R^i{}_{jik}$
    and $\widetilde R \equiv \widetilde R^k{}_k$.
    The modified Riemann tensor is antisymmetric
    in the last two indices, as usual,
    but need {\em not} be antisymmetric in the first two indices,
    or symmetric under $(ij)\leftrightarrow (kl)$ pair exchange.
    Instead,
    $
	\widetilde R_{ijkl}
	=
	\widehat R_{ijkl} + \Delta\widetilde R_{(ij)kl}
    $
    where
    $
	\widehat R_{ijkl}
    $
    obeys the usual symmetries
    [odd under $i\leftrightarrow j$ or $k\leftrightarrow l$,
    even under $(ij)\leftrightarrow(kl)$],
    while
    $
	\Delta\widetilde R_{ijkl}
	=
	\half G'_{ij} \, \Omega_{kl}
	+ \fourth
	    \left[
		 G'_{ik} \, \Omega_{jl}
		 - G'_{il} \, \Omega_{jk}
		 + G'_{jl} \, \Omega_{ik}
		 - G'_{jk} \, \Omega_{il}
	    \right]
    $.
    The two-form $\Omega$, defined in eq.~(\ref{eq:Omegadef}),
    is the ``magnetic'' field strength associated with the radial
    shift symmetry.
    The extra piece $\Delta \widetilde R_{ijkl}$ of the modified spatial
    Riemann tensor leads to a corresponding term
    $
	\Delta \widetilde R_{ij} =
	\fourth
	\left[ G'\cdot \Omega - \Omega\cdot G' + \Omega \, (\tr G') \right]
    $
    in the modified spatial Ricci tensor which is antisymmetric.
    \label{fn:modR}
    }

With these preliminaries in hand, we now examine the
resulting Einstein equations.
The $D{+}1$ dimensional set of equations (\ref{eq:Einstein}) must decompose
into one symmetric rank two spatial tensor equation, two 
spatial vector equations, and three spatial scalar equations.
After tedious work, one finds the following simple results.
The three scalar equations may be written as:%
\footnote
    {%
    We use a mixture of index-free notation
    (for simple factors like $\tr G'$, $F\cdot F$, or $\grad\cdot F'$),
    together with indices on more involved expressions;
    this makes the results most concise.
    Factors of the inverse spatial metric $G^{-1} = \| G^{ij} \|$
    are implicitly present in raised spatial indices.
    Be aware that raising of indices does {\em not} commute
    with radial or temporal differentiation.
    }
\begin{align}
    0 &= \tr \!\left( G'' - \half G'^{\,2}\right) ,
\label{eq:Sigrr}
\\[4pt]
    0 &= A''
    + \half \grad\cdot F'
    + \half F' \cdot F'
    + \half (\tr \, d_+G)'
    + \fourth\, \tr (G' \, d_+G)
    + {2\Lambda}/\dim
    \,,
\label{eq:Arr}
\\[4pt]
    0 &= \tr[d_+(d_+G)
    - A' \, (d_+G)
    - \half (d_+G)^2]
    + 2 \, \grad\cdot E
    + \half \tr (\Omega^2)
    \,.
\label{eq:Sigddot}
\end{align}
The dot products appearing here and in subsequent expressions
are defined using the spatial metric $G_{ij}$.
The spatial tensors $G'$ and $F'$ are defined as the radial derivatives of
{\em covariant} components,
$(G')_{ij} \equiv (G_{ij})'$ and $(F')_i \equiv (F_i)'$.
Likewise for $G''$, $d_+G$, $d_+F$, etc.
Hence, $G'^{\,i}{}_{\!j} = G^{ik} G'_{kj}$ and $F'^{\,i} = G^{ij}F'_j$.
Therefore
$\tr (G') \equiv G'^{\,i}{}_{\!i} = G^{ij}G'_{ji}$ and
$F\cdot F = F^i F_i = F_i\, G^{ij} F_j$.
In equation (\ref{eq:Sigddot}), the last term involves the square
of the two-form
\begin{equation}
    \Omega_{ij} \equiv
    F_{j|i} -F_{i|j} =
    F_{j,i} -F_{i,j} + F_i F'_j - F_j F'_i \,,
\label{eq:Omegadef}
\end{equation}
which is the spatial (or ``magnetic'')
part of the field strength associated with the 
radial shift symmetry.
The penultimate term involves the
corresponding time-space (or ``electric'')
part of the radial shift field strength,
\begin{equation}
    E_i \equiv d_+F_i - d_i \, A
    = F_{i,t} - A_{,i} + A \, F'_i - F_i \, A'
    \,.
\label{eq:Edef}
\end{equation}
The two vector equations are:
\begin{align}
    0 &= G_{ik} \big[G^{1/2} \, F'^{\,k} \big]' \, G^{-1/2}
    - G'^{\,k}{}_{i|k} + (\tr G')_{|i} \,,
\label{eq:Frr}
\\[6pt]
    0 &=
    d_+F'_i
    + (d_+G)^k{}_{i|k}
    - (\tr d_+G)_{|i}
    + \half (\tr \, d_+G) F'_i
    - 2 A'_{|i}
    - G'_i{}^k E_k
    + \Omega^k{}_{i|k}
    + F'_k \, \Omega^k{}_i
    \,,
\label{eq:Fdot}
\end{align}
with $G^{1/2} \equiv (\det G)^{1/2}$.
And the symmetric tensor equation is:
\begin{align}
    0 &= \Bigl\{
      G_{ik}  \big[G^{1/4} (d_+G)^k_{\;j} \big]' \, G^{-1/4}
    + \fourth G'_{ij} \, \tr(d_+G) 
    - \widetilde R_{ij}
    + \tfrac 2\dim \, \Lambda \, G_{ij}
    + F'_{\,i|j}
    + \half F'_i F'_j
    \Bigr\}
    + (i \leftrightarrow j) \,,
\label{eq:Gdot}
\end{align}
with $\widetilde R_{ij}$ the modified spatial Ricci tensor.
The trace of this equation separates from the traceless part, and reads
\begin{align}
    0 &= 
      \big[G^{1/2} \, \tr (d_+G) \big]' \, G^{-1/2}
    - \widetilde R + 2\Lambda
    + \grad \cdot F'
    + \half F' \cdot F'
    \,,
\label{eq:Sigdot}
\end{align}
with $\widetilde R$ the modified spatial Ricci scalar.
Every term appearing in eqs.~(\ref{eq:Sigrr})--(\ref{eq:Sigddot})
and (\ref{eq:Frr})--(\ref{eq:Sigdot})
is invariant under the radial shift symmetry.

\subsection	{Propagating fields, auxiliary fields, and constraints}
\label		{sec:structure}

To elucidate the structure of equations (\ref{eq:Sigrr})--(\ref{eq:Sigdot})
it is helpful to write them in a more schematic form after
extracting an overall scale factor $\Sigma$ from
the spatial metric $G_{ij}$. Let
\begin{equation}
    G_{ij}(X) = \Sigma(X)^2 \> \ghat_{ij}(X) \,,
\label{eq:Ghat}
\end{equation}
with the rescaled metric $\ghat \equiv \| \ghat_{ij} \|$
defined to have unit determinant,%
\footnote
    {%
    The spatial scale factor $\Sigma$ must be non-zero throughout the computational domain,
    as any zero in $\Sigma$ implies a coordinate singularity at which the metric degenerates.
    The determinant of the spatial metric (\ref{eq:Ghat}) coincides (up to a sign) with the
    determinant of the complete bulk metric (\ref{eq:ansatz}),
    $
	\det \| G_{ij} \| = - \det \| g_{MN}\| = \Sigma^{2\nu}
    $.
    }
\begin{equation}
    \det \ghat(X) = 1 \,.
\label{eq:unitdet}
\end{equation}
Equations (\ref{eq:Sigrr}), (\ref{eq:Frr}), and (\ref{eq:Arr}) are
linear second order radial ordinary differential equations (ODEs) for
$\Sigma$, $F$, and $A$, respectively, having the forms%
\footnote
    {%
    To convert eq.~(\ref{eq:Sigrr}) to the form (\ref{eq:Sigeqn}),
    note that $\det \hat g =1$ implies that $\tr(\hat g') = 0$ and
    $\tr(\hat g'') = \tr(\hat g'^{\,2})$.
    Hence,
    $
    \tr(G') = 2\dim \, \Sigma'/\Sigma
    $,
    while
    $
    \tr(G'') =
	2\dim\left[ \Sigma''/\Sigma + (\Sigma'/\Sigma)^2 \right]
	+ \tr(\ghat'^{\,2})
    $
    and
    $
    \tr(G'^{\,2}) = 4\dim \, (\Sigma'/\Sigma)^2
	+ \tr(\ghat'^{\,2})
    $.
    The conversion of eq.~(\ref{eq:Sigddot}) to the form (\ref{eq:Sigddoteqn})
    below uses the analogous relations
    $
    \tr(d_+G) = 2\dim \, (d_+\Sigma)/\Sigma
    $,
    $
    \tr(d_+(d_+ G)) =
	2\dim\left[ (d_+(d_+\Sigma))/\Sigma + (d_+\Sigma)^2/\Sigma^2 \right]
	+ \tr((d_+ \ghat)^2)
    $,
    and
    $
    \tr((d_+ G)^2) = 4\dim \, (d_+\Sigma)^2/\Sigma^2
	+ \tr((d_+ \ghat)^2)
    $.
    }
\begin{equation}
    \left( \partial_r^2 + Q_\Sigma[\ghat] \right) \Sigma = 0 \,,
\label{eq:Sigeqn}
\end{equation}
\begin{equation}
    \left(
	\delta^j_i \, \partial_r^2
	+ P_F[\ghat,\Sigma]^j_i \, \partial_r
	+ Q_F[\ghat,\Sigma]^j_i
    \right) F_j
    =
    S_F[\ghat,\Sigma]_i \,,
\label{eq:Feqn}
\end{equation}
\begin{equation}
    \vphantom{\bigg|}
    \partial_r^2 \, A = S_A[\ghat,\Sigma,F,d_+\Sigma,d_+{\ghat}] \,.
\label{eq:Aeqn}
\end{equation}
The trace (\ref{eq:Sigdot}) and traceless parts of the tensor
equation (\ref{eq:Gdot}),
and the second vector equation (\ref{eq:Fdot}), 
are first order radial ODEs for the (modified) time derivatives
$d_+ \Sigma$, $d_+ \ghat$, and $d_+ F$, respectively,
with the schematic forms
\begin{align}
    \left(
	\partial_r + Q_{{d_+\Sigma}}[\Sigma]
    \right) d_+\Sigma
    &=
    S_{d_+\Sigma}[\ghat,\Sigma,F] \,,
\label{eq:Sigdoteqn}
\\[6pt]
    \left(
	\delta^k_{(i} \delta^l_{j)} \, \partial_r
	+ Q_{d_+ {\ghat}}[\ghat,\Sigma]^{kl}_{ij}
    \right) d_+\ghat_{kl}
    &=
    S_{d_+{\ghat}}[\ghat,\Sigma,F,d_+\Sigma]_{ij} \,,
\label{eq:gdoteqn}
\\[6pt]
    \left(
	\delta^j_i \, \partial_r + Q_{d_+ F}[\ghat,\Sigma]^j_i
    \right) d_+F_j
    &=
    S_{d_+ F}[\ghat,\Sigma,F,d_+\Sigma,d_+{\ghat},A]_i \,.
\label{eq:Fdoteqn}
\end{align}
The final scalar equation (\ref{eq:Sigddot}) directly expresses
the (modified) second time derivative of $\Sigma$
in terms of the fields $\ghat$, $\Sigma$, $F$, and $A$,
plus the first $d_+$ derivatives of $\Sigma$ and $\ghat$,
\begin{equation}
    d_+ (d_+\Sigma)
    = S_{d_+^2 \Sigma}[\ghat,\Sigma,F,d_+\Sigma,d_+{\ghat},A] \,.
\label{eq:Sigddoteqn}
\end{equation}
The coefficient functions appearing in the above linear operators are
\begin{subequations}
\begin{align}
    Q_\Sigma[\ghat] &\equiv
	    \tfrac 1{4\dim} \, \tr\bigl( \ghat'^{\,2} \bigr) ,
\\[6pt]
    P_F[\ghat,\Sigma]^j_i
	&\equiv
	-G'^{\;j}_i + \dim \, (\Sigma'/\Sigma) \, \delta^j_i \,,
\\[6pt]
    Q_F[\ghat,\Sigma]^j_i
	&\equiv
	-G''^{\;j}_i
	+ (G'^{\,2})^{\;j}_i 
	-\dim (\Sigma'/\Sigma) \, G'^{\;j}_i
	+ \tr (G'' {-} \half G'^{\,2})\, \delta_i{}^j
	\,,
\\[6pt]
    Q_{d_+\Sigma}[\Sigma] &\equiv (\dim{-}1) \, \Sigma'/\Sigma \,,
\\[6pt]
    Q_{d_+ {\ghat}}[\ghat,\Sigma]^{kl}_{ij}
	&\equiv
	-G'^{\;k}_{(i} \delta^l_{j)}
	+ \tfrac 1\dim\, G'^{kl} G_{ij}
	+ (2{+}\tfrac\dim 2)(\Sigma'/\Sigma) \,
	    (\delta^k_{(i} \delta^l_{j)} - \tfrac 1\dim G^{kl} G_{ij} )
	\,,
\\[6pt]
    Q_{d_+ F}[\ghat,\Sigma]^j_i
	&\equiv
	- G'^{\,j}_i \,.
\end{align}
\end{subequations}
The various source functions
$S_F[\ghat,\Sigma]$,
$S_{d_+\Sigma}[\ghat,\Sigma,F]$,
$S_{d_+{\ghat}}[\ghat,\Sigma,F,d_+\Sigma]$,
$S_A[\ghat,\Sigma,F,d_+\Sigma,d_+{\ghat}]$,
$S_{d_+ F}[\ghat,\Sigma,F,d_+\Sigma,d_+{\ghat},A]$ and
$S_{d_+^2 \Sigma}[\ghat,\Sigma,F,d_+\Sigma,d_+{\ghat},A]$
appearing in the inhomogeneous ODEs
(\ref{eq:Feqn})--(\ref{eq:Sigddoteqn})
depend only on the indicated fields
(and their radial and spatial derivatives).
Explicit forms of these source functions
may be easily extracted from
eqs.~(\ref{eq:Frr}),
(\ref{eq:Sigdot}),
(\ref{eq:Gdot}),
(\ref{eq:Arr}),
(\ref{eq:Fdot}) and
(\ref{eq:Sigddot}), respectively.

The function $A$ is an auxiliary field;
no time derivative of $A$ appears in any of the above equations.
One must integrate the second order radial ODE (\ref{eq:Aeqn})
on every time slice (after determining the fields appearing in
the source term for this equation) to find $A$.%
\footnote
    {%
    The specification of appropriate integration constants for this,
    and all the other, radial ODEs
    will be discussed in subsection~\ref{sec:strategy}.
    }

The first order radial ODEs
(\ref{eq:Sigdoteqn}),
(\ref{eq:gdoteqn}) and (\ref{eq:Fdoteqn})
determine the modified time derivatives
of $\Sigma$, $\ghat$, and $F$.
One may regard the functions $\Sigma$, $\ghat$, and $F$
as propagating fields, with
the second order ODEs (\ref{eq:Sigeqn}) and (\ref{eq:Feqn})
serving as constraints on initial data for $\Sigma$ and $F$.
If these constraints hold at one time, then the dynamical equations
(\ref{eq:Sigdoteqn})--(\ref{eq:Fdoteqn})
ensure that these constraints will be satisfied at all later times.

Alternatively, one may choose to regard $\Sigma$ and $F$ as
auxiliary fields which are determined on each time slice by integrating
the second order ODEs (\ref{eq:Sigeqn}) and (\ref{eq:Feqn})
(with appropriate boundary conditions).
These auxiliary field equations are completely local in time.
With this choice of perspective, only the rescaled spatial metric
$\ghat$ encodes propagating information.

The final equation (\ref{eq:Sigddoteqn}) for $d_+(d_+ \Sigma)$,
as well as eqs.~(\ref{eq:Sigdoteqn}) and (\ref{eq:Fdoteqn})
for $d_+\Sigma$ and $d_+F$,
may be viewed as boundary value constraints.
If these equations hold at one value of $r$, then the other equations ensure
that eqs.~(\ref{eq:Sigdoteqn}), (\ref{eq:Fdoteqn}) and (\ref{eq:Sigddoteqn})
hold at all values of $r$.
This follows from the Bianchi identities.
It is eqns.~(\ref{eq:Fdoteqn}) and (\ref{eq:Sigddoteqn})
which impose the condition (\ref{eq:conservation})
of boundary stress-energy conservation.

\subsection	{Residual gauge fixing}
\label		{sec:residualdiffeo}

The residual reparameterization freedom associated with
radial shifts (\ref{eq:radialshift}) is apparent in the
asymptotic near-boundary behavior of solutions to Einstein's
equations.
After choosing the boundary metric (\ref{eq:leadingasymp})
one finds the asymptotic behavior:%
\footnote
    {%
    These asymptotic expansions hold for $D > 2$.
    For $D=2$ (three-dimensional gravity),
    expansions in $1/r$ terminate.
    Exact solutions to Einstein's equations (\ref{eq:Einstein}),
    with a flat boundary metric (\ref{eq:leadingasymp}),
    are given by
    $\Sigma = r + \lambda$,
    $A = \half (r+\lambda)^2 - \partial_0 (\lambda + \chi)$,
    and $F = -\partial_1 (\lambda + \chi)$,
    with $\lambda = \lambda(x^0,x^1)$ completely arbitrary and
    $\chi = \chi(x^0,x^1)$ an arbitrary solution of the
    free wave equation, $\partial^2 \chi = 0$.
    }
\begin{subequations}
\begin{align}
    &A = \half (r{+}\lambda)^2 - \, \partial_t\lambda +
    a^{(D)} \, r^{2-D} + O(r^{1-D})\,,
&
    &F_i = -\partial_i \lambda + f^{(D)}_i \, r^{2-D} + O(r^{1-D}) \,,
\\
    &\Sigma = r {+} \lambda + O(r^{1-2D})\,,
&
    &\ghat_{ij} = \delta_{ij} + \hat g_{ij}^{(D)} \, r^{-D}
    + O(r^{-D-1}) \,,
\\
    &d_+\Sigma = \half (r {+} \lambda)^2
    + a^{(D)} \, r^{2-D} + O(r^{1-D}) \,,
&
    &d_+\ghat_{ij} = -\tfrac D2 \, \hat g_{ij}^{(D)} \, r^{1-D} + O(r^{-D}) \,,
\end{align}\label{eq:asymp}%
\end{subequations}
where $\lambda = \lambda(x)$ is completely undetermined.
Here and henceforth we have, for convenience,
set the curvature scale $L = 1$.
(Factors of $L$ can be restored using dimensional analysis.)

As mentioned earlier, asymptotic analysis also cannot determine
the values of the subleading order-$D$ coefficients in the metric
which, after the renaming (\ref{eq:names}) of metric functions,
are the coefficients $a^{(D)}$, $f^{(D)}_i$, and $\hat g^{(D)}_{ij}$
(each of which is a function of $x$).
Reexpressing the result (\ref{eq:Tmunu}) for the stress-energy tensor
using our renamed metric functions, we have%
\footnote
    {%
    Because $\ghat$ has unit determinant,
    the sub-leading coefficient $\hat g_{ij}^{(D)}$
    is automatically traceless (as well as symmetric).
    So the full stress-energy tensor
    (\ref{eq:TMN}) of the dual field theory
    is automatically traceless as well.
    }
\begin{equation}
    \langle \widetilde T^{00} \rangle =
	-2 \tfrac {D-1}{D}
	\, a^{(D)} \,,\quad
    \langle \widetilde T^{0i} \rangle =
	f^{(D)}_i \,,\quad
    \langle \widetilde T^{ij} \rangle =
	\hat g^{(D)}_{ij} -\tfrac 2D \, a^{(D)} \, \delta_{ij} \,.
\label{eq:TMN}
\end{equation}
One must solve Einstein's equations throughout the bulk to determine
the coefficients $a^{(D)}$, $f_i^{(D)}$, and $\hat g_{ij}^{(D)}$;
our procedure for doing so will be discussed in the next subsection.
But $\lambda(x)$ is determined by fiat --- one must simply adopt
some scheme for fixing~$\lambda$.

One seemingly natural approach is to demand that $\lambda$ vanish
identically.
That is, one could require that $\Sigma(x,r) - r$ vanish,
for all $x$, as $r \to \infty$.
However, this turns out to be a bad choice as it leads to apparent horizons
whose radial positions vary rapidly with $x$.
Such variation causes greater difficulty with numerical loss of precision
due to cancellations between terms which grow large deep in the bulk.
And it can lead to situations where the radial coordinate $r$ decreases
to zero and turns negative before the apparent (or Poincar\'e) horizon
is reached --- which is a nuisance since it makes the inverted radial
coordinate $u \equiv 1/r$, which is otherwise convenient for numerical work,
singular within the computational domain of interest.

A much preferable choice is to use the residual reparameterization freedom
to put the apparent horizon at a fixed radial position,
\begin{equation}
    \rh(x) = \rhbar
\label{eq:fixedhorizon}
\end{equation}
for all $x$.
This choice makes the computational domain a simple rectangular region.
If the surface $r = \rhbar$ is an apparent horizon,
then an 
outgoing null geodesic congruence, normal to the surface and restricted
to a $t = \mathrm{const.}$ slice, will have vanishing expansion
\cite{Wald:1984rg,Booth:2005qc}.
This translates, in our metric ansatz,
to a condition on $d_+\Sigma$ at the apparent horizon.%
\footnote
    {%
    Ref.~\cite{Poisson} has a particularly nice treatment of
    null congruences.
    The congruence may be defined as
    $k_\alpha(x) = \mu(x) \phi(x)_{,\alpha}$
    where, within the time-slice of interest,
    the surface $\phi(x) = C$ for some value of the constant $C$
    will define the apparent horizon.
    Requiring that $k$ be null fixes the time derivative
    $\partial_t \phi$ in terms of spatial derivatives of $\phi$.
    Requiring the congruence to satisfy the (affinely parameterized)
    geodesic equation
    $k^\alpha k_{\beta;\alpha}=0$
    determines the time derivative
    of the multiplier function $\mu$ in terms of its spatial derivatives.
    Given these time derivatives, one may then compute
    the expansion via $\theta = \nabla \cdot k$.
    Demanding that the result vanish on the surface $\phi = C$
    gives the condition that this surface be an apparent horizon.
    eq.~(\ref{eq:Sigdothor}) is the result of specializing this
    condition to the case $\phi = r$,
    so that the surface under consideration
    lies at a fixed radial position.
    }
One finds:
\begin{equation}
    \left. d_+\Sigma \right|_{\rhbar}
    =
    S_{d_+\Sigma_{\rm h}} [\hat g, \Sigma, F] \,,
\label{eq:Sigdothor}
\end{equation}
with
\begin{equation}
    S_{d_+\Sigma_{\rm h}} [\hat g, \Sigma, F]
    \equiv
    - \half \, \Sigma' \, F^2
    - \tfrac 1\nu \, \Sigma \, \nabla \cdot F
    \,.
\end{equation}
and all fields evaluated at radial position $\rhbar$.%
\footnote
    {%
    This expression and the subsequent horizon stationarity condition
    (\ref{eq:apphoreq})
    are written using ordinary spatial covariant derivatives,
    not our modified derivatives (\ref{eq:modderiv}).
    These gauge fixing conditions are, by necessity, not invariant
    under radial shifts and do not have simpler forms when written
    using the modified derivative $\grad$.
    }

We want condition (\ref{eq:Sigdothor}) to hold at all times.
It is convenient to regard this as the combination of a constraint
on initial data 
(which is implemented by finding the radial shift
(\ref{eq:radialshift}) needed to satisfy condition (\ref{eq:Sigdothor})
at the initial time $t_0$),
together with the condition that the horizon position be time-independent,
$
    {\partial \rh}/{\partial t} = 0
$,
which requires that the time derivative of condition (\ref{eq:Sigdothor})
hold at all times,
\begin{equation}
    \left. \partial_t \, d_+\Sigma \right|_{\rhbar}
    =
    \partial_t \, S_{d_+\Sigma_{\rm h}} [\hat g, \Sigma, F] \,.
\label{eq:drh/dt}
\end{equation}
Evaluating this horizon stationarity condition
[and using eqs.~(\ref{eq:Sigdoteqn}),
(\ref{eq:gdoteqn}), and (\ref{eq:Sigddoteqn})
to simplify]
leads to a second order linear elliptic differential equation for $A$
on the horizon.
Explicitly, one finds
\begin{align}
    0 &= \nabla^2 A
	- \nabla A \cdot (F' - G' F)
	+ \half A \Bigl[
			- R^{(\dim)} + 2\Lambda
			+ \half (F' {-} G'F) \cdot (F' {-} G'F)
			- \nabla \cdot (F' {-} G'F)
		\Bigr]
    \nonumber\\ & \quad {}
	+ \half F \cdot F \Bigl[
			- \half \tr[(d_+G)']
			- (\nabla\cdot F)'
			- F_{i;j} G'^{\,ji}
			- \fourth (F \cdot F)' \, \tr G'
		    \Bigr]
	- \fourth (F_{i;j}{-}F_{j;i})(F^{j;i}{-}F^{i;j})
    \nonumber\\ & \quad {}
	- \fourth \tr [(d_+G)^2]
	- (d_+G)^{ji} F_{i;j}
	+ F \cdot \nabla^2 F
	- \half (F'-G'F) \cdot \nabla(F\cdot F)	
	\;\Bigr|_{r=\rh}\,,
\label{eq:apphoreq}
\end{align}
with $R^{(\dim)}$ the spatial Ricci scalar.

\subsection	{Integration strategy}
\label		{sec:strategy}

The set of equations (\ref{eq:Sigeqn})--(\ref{eq:Sigddoteqn})
have a remarkably convenient nested structure, which permits
a simple and efficient integration strategy.  

On some given time slice $t_0$,
eq.~(\ref{eq:Sigeqn}) is a linear (in $\Sigma$) second order
radial ODE which may be integrated to determine $\Sigma(t_0,\x,r)$,
provided $\hat g$ is already known on the time slice $t_0$.
Linearly independent homogeneous solutions behave as $r^1$ and $r^0$
as $r\to\infty$.
Consequently, the two needed integration constants may be fixed using
the leading and first sub-leading terms in the asymptotic behavior,
$
    \Sigma \sim r + \lambda + \cdots
$
[c.f. eq.~(\ref{eq:asymp}b)].
However, this implies that $\lambda(t_0,\x)$ must be known,
in addition to $\hat g_{ij}(t_0,\x,r)$,
to determine $\Sigma(t_0,\x,r)$.

Once $\Sigma$ and $\hat g$ are known at time $t_0$,
the set (\ref{eq:Feqn}) of second order radial ODEs can be
integrated to determine the $D{-}1$ components $F_i(t_0,\x,r)$.%
\footnote
    {%
    In addition to the manifest dependence on $F'$ in the first
    terms of equation (\ref{eq:Frr}),
    the second and third terms in the equation generate,
    through the modified covariant derivatives,
    terms which depend linearly on $F$.
    To solve for $F$, it is convenient to use the equivalent form
    (\ref{eq:uglyFrrT}) which uses ordinary covariant derivatives.
    In the absence of bulk sources,
    one may decouple the equations for different components of 
    $F$ by integrating first to find $G^{1/2} \, G_{ik} (F^k)'$,
    extracting $(F^k)'$, and then re-integrating to find the
    contravariant components of $F$.
    }
Linearly independent homogeneous solutions behave as $r^2$ and $r^{2-D}$
as $r\to\infty$.
Consequently,
the needed integration constants may once again be fixed 
from the leading and first sub-leading terms in the asymptotic
behavior,
$
    F_i \sim -\partial_i \lambda + f_i^{(D)} \, r^{2-D} + \cdots
$
[c.f.~eq.~(\ref{eq:asymp}a)].
This assigns a vanishing coefficient to the $r^2$ homogeneous solution,
and a specified coefficient $f_i^{(D)}$ to the other homogeneous solution.
Hence, in addition to $\hat g$ and $\lambda$ at time $t_0$,
one must also know the subleading coefficient $f_i^{(D)}(t_0,\x)$
before integrating the $F$ equations; how to accomplish
this will be discussed momentarily.

Next up is eq.~(\ref{eq:Sigdoteqn}), which is a
first order linear radial ODE for $d_+\Sigma$,
whose coefficients and source term depend on the already-determined
values of $\hat g$, $\Sigma$, and $F$ at time $t_0$.
Note that, with time derivatives rewritten in terms of $d_+$,
this equation has no explicit dependence on $A$.
The homogeneous solution behaves as $r^{2-D}$ as $r \to \infty$,
so the single needed integration constant may be fixed by the coefficient
of the sub-leading asymptotic term,
$
    d_+\Sigma \sim \half (r{+}\lambda)^2 + a^{(D)} \, r^{2-D}
    + \cdots
$,
[c.f.~eq.~(\ref{eq:asymp}c)].
Hence, in addition to $\hat g$, $\lambda$, and $f_i^{(D)}$ at time $t_0$,
we also require that the subleading coefficient $a^{(D)}(t_0,\x)$
be known before integrating the $d_+\Sigma$ equation; how to accomplish
this will also be discussed momentarily.

Now consider eq.~(\ref{eq:gdoteqn}).
This is, in general, a set of coupled first order
linear radial ODEs for the $\half D(D{-}1)-1$ components of the traceless
symmetric spatial tensor $d_+\hat g_{ij}$.
The coefficients and source terms of these equations
again depend only on the already-determined
values of $\hat g$, $\Sigma$, $F$, and $d_+\Sigma$ at time $t_0$.
The homogeneous solution to this equation behaves as $r^{(1-D)/2}$
as $r \to \infty$;
the needed integration constant just corresponds to demanding the
absence of any such homogeneous piece, so that
$
    d_+\hat g_{ij} = O(r^{1-D})
$
as $r \to \infty$.

Next turn to eq.~(\ref{eq:Aeqn}), which is a trivial second-order
linear radial ODE for $A$, with a source term depending on the
already-determined values of $\hat g$, $\Sigma$, $F$,
$d_+\Sigma$, and $d_+\hat g$.
Linearly independent homogeneous solutions are $r^1$ and $r^0$.
The asymptotic behavior
$
    A \sim \half(r {+} \lambda)^2 -\, \partial_t \lambda + \cdots
$
[c.f.~eq.~(\ref{eq:asymp}a)],
shows that knowledge of $\lambda$ and $\partial_t \lambda$
(at time $t_0$) determines these integration constants.
If one fixes the residual reparameterization invariance
(\ref{eq:radialshift})
by choosing, a-priori, the value of $\lambda$
as a function of both $t$ and $\x$, then this choice determines
the two constants needed to integrate eq.~(\ref{eq:Aeqn})
for $A$.

However, as discussed in section \ref{sec:residualdiffeo},
it is preferable to adjust $\lambda$ dynamically so as to fix the
radial position of the apparent horizon,
which forms the IR boundary of the computational domain.
As described above,
the horizon position invariance condition, $d\rh/dt = 0$,
reduces to the second order linear elliptic differential equation
(\ref{eq:apphoreq}) for $A$ on the horizon.
The functions (evaluated at a given time $t_0$ and radius $\rhbar$)
appearing in the coefficients and source term of
this linear elliptic PDE have all been determined in earlier steps
of the integration procedure.
Solving the linear PDE (\ref{eq:apphoreq})
(with appropriate boundary conditions in the spatial directions)
will determine the IR boundary value $A(t_0,\x,\rhbar)$.
This provides one of the two integration constants needed to 
integrate eq.~(\ref{eq:Aeqn}) and determine $A$ everywhere on the
$t = t_0$ time slice;
the second integration constant is fixed by the
asymptotic behavior $A \sim \half r^2 + \lambda r +O(1)$ as $r \to \infty$,
showing that $\lambda$ is the coefficient of the term linear in $r$.

After the determination of $A$ in this manner,
using the horizon-invariance condition,
one may extract the time derivative of $\lambda$ from the
subleading asymptotic behavior (\ref{eq:asymp}a) of $A$.
The needed term may be isolated most conveniently
by combining $A$ with $d_+\Sigma$, as
\begin{equation}
\label{eq:lambdaevo}
    \partial_t \lambda = \lim_{r \to \infty} \left( d_+\Sigma - A \right) ,
\end{equation}
with corrections to the limit vanishing as $O(r^{2-2D})$.
The determination of $A$ also allows one to extract $t$-derivatives
from $d_+$ derivatives so that,
on the $t = t_0$ time slice, one can now evaluate
\begin{equation}
\label{eq:timeevo}
    \partial_t \, \hat g_{ij}
    =
    d_+ \hat g_{ij} - A \, \partial_r \, \hat g_{ij} \,.
\end{equation}

To recap,
having started at time $t = t_0$ with $\hat g_{ij}$, $\lambda$,
$f_i^{(D)}$, and $a^{(D)}$,
the above procedure allows one to evaluate
the time derivatives of $\hat g_{ij}$ and $\lambda$.
Using a suitable integration method (such as fourth-order Runge-Kutta),
these time derivatives
provide the information needed to determine
$\hat g_{ij}$ and $\lambda$ on the next time slice
at $t = t_0 + \epsilon$,
up to an error vanishing as a power of the time step $\epsilon$
(e.g., $\epsilon^5$ for fourth-order Runge-Kutta).
Appropriate choices for time integration methods are discussed below
in subsection \ref{sec:timeint}.

However, before one can repeat the entire procedure above on the
$t_0 + \epsilon$ time slice, 
one must also evaluate the time derivatives of asymptotic coefficients
$f_i^{(D)}$ and $a^{(D)}$,
as these are needed to determine the values of
these coefficients on the subsequent time slice.
The time derivative of $f_i^{(D)}$
could be obtained by integrating the linear radial
ODE (\ref{eq:Fdoteqn}) to find $d_+ F$, converting the $d_+$ derivative
to a $t$ derivative, and then extracting $\partial_t f_i^{(D)}$ from the
first subleading term in the large $r$ asymptotic behavior of
$\partial_t F$.
Likewise, $\partial_t \, a^{(D)}$ could be obtained by integrating
the final radial ODE (\ref{eq:Sigddoteqn}) to find $\partial_t \, d_+\Sigma$,
and then extracting $\partial_t \, a^{(D)}$ from its subleading asymptotic behavior.
However, there is a simpler, far more efficient approach: direct use
of boundary stress-energy conservation (\ref{eq:conservation}).
As indicated in eq.~(\ref{eq:TMN}),
up to a common overall factor,
$-\frac {2D-2}D \,a^{(D)}$ is the energy density
(and the trace of the spatial stress tensor),
$f_i^{(D)}$ are the components of the momentum density,
and $\hat g_{ij}^{(D)}$
is the traceless part of the spatial stress tensor.
Hence, the needed time derivatives of $f_i^{(D)}$ and $a^{(D)}$ are given by
\begin{equation}
\label{eq:bndevo}
    \partial_t \, a^{(D)}
    = \tfrac {D}{2D-2} \, \partial_i \, f_i^{(D)} \,,
\qquad
    \partial_t \, f_i^{(D)} =
    \tfrac 2D \, \partial_i \, a^{(D)}
    - \partial_j \, \hat g_{ji}^{(D)} \,,
\end{equation}
where all quantities on the right hand sides are already known
known on the $t_0$ time slice.
(The traceless stress coefficient $\hat g_{ij}^{(D)}$ must be extracted
from the leading large $r$ behavior of $\hat g_{ij}$.)
Given these time derivatives,
updated values for $a^{(D)}$ and $f_i^{(D)}$
on the next $t_0 + \epsilon$ time slice are computed using the same
time integration method employed for $\hat g_{ij}$ and $\lambda$.

This completes the series of steps needed to turn Einstein's equations
into an algorithm for evolving information from a given $t = t_0$ null slice
to a subsequent slice at $t_0 + \epsilon$.
It should be emphasized that although one is solving the highly non-linear
Einstein equations,
this approach breaks the central time-evolution process down into a
sequence of steps which only require solving the linear first or second
order radial ODEs (\ref{eq:Sigeqn})--(\ref{eq:gdoteqn}),
plus the linear elliptic horizon PDE (\ref{eq:apphoreq}).
The specific procedure described above is not, however, unique.
Instead of treating $\Sigma$ as an auxiliary field,
to be computed anew on each time slice using the Schrodinger-like
eq.~(\ref{eq:Sigeqn}),
as mentioned earlier one could choose to treat $\Sigma$ as a dynamical field
which is evolved by computing $d_+\Sigma$ and then extracting
$\partial_t \Sigma$.
Likewise, the vector $F$ could be treated as a dynamical field and evolved using
eq.~(\ref{eq:Fdoteqn}), instead of computing it as an auxiliary field
from the second order eq.~(\ref{eq:Feqn}).
One could fix the integration constant in eq.~(\ref{eq:Sigdoteqn})
for $d_+\Sigma$ using the planar horizon condition
(\ref{eq:Sigdothor}) directly on every time slice,
instead of using (and evolving) $a^{(D)}$ to fix the subleading large $r$
asymptotic behavior of $d_+\Sigma$.
These are just a few of the possibilities.

Different choices, while formally equivalent, have differing
sensitivities to discretization effects and lead to
algorithms with quite different numerical stability.
Our experience is that stability is improved by computing
auxiliary fields afresh on each time slice (instead of
dynamically evolving these fields), and by using boundary
stress-energy conservation to evolve the relevant
subleading asymptotic coefficients directly, as described 
in the above scheme.

\subsection	{Initial data}
\label		{rec:initialdata}

To start the integration procedure,
one must specify the spatial dependence of the asymptotic coefficients
$a^{(D)}(t_0,\x)$ and $f_i^{(D)}(t_0,\x)$
on some initial $t = t_0$ time slice.
And one must specify the radial and spatial dependence of the
rescaled spatial metric $\hat g_{ij}(t_0,\x,r)$.
The asymptotic behavior of $\hat g_{ij}$ (specifically the coefficient
$\hat g_{ij}^{(D)}$) determines the initial traceless stress tensor,
while $a^{(D)}$ and $f_i^{(D)}$
fix the initial energy and momentum density
[c.f., eq.~(\ref{eq:TMN})].
Changes in the radial dependence of $\hat g_{ij}$
(for some prescribed asymptotic behavior)
encode changes in multi-point correlations in the
dual field theory state, but do not affect one point expectation values
of operators local in time, evaluated at time $t_0$.
(Different choices for $\hat g_{ij}$ in the bulk,
on the initial slice, may of course alter one point expectation
values at later times.)

In practice, there are several options for selecting initial data.
One can choose to study ``incoming'' scattering states
which, at time $t_0$, contain well-separated excitations
that, if considered in isolation, would have simple known evolution.
Our case study of colliding planar shock waves in section
\ref{sec:shocks} is an example of this type.
Alternatively, one can start with a known static
(or stationary) geometry describing an equilibrium state in the
dual theory and then, after the initial time $t_0$, drive the
system out of equilibrium using time-dependent external sources.
This was the approach used in
refs.~\cite{CY:isotropize,CY:boostinvar}, where specified
time-dependent boundary geometries represent sources
coupled to $T^{\mu\nu}$.
Finally, one can simply make an arbitrary choice for the radial
dependence of $\hat g_{ij}$ on the initial time slice.
To a large extent, features in $\hat g_{ij}$ deep in the bulk
quickly disappear behind the horizon and have little influence
on the future geometry; they reflect initial transients. 

Given some choice of initial data,
before proceeding with the integration strategy outlined above
one must first find the value of the radial shift $\lambda(t_0,\x)$
which leads to an apparent horizon at the desired location $r = \rhbar$.
This requires integrating eqs.~(\ref{eq:Sigeqn}),
(\ref{eq:Feqn}) and (\ref{eq:Sigdoteqn})
with $\lambda$ set to zero (or some other arbitrary choice),
to obtain provisional solutions for $\Sigma$, $F_i$ and $d_+\Sigma$
on the initial slice.
Using these functions,
one can locate the outermost value of $r$ (for each $\x$)
at which the apparent horizon condition (\ref{eq:Sigdothor})
is satisfied,
and then adjust $\lambda(\x)$, at time $t_0$, to shift this radial position to
the prescribed value.%
\footnote
    {%
    More precisely, one must use an iterative root-finding scheme,
    as the condition (\ref{eq:Sigdothor}) is satisfied \emph{when}
    there is an apparent horizon at $r = \rhbar$, but is
    not covariant under radial shifts.
    We use a simple Newton iteration procedure based on the value
    and first radial derivative of
    $d_+\Sigma - S_{d_+\Sigma_{\rm h}}[\ghat,\Sigma,F]$
    at $r = \rhbar$.
    }


\subsection	{Finite spatial volume}
\label		{sec:finitevol}

With finite computational resources,
one needs a finite computational domain in all directions,
including the $D{-}1$ spatial directions.%
\footnote
    {%
    Problems with translation symmetry in one or more spatial directions,
    such as our first two examples below, are
    trivial exceptions to this assertion.
    One needs a finite computation domain in all directions in which
    solutions of interest have non-trivial variation.
    }
One could make an $r$-independent change of variables
which would map the range of the spatial $\{ x_i \}$ coordinates
to a finite interval, while preserving the form of the metric
ansatz (\ref{eq:ansatz}).
Such $r$-independent transformations are part of the
residual diffeomorphism freedom.
However, we have not found such remapping to be desirable, as this leads
to equations which are singular and ill-behaved at the ends of the
spatial interval.

A simple alternative which does not degrade
numerical accuracy or stability is compactification of
the spatial directions.
We impose simple cubic periodic boundary conditions in
spatial directions, with period $L_s$.
This should be viewed as a complementary part of the IR cutoff
needed for computation.
This spatial compactification also dictates the appropriate boundary
conditions to use in solving the horizon invariance condition
(\ref{eq:apphoreq}), namely spatial periodicity of $A_{\rm h}$.

Of course, compactification of spatial directions can have
undesirable consequences.
In scattering problems, as outgoing excitations separate there
will be a limited time duration before the evolution is
polluted by ``wrap-around'' effects caused by the compactification.
If one is interested in exploring the uncompactified dynamics
for some time duration $\tau$, then one will generally need a
spatial compactification with size $L_s \ge c \, \tau$.

\subsection	{Field redefinitions}
\label		{sec:fieldredef}

For numerical work, it is helpful to make a change of variable
which maps the unbounded radial coordinate $r$ to a finite interval.
We just invert, and define
\begin{equation}
    u \equiv 1/r \,.
\end{equation}
In all the radial ODEs (\ref{eq:Sigeqn})--(\ref{eq:gdoteqn}),
the endpoint $u=0$ (or $r=\infty$) is a regular singular point.
As shown in eq.~(\ref{eq:asymp}), the metric functions $A$ and
$\Sigma$, as well as the time derivative $d_+\Sigma$,
diverge as $u \to 0$.
For numerical purposes,
it is very helpful to define subtracted functions in which
the (known) leading pieces which diverge as $u \to 0$ are removed,
and to rescale the subtracted
functions by appropriate powers of $u$ so that the
resulting functions vanish linearly,
or approach a constant, as $u \to 0$.
This diminishes the substantial loss of precision 
which can occur due to large cancellations
between different terms near $u = 0$.
Altogether, this has lead us to use the following redefined fields
in much of our numerical work:%
\footnote
    {%
    If one introduces an explicit parameterization for $\hat g_{ij}$
    which solves the unit determinant constraint,
    as we do below in the examples discussed in section~\ref{sec:examples},
    then the redefinitions (\ref{eq:redefs}) for $\hat g_{ij}$
    and $d_+\hat g_{ij}$
    are replaced by analogous rescaling of the individual functions
    parameterizing $\hat g_{ij}$ and their time derivatives.
    }
\begin{subequations}\label{eq:redefs}
\begin{align}
    \sigma(x,u) &\equiv \Sigma(x,1/u) - 1/u \,,
&
    \gamma_{ij}(x,u) &\equiv
    u^{1-D} \left[ \hat g_{ij}(x,1/u) - \delta_{ij} \right],
\label{eq:ghatredef}
\\
    a(x,u) &\equiv A(x,1/u) - \half\Sigma(x,1/u)^2 ,\!
&
    \dot\gamma_{ij}(x,u) &\equiv
    u^{2-D} \left[ d_+\hat g_{ij}(x,1/u) \right],
\\
    f_i(x,u) &\equiv F_i(x,1/u) \,,
&
    \dot\sigma(x,u) &\equiv
    u^{3-D} \left[ d_+\Sigma(x,1/u) - \half \, \Sigma(x,1/u)^2 \right] .
\end{align}
\end{subequations}
Writing $\Sigma^2$, and not just $(u^{-1}{+}\lambda)^2$,
in the subtraction terms for $A$ and $d_+\Sigma$ is an arbitrary choice
which makes no practical difference as $\Sigma$ coincides with
$u^{-1} {+} \lambda$
up to $O(u^{2D-1})$ terms which are negligible near the boundary.
The resulting $u \to 0$ boundary conditions for these redefined fields are:
\begin{subequations}
\begin{align}
    \sigma(x,u) &\to \lambda(x) \,, 
&
    \gamma_{ij}(x,u) &\sim u \, \hat g^{(D)}_{ij}(x) \,,
&
    a(x,u) &\mbox{ remains regular}\,,
\\
    \dot\sigma(x,u) &\sim u \, a^{(D)}(x) \,,
&
    \dot\gamma_{ij}(x,u) &\to 0 \,,
&
    f_i(x,u) &\sim -\partial_i \lambda + u^{D-2} f^{(D)}_i(x) \,.
\end{align}
\end{subequations}

\subsection	{Discretization}
\label		{sec:discretization}

To integrate the radial ODEs (\ref{eq:Sigeqn})--(\ref{eq:gdoteqn}),
and the horizon equation (\ref{eq:apphoreq}),
one must discretize the radial and spatial coordinates,
represent functions as finite arrays of function values on some
specified set of points,
and replace derivatives with suitable finite difference approximations.

Complications arise from the fact that $u = 0$ is a
singular point in all the radial ODEs.
Typical numerical ODE integrators
(involving short-range finite difference approximations)
do not tolerate such a singular point at the endpoint of the
computational interval.
One must introduce some finite separation scale $u_{\rm min}$,
use truncated (analytically derived) asymptotic expansions to
approximate functions in the near-boundary region $0 < u < u_{\rm min}$,
and only use numerical integration for $u > u_{\rm min}$.
To achieve accurate results one must carefully select $u_{\rm min}$,
and the order of the asymptotic expansion, so that the (in)accuracy
of the truncated asymptotic expansion is comparable to that of the
numerical integration.
As one uses progressively finer discretizations
(together with suitably matched improvements in the treatment of
the asymptotic region), 
the gain in accuracy scales, at best, as a power of the radial discretization,
error${} \sim (\Delta u)^k$, with the exponent $k$ depending on the
range of the chosen finite difference approximation.

For many differential equations,
substantially improved numerical accuracy
can be obtained by using {\em spectral} methods.%
\footnote
    {%
    For a good introduction to spectral methods, see ref.~\cite{Boyd:2001}.
    }
This approach entails the use of very long-range approximations to derivatives.
In essence, one represents functions as linear combinations of
a (truncated) set of basis functions,
and then exactly evaluates derivatives of these functions.
For functions periodic on an interval of length $L_s$,
the natural basis functions are complex exponentials,
$e^{i k_n x}$ with $k_n \equiv 2 \pi n / L_s$
(or the equivalent sines and cosines),
and the expansion is just a truncated Fourier series,
\begin{equation}
    f(x) = \sum_{n = -M}^M \> \alpha_n \, e^{i k_n x} \,.
\label{eq:spectral1}
\end{equation}
For aperiodic functions on an interval, 
convenient basis functions are Chebyshev polynomials,
$T_n(z) \equiv \cos(n \cos^{-1}z)$.
For functions on the interval $0 < u < 1$, the appropriate expansion reads
\begin{equation}
    g(u) = \sum_{n = 0}^{M} \> \alpha_n \> T_n(2u - 1) \,.
\label{eq:spectral2}
\end{equation}
This is nothing but a Fourier cosine series in the variable
$\theta \equiv \cos^{-1}(2u {-}1)$.

In so-called \emph{pseudospectral} or \emph{collocation} approaches,
one determines the expansion coefficients $\{ \alpha_n \}$ by 
inserting the truncated expansion (\ref{eq:spectral1})
or (\ref{eq:spectral2}) into the differential equation of interest
and demanding that the residual vanish exactly at a selected set of points
whose number matches the number of expansion coefficients.
For the Fourier series (\ref{eq:spectral1}), these grid
points should be equally spaced around the interval,
\begin{equation}
    x_m = L_s \left(\frac {m}{2M{+}1}\right) + \text{const.},
    \label{eq:fouriergrid}
\end{equation}
for $m = -M,{\cdots}, M$.
Knowledge of the expansion coefficients $\{ \alpha_n \}$ is completely
equivalent to knowledge of the function values $\{ f_m \}$
on the collocation grid points,
\begin{equation}
    f_m \equiv f(x_m) \,.
\end{equation}
For the Chebyshev case (\ref{eq:spectral2}),
appropriate grid points are given by the extrema and endpoints of
the $M$'th Chebyshev basis function.%
\footnote
    {%
    The Chebyshev grid points (\ref{eq:chebygrid}) are simply the image,
    under the mapping $u = \half (1 + \cos \theta)$, of equally spaced
    points in $\theta$ which would be appropriate for a Fourier cosine
    expansion.
    This choice of grid points, which include the interval endpoints,
    is most convenient when dealing with the imposition of boundary conditions.
    }
With the $[0,1]$ interval used in expansion~(\ref{eq:spectral2}),
these are
\begin{equation}
    u_m =  \half \left( 1 - \cos \frac{m\pi}{M} \right) ,
\label{eq:chebygrid}
\end{equation}
for $m = 0,{\cdots},M$.
Again,
knowledge of the expansion coefficients $\{ \alpha_n \}$ is completely
equivalent to knowledge of the function values $\{ g_m \equiv g(u_m) \}$
on the collocation grid points.
In practice, one uses these function values, plus interpolation formula,
which together exactly reproduce the truncated basis expansions
(\ref{eq:spectral1}) or (\ref{eq:spectral2}).%
\footnote
    {%
    In brief, for each truncated basis expansion,
    one reexpresses the expansion in the form
    $
	f(x) = \sum_m f_m \, C_m(x)
    $
    where the ``cardinal'' function $C_m(x)$ is the unique 
    function which ({\it i}) can be represented in terms of
    the same truncated basis expansion, and ({\it ii}) vanishes identically
    at all collocation grid points except the $m$'th point,
    where it equals unity [so that $C_m(x_n) = \delta_{mn}$].
    Cardinal functions are essentially regularized delta functions.
    See ref.~\cite{Boyd:2001} for more discussion including
    (in appendix E of that reference) explicit formulas for the
    appropriate cardinal functions for the Fourier expansion
    (\ref{eq:spectral1}) and the Chebyshev expansion (\ref{eq:spectral2}).
    }

For linear differential equations, spectral methods convert the
differential equation into a straightforward linear algebra problem
(albeit one with a dense coefficient matrix, not a banded or sparse matrix as
would be the case when using short-range finite difference approximations).
One key advantage of spectral methods is improved convergence.
For sufficiently well-behaved functions, accuracy improves
{\em exponentially} as the number of basis functions is increased.
A second advantage is that one can directly apply spectral methods
to differential equations with regular singular points,
as long as the specific solution of interest is well-behaved
at the singular point.
See ref.~\cite{Boyd:2001} for further detail.

We have found the use of (pseudo)spectral methods to be quite advantageous.
We use the Fourier series form (\ref{eq:spectral1}) to represent
functional dependence on periodic spatial coordinates, and the
Chebyshev form (\ref{eq:spectral2})
to represent functional dependence in the radial direction
(using the inverted radial variable $u$).%
\footnote
    {%
    Convergence of the spectral approximation (\ref{eq:spectral2}) with
    increasing order $M$ is naturally related to analytic properties
    of the functions under consideration.
    For problems involving a flat boundary geometry,
    all metric functions have expansions about $u = 0$
    in integer powers of $u$.
    After applying the field redefinitions discussed above, 
    expansions of our unknown functions only involve \emph{non-negative}
    powers of $u$.
    As noted in footnote \ref{fn:nonflat},
    for problems involving a non-flat boundary geometry,
    and an even dimension $D$,
    the near-boundary expansion necessarily includes logarithmic terms.
    One can still usefully apply spectral methods in this case,
    provided one subtracts these log terms (to reasonably high order)
    in the field redefinitions.
    Convergence of the spectral expansion will be
    degraded and non-exponential, but the performance of spectral
    methods can still be superior to traditional
    short range discretization methods.
    }

\subsection	{Time integrator}
\label		{sec:timeint}

As outlined above in Section \ref{sec:strategy}, in our evolution scheme 
we choose to evolve the minimal set of fields  $\Phi \equiv \{\hat g_{ij}, a^{(D)}, f_i^{(D)}, \lambda\}$.
Discretizing the geometry with $N_i$ grid points in the $x^i$ spatial direction
and $N_u$ points in the radial direction,
the fields in
$\Phi$ constitute a total of
$
     [ \half (\nu {-}1) \, N_u + 1 ](\nu {+} 2 )
    \prod_{i = 1}^\nu N_i
$ 
independent degrees of freedom.  
The time evolution portion of the spatially discretized Einstein equations
then takes the schematic form
\begin{equation}
\label{eq:odes}
\frac{d \Phi}{dt} = \mathcal F[ \Phi] \,.
\end{equation}
In other words, after discretizing the spatial and radial directions,
Einstein's equations reduce to a large system of simple,
first-order ODEs describing the time-evolution of $\Phi$.
Evaluating $\mathcal F[\Phi]$ is tantamount
to first solving the nested system of radial equations
(\ref{eq:Sigeqn})--(\ref{eq:gdoteqn}) 
to find  $d_+ \hat g_{ij}$,
then using eq.~(\ref{eq:timeevo}) to extract the discretized
field velocities $\partial_t \hat g_{ij}$ from $d_+ \hat g_{ij}$,
and finally using eqns.~(\ref{eq:lambdaevo}) and (\ref{eq:bndevo})
to compute $\partial_t a^{(D)}$,
$\partial_t f_i^{(D)}$, and $\partial_t \lambda$.

The first order system (\ref{eq:odes}) of simple ODEs
can integrated using a variety of numerical ODE solvers.
For simplicity, we limit our discussion to non-adaptive constant
time step schemes.%
\footnote
    {%
    Employing adaptive time-step schemes
    is clearly advantageous for some problems.
    However, all the issues discussed below, involving
    trade-offs between stability, accuracy, and computational
    efficiency, remain relevant for more complicated adaptive schemes.
    For more extensive discussion of numerical methods 
    for solving ODEs see ref.~\cite{Press:2007zz},
    or most any other book on scientific computing.
    }
We have used both implicit and explicit evolution schemes.
When using explicit time evolution schemes, 
stability of the resulting numerical evolution requires that one use
a suitably small time step.
The Courant-Friedrichs-Lewy (CFL) condition \cite{CFL}, required for stability,
imposes an upper limit on the time step.
For diffusive equations, the time step $\Delta t$
must satisfy $D \Delta t \ll \Delta x^2$, where 
$\Delta x$ is the minimum spatial grid spacing
and $D$ is the relevant diffusion constant.
For wave equations with unit propagation velocity,
the time step must satisfy $\Delta t \ll \Delta x$.  
(In general, the relevant condition is that
the numerical domain of dependence of new field values must
encompass the appropriate physical domain of dependence.)
Gravitational evolution in asymptotically AdS spacetime contains
both diffusive and propagating modes.
Diffusive gravitational modes are holographically related to
diffusive modes in the dual quantum field theory which describe
the spreading of (transverse) momentum density or other conserved
charge densities.
In the gravitational description, diffusive modes characterize
the behavior of conserved densities near the horizon,
as seen in the membrane paradigm \cite{Thorne:1986iy}
for horizon dynamics.
Consequently, diffusive behavior of gravitational modes predominantly
occurs in the spatial directions, and not in the radial direction.
Therefore, one CFL condition for the time step is
$D \Delta t \ll  \Delta x^2$.
(Near equilibrium, with some effective temperature $T$,
the diffusion constant $D = (2\pi T)^{-1}$
\cite{hep-th/0205052}.)
Gravitational waves can propagate in both radial and spatial directions,
and outward-going radial waves propagate near the boundary with
$\partial u/\partial t \simeq 1$.
Hence, the time step $\Delta t$ must also satisfy the propagating wave
CFL conditions $\Delta t \ll \Delta x$ and $\Delta t \ll \Delta u$.%
\footnote
    {%
    The grid spacing relevant for the radial CFL condition is
    the spacing near the middle of the non-uniform Chebyshev grid
    (\ref{eq:chebygrid}), or $\Delta u \simeq 1/N_u$
    if $N_u$ points are used in the radial discretization.
    The radial grid is much denser near the endpoints
    (where $\Delta u \sim 1/N_u^2$),
    and one might expect the finer near-endpoint spacing
    to mandate a far more stringent CFL bound on the timestep.
    Fortunately, this is not the case.
    Near the boundary, the amplitudes of propagating modes decrease
    rapidly [as $O(u^D)$], and do not perturb the boundary geometry.
    And near the horizon, the relevant propagation speed vanishes,
    $\partial u/\partial t \sim 1{-}u$, reflecting the asymptotic
    slowing down of infalling perturbations as seen by a boundary observer.
    }

In various applications,
we have obtained good results using a third order Adams-Bashforth method
as well as both implicit and explicit fourth order Runge-Kutta methods.
Which solver is best depends on available computing resources,
desired accuracy, and stability.
Adams-Bashforth methods have the advantage that only one evaluation
of $\mathcal F[\Phi]$ is needed per time step.
However, stability  can require a very small time step.
Explicit fourth-order Runge-Kutta methods
require four evaluations of $\mathcal F[\Phi]$ per time step,
but are more stable than Adams-Bashforth methods and allow use 
of a larger time step.  
Implicit fourth-order Runge-Kutta methods are much more stable
than explicit evolution.
Moreover, with implicit evolution 
the time step need not satisfy the CFL condition.
However, as we discuss below, implicit evolution requires many
evaluations of $\mathcal F[\Phi]$ per time step, which is costly.

Runge-Kutta methods, either implicit or explicit, require the
computation of a set of ``field velocities'' $ \{ k_i \}$,
$i = 1,2, {\cdots},M$,
defined by
\begin{equation}
\label{eq:rk4}
k_i \equiv
\mathcal F[ \Phi_n + \sum_{j=1}^M \> \alpha_{ij} \, k_{j} \, \Delta t ]\,,
\end{equation}
where $\Delta t$ is the time step, $\Phi_n \equiv \Phi(t_n)$,
and $\| \alpha_{ij} \|$ is an $M \times M$ matrix 
which determines the particular Runge-Kutta method.  
Once the set of $M$ velocities $\{k_i\}$ have been evaluated at time $t_n$,
the new fields at time $t _{n+1} \equiv t_n+ \Delta t$ are given by
\begin{equation}
\Phi_{n+1} \equiv \Phi_n + \Delta t \> \sum_{i=1}^M \> b_i \, k_i,
\end{equation}
for a set of coefficients $\{ b_i \}$ which again depend on
the particular Runge-Kutta method employed.

For explicit evolution, we use the classic fourth order
Runge-Kutta (RK4) method for which
\begin{equation}
\| \alpha_{ij} \|  = \left [ \begin{array}{@{\extracolsep{5pt}}cccc}
 \>0 & 0 & 0 & 0 \\
\>\frac 12 & 0 &0 & 0 \\
\>0 & \frac 12 & 0 & 0   \\
\>0 & 0 & 1  &0  \\
\end{array} \right],
\quad
\|b_i\| = \left[\tfrac 16,\,  \third,\, \third,\, \tfrac 16 \right] .
\end{equation}  
Since the matrix $\|\alpha_{ij}\|$ is lower triangular,
the field velocities $k_i$ can be computed sequentially,
with one evaluation of $\mathcal F[\Phi]$ for each $k_i$.
Hence, overall, RK4 requires
four evaluations of $\mathcal F[ \Phi]$ per time step.
Deviations of the numerical RK4 solution from the exact solution to
eq.~(\ref{eq:odes}) scale as $O\big((\Delta t)^5\big)$.

For implicit Runge-Kutta methods, the matrix $\| \alpha_{ij} \|$
is not lower triangular and the set of equations (\ref{eq:rk4})
implicitly define the different field velocities.
When using implicit evolution, we compute the $k_i$ iteratively.
Specifically, we start with a guess for the $k_i$
(\textit{e.g.}, the values of $k_i$ at the previous time step)
and compute
$k_i' \equiv \mathcal F[ \Phi_n + \alpha_{ij} \, k_{j} \, \Delta t]$.
After evaluating an error norm $\Delta \equiv |k_i - k_i'|$,
we set $k_i = k_i'$,
reevaluate
$k_i' \equiv \mathcal F[ \Phi_n + \alpha_{ij} \, k_{j} \, \Delta t]$,
and repeat the processes until $\Delta$ approaches zero
to within a chosen accuracy threshold.
For this very simpleminded iterative process,
the maximum time step is limited by convergence of the iterative scheme,
and not by stability of the actual numerical evolution in time.
The particular implicit Runge-Kutta method we employ is a
fourth-order method known as Lobatto IIIC for which
\begin{equation}  
\|\alpha_{ij}\| =
\left [ \begin{array}{ccc}
\phantom.\frac 16 & -\frac 13 & \phantom-\frac 16  \\[3pt]
\phantom.\frac 16 & \phantom-\frac 5{12} & -\frac 1{12}  \\[3pt]
\phantom.\frac 16 & \phantom-\frac 23 & \phantom-\frac 16  
\end{array} \right],
\quad
\| b_i \| = \left[\tfrac 16,\, \tfrac 23,\, \tfrac 16 \right].
\end{equation}  
This implicit RK method
is also a fourth-order scheme, with errors scaling as
$O\big((\Delta t)^5\big)$.

The third order Adams-Bashforth (AB3) method we employ 
uses prior values of $\mathcal F[\Phi]$ on the previous two time slices.
The fields on time slice $t_n$ are given by
\begin{equation}
\Phi_n = \Phi_{n - 1} + \Delta t \left \{ \strut
      \tfrac{23}{12}\, \mathcal F[ \Phi_{n - 1}]
    - \tfrac{4}{3}  \, \mathcal F[ \Phi_{n - 2}]
    + \tfrac{5}{12} \, \mathcal F[ \Phi_{n - 3}]
    \right \}.
\end{equation} 
With this third-order method, errors scale as $O\big((\Delta t)^4\big)$.
Since the AB3 method requires knowledge of
$\mathcal F[\Phi]$ on three consecutive time slices,
one must use some other scheme to compute $\Phi$ for the
first two steps.
This initialization can be performed using
the above-described explicit fourth-order Runge-Kutta method.

In general, when using non-adaptive integrators
we recommend either implicit RK4, or explicit RK4 with
suitably small time step, if computational resources (and patience) allows, 
and using AB3 if anything better is too slow.
For problems where characteristic time scales lengthen
as the evolution proceeds, use of an adaptive integrator
(such as one which incorporates and compares RK4 and RK5 steps)
is a reasonable choice.
For more discussion of performance, see section~\ref{sec:performance}.

\subsection	{Filtering}
\label{sec:filtering}

In addition to the CFL instabilities discussed above,
discretization of \textit{non-linear} PDEs can create spurious mechanisms,
absent in the continuum limit,
that cause artificial, unphysical growth in the amplitudes
of short wavelength modes.
This unphysical excitation of short wavelength modes
leads to a progressive loss of accuracy and may eventually cause complete
breakdown of the numerical evolution.

This problem is referred to as ``aliasing,'' or ``spectral blocking''.%
\footnote
    {%
    For more extensive discussion of spectral blocking
    see, for example, ref.~\cite{Boyd:2001}.
    }
To understand how short wavelength modes can become artificially excited
in discretizations of non-linear equations consider, for example,
the product of two functions, $f(x)g(x)$, defined on the periodic
interval $[-\pi,\pi]$,
when both $f$ and $g$ are approximated by truncated Fourier expansions
with $2M{+}1$ terms,
$f(x)= \sum_{k = -M}^{M}  \hat f_k \, e^{i k x}$ and
$g(x) = \sum_{k = -M}^{M}  \hat g_k \, e^{i k x}$.
The product $f(x)g(x)$ takes the form  
\begin{equation}
    f(x) \, g(x)
    = \sum_{k = -M}^{M} \sum_{q = -M}^{M}
	\hat f_k \, \hat g_q \, e^{i (k + q)x}
    = \sum_{p = -2M}^{2M} \hat h_p \, e^{i p x},
\end{equation}
with
$
    \hat h_p \equiv
    \sum_{k=-M}^M \sum_{q=-M}^M \delta_{k+q,p} \, \hat f_k \, \hat g_q
$.
The Fourier expansion of the product contains modes with wavenumber $p$ lying
outside the truncated domain $|p| \leq M$.  
When sampled on a grid with spacing $\Delta x = 2\pi/(2M{+}1)$
(which is the appropriate collocation grid for the truncated expansion
with $|p| \leq M$),
a mode with wavenumber $|p| > M$ is indistinguishable from the mode
with wavenumber $k \in [-M,M]$ for which $p{-}k$ is an integer multiple
of $2\pi M$.
One says that the high momentum mode with $|p| > M$ has been ``aliased''
to the low momentum mode with $k = p - 2 m M$ (for some integer $m$).

When computing the time evolution of non-linear PDEs,
spectral aliasing typically leads to continuing unphysical growth
in the amplitudes of modes near the $|p| = M$ UV cutoff.%
\footnote
    {%
    The power spectrum of Fourier coefficients of the exact solution
    will fall with
    increasing magnitude of the wavenumber for $|p| \geq M$,
    provided the solution is smooth on the scale of
    $\Delta x = 2\pi/(2M{+}1)$.
    Consequently, of the modes which suffer from aliasing,
    the largest amplitude modes are those just slightly above the
    UV cutoff at $|p| = M$,
    and these modes are aliased to modes lying just slightly below
    $|p| = M$.
    Therefore, aliasing predominantly transfers power which should
    have appeared in modes above the UV cutoff to modes just below
    the cutoff --- amplitudes of these modes
    receive the the greatest damage due to aliasing.
    }
This growth of short wavelength modes due to aliasing is called
spectral blocking.  
The same phenomena occurs when
employing a basis of Chebyshev polynomials.
Spectral aliasing can cause
truncation error to grow unboundedly, and lead to
time evolution becoming numerically unstable \cite{Boyd:2001}.
To make numerical evolution stable,
for many PDEs,
it is necessary to introduce some form of artificial dissipation
which damps short wavelength modes.
This can take the form of explicit addition of higher derivative
terms (``numerical viscosity'') to the equations of motion
(as we did in ref.~\cite{CY:isotropize}).
Or one can just selectively filter high $k$ modes whose amplitudes
are badly affected by spectral blocking \cite{Boyd:2001}.

For gravity, which is highly non-linear, one might expect significant
aliasing and resultant spectral blocking.  However, black branes in
asymptotically AdS spacetime allow rapid dissipation of short
wavelength modes, with an attenuation scale set by the infall time
into the black brane's horizon.%
\footnote
  {%
  For example, high momentum quasinormal modes of black branes in
  asymptotically AdS spacetime decay on a time scale
  set by the gravitational infall time \cite{Starinets:2002br,Kovtun:2005ev}.
  In this context, ``high-momentum'' applies to modes with rapid
  radial and/or spatial variations.
  }
Moreover, in infalling Eddington-Finkelstein coordinates, where lines
of constant time $t$ are infalling null geodesics, short wavelength
modes can propagate into the black brane horizon instantaneously in
coordinate time $t$, and hence need not persist and pollute the
subsequent numerical evolution.  
As a result, for characteristic evolution
of black brane geometries in asymptotically AdS spacetime,
instabilities resulting from spectral blocking are less serious
than might be expected.

Nevertheless, to ameliorate spectral blocking
we have found it useful and often necessary
either to introduce numerical viscosity,
or to selectively filter short wavelength modes.
For spatial directions, applying a sharp low-pass filter
which sets to zero all Fourier components for which
$\frac 23 k_{\rm max} < |k_i| \le k_{\rm max}$,
for any spatial direction $i$,
is a simple, computationally efficient choice.%
\footnote
    {%
    This is the ``$2/3$'s rule''
    \cite {Orszag:1971a,Boyd:2001}.
    For equations with only quadratic non-linearity,
    removing the uppermost third of Fourier components is
    sufficient to prevent aliasing from corrupting the
    components which are retained.
    Einstein's equations have higher order (cubic, quartic, and worse)
    non-linearities.
    Nevertheless, our experience is that filtering using the 2/3 rule
    is very effective in removing spectral blocking artifacts.
    The sufficiency of 2/3's rule filtering, despite high order
    non-linearities in the equations, is undoubtedly a reflection
    of the above-mentioned dissipation of short wavelength modes
    which is an intrinsic feature of black-brane geometries
    in asymptotically AdS spacetimes.
    }%
$^,\,$%
\footnote
    {%
    To implement this low-pass filter, one can use fast Fourier
    transforms (FFTs) to transform from real space to momentum space
    and back.
    Or one can construct the one-dimensional real space filter
    which is exactly equivalent to the desired momentum cutoff,
    and apply this filter as a convolution in real space.
    Asymptotically, for very fine spatial discretizations,
    using FFTs is most efficient.
    However, given matrix multiplication and convolution routines
    which are optimized for modern multi-core processors, 
    the break-even point beyond which FFTs become preferable to
    real-space convolution can lie at surprisingly large values
    of the number of points $N_i$ used in the discretization
    of a given spatial direction.
    }
For controlling spectral blocking in the radial direction,
our preferred method is filtering in real-space.   
Let $\{u_{\rm fine} \}$ represent the radial spectral grid
(\ref{eq:chebygrid}) used to solve Einstein's equations,
and let $\{u_{\rm coarse} \}$ represent a
spectral grid with two thirds as many points in the radial direction.
At each time step we interpolate the geometry from the fine grid
to the coarse grid.
The interpolation can be done without losing spectral accuracy
by employing the spectral representation (\ref{eq:spectral2})
to evaluate a function at off-grid locations.
We then reinterpolate the geometry from
the coarse grid back to the fine grid.
As the coarse grid is also a spectral grid, the interpolation back to
the fine grid can also be performed without losing spectral accuracy.
The process of interpolation from fine to coarse and back to the fine grid
has the effect of filtering short wavelength modes
(albeit with a soft cutoff, instead of a sharp momentum space cutoff).
Filtering in real-space with the Chebyshev grid (\ref{eq:chebygrid}),
which includes the boundary $u = 0$, has the advantage that
Dirichlet boundary conditions at $u = 0$ are completely
unaffected by filtering.

\subsection	{Parallelization}

The characteristic formulation Einstein's equations presented above
is easily amenable to parallelization.
Imposition of the fixed horizon condition (\ref{eq:Sigdothor})
makes the computation domain a simple rectangular box, with the
resulting discretized spatial lattice a tensor product grid.
For such a grid, spatial and radial derivatives of all
functions can easily be computed in parallel.
When computing, for example, the radial derivative of a function,
with a tensor product grid one can evaluate the radial derivative
independently at each point in space.
Computation of the radial derivatives at a given spatial point
(or set of points) can be performed independently by different processors.

Moreover, as discussed above, Einstein's equations in the
characteristic formulation take the form of linear ODEs
in the radial coordinate.
After all needed spatial and radial derivatives have been computed,
these radial ODEs can be solved independently at each spatial point.
In other words, the radial ODEs can be integrated, in parallel,
using independent CPUs for each point in space.
Solving the radial ODEs in parallel greatly increases computation speed;
specific performance results will be discussed below
in section~\ref{sec:performance}.

\subsection	{Domain decomposition}

\label{sec:domaindecomp}

Consider a discretization with $N_u$ points in the radial direction
and $N_i$ points in the spatial $x^i$ direction
(so the total number of grid points on a timeslice is
$N_u N_s$, with
$N_s \equiv \prod_{i=1}^{D-1} N_i$ the number of spatial
discretization points).
For a sufficiently fine spatial discretization,
the rate limiting step in our time-evolution procedure 
is the solution of the linear elliptic PDE (\ref{eq:apphoreq})
for the function $A$ at the apparent horizon.
Employing spectral methods,
solving eq.~(\ref{eq:apphoreq}) requires the solution of a
linear system with a dense coefficient matrix with
order $(N_s)^2$ elements, which is the discretization of
the linear operator appearing in eq.~(\ref{eq:apphoreq}).
Besides requiring extensive computation time, which scales as $O(N_s^3)$,
memory consumption can become problematic for large $N_s$.
Fortunately, it is easy to ameliorate these difficulties.

Linear elliptic PDEs such as eq.~(\ref{eq:apphoreq}) can
be efficiently solved using domain decomposition \cite{Boyd:2001}.
In this procedure,
the spatial interval in each $x_i$ direction is broken
up into $m_i$ separate subintervals,
thereby decomposing the spatial computational domain into
a total of $m_s \equiv \prod_i m_i$ subdomains.
Let $\ell = 1,{\cdots},m_s$ index these subdomains.
The $x_i$ dependence of functions within some subdomain $\ell$ can be
represented as a sum of $n^{(\ell)}_i$ Chebyshev polynomials.
Boundary points of the collocation grids in adjacent subdomains coincide.
In each subdomain, the solution to the linear equation
(\ref{eq:apphoreq}) can be decomposed in terms of a particular
solution $P^{(\ell)}(x)$ and a set of homogeneous solutions
$H^{(\ell)}_j(x)$,
\begin{equation}
    A(x) = P^{(\ell)}(x) + \sum_j C^{(\ell)}_j \, H^{(\ell)}_j(x),
\end{equation}
with the summation index $j$ running from 1 up to the number of
boundary points in the collocation grid for subdomain $\ell$.
Each homogeneous solution may be chosen to 
vanish at all but one boundary point of the subdomain, so
$H^{(\ell)}_j(x) \equiv 0$ for all boundary points 
except the $j^{\rm th}$ point, at which $H^{(\ell)}_j(x) \equiv 1$.
In other words, the homogeneous solutions $\{ H^{(\ell)}_j(x) \}$ represent
the discretized boundary Green's functions of the linear differential
operator on the subdomain $\ell$.
The coefficients $\{ C^{(\ell)}_j \}$ are computed by demanding that $A$
and $\nabla A$ be continuous across adjacent subdomains.
The resulting linear equations for the coefficients $C^{(\ell)}_j$
form a very sparse linear system whose solution can be efficiently
computed using sparse matrix routines in standard numerical
linear algebra packages.

The particular solution $P^{(\ell)}(x)$,
and the homogeneous solutions $\{ H^{(\ell)}_j(x) \}$, can be computed
independently in each subdomain.  Therefore the computation of the
set of functions $\{P^{(\ell)}(x) ,H^{(\ell)}_j(x) \}$ can easily
be performed in parallel.
Moreover, one may choose the numbers of subintervals $\{ m_i \}$
so that the total number of grid points within each subdomain is
easily manageable, i.e., small enough that the computation of 
$\{P^{(\ell)}(x), H^{(\ell)}_j(x) \}$ requires comparatively
little memory.

Using domain decomposition to solve the elliptic equation
(\ref{eq:apphoreq}) requires interpolating the coefficient functions
of the linear operator and the source terms appearing in
eq.~(\ref{eq:apphoreq})
from the global grid with $N_i$ points in each direction to the
local grid in element $\ell$ with $n^{(\ell)}_i$ points in each
direction.
Once the solution in element $\ell$ is obtained,
one must interpolate the local solution back to the global grid.
As noted above in section \ref{sec:filtering}, these interpolations 
can be performed without losing spectral accuracy
by employing the spectral representations (\ref{eq:spectral1})
and (\ref{eq:spectral2}) for the functions being interpolated.

Domain decomposition can also be usefully employed in the radial direction.
This entails breaking the radial interval up into $m_u$ subdomains and
coupling adjacent subdomains via boundary conditions.
There are several reasons to employ domain decomposition
in the radial direction.
First, the global nature of spectral methods implies that if the
metric happens to be badly behaved deep in the bulk,
the spectral representation of the metric will converge
poorly everywhere, including near the boundary.
However, for many situations, including the planar shock collisions
discussed below in section \ref{sec:shocks}, as time
progresses features in the geometry deep in the bulk can rapidly fall
through the event horizon.
In such situations, there is little practical value in knowing
the metric with high spectral accuracy in such regions.
By employing domain decomposition in the radial direction,
spectral convergence in one subdomain is only weakly dependent,
via boundary conditions, on spectral convergence in other subdomains.
Consequently, domain decomposition helps improve convergence near the
boundary when convergence is poor deep in the bulk.

Domain decomposition in the radial direction can also be helpful
in controlling the effects of round-off error.
Near the boundary, Einstein's equations contain $1/u^2$ singularities.
Since the grid is clustered around $u = 0$, round-off error coming from
points close to $u = 0$ can be greatly amplified by the presence of the
nearby singularity.
Domain decomposition helps with this simply because it allows
the grid spacing $\Delta u$ near $u = 0$ to be much larger 
than it will be if one uses a single domain with Chebyshev grid points.

The implementation of domain decomposition in the radial direction
is completely analogous to the spatial decomposition of elliptic PDEs
discussed above.
The radial domain is split up into
$m_u$ subdomains with $n_u$ Chebyshev grid points in each subdomain.
The endpoints of adjacent subdomains coincide.  The boundary 
conditions on the second order equations
(\ref{eq:Sigeqn})--(\ref{eq:Aeqn})
for $A$, $\Sigma$, and $F$ are simply that these functions, and
their radial derivatives, are continuous across interfaces.
When solving the first order radial equations (\ref{eq:Sigdoteqn}) and
(\ref{eq:gdoteqn}) for $d_+ \Sigma$ and $d_+ \hat g_{ij}$,
one constructs solutions for these functions
which are continuous along infalling geodesics.
When computing the time derivative $\partial_t \hat g_{ij}$
via eq.~(\ref{eq:timeevo}), one requires that
$\partial_t \, \hat g_{ij}$ also be continuous along outgoing geodesics.

\subsection	{Performance}
\label{sec:performance}

Key parameters controlling performance of a numerical calculation using
our approach are, naturally, the number of points used
in the spectral grids in the radial and spatial directions,
the number of time steps which are taken,
plus the speed (and memory capacity) of available computing resources.

With appropriate use of domain decomposition, 
both memory requirements and computational cost per time step are
essentially linear in the total number of grid points (radial times spatial).
For many problems, such as our homogeneous isotropization and 2D turbulence
examples below,
20--25 points in the radial grid are sufficient.
For our colliding shock example, we use up to 80 radial points
(partitioned into multiple subdomains).
The size and spacing of the spectral grid used for spatial directions
inevitably depends on the nature of the chosen problem.
For colliding shocks, we have used a Fourier grid with just over 800 points
in the longitudinal direction; this allows us to evolve the outgoing 
remnants of the collisions quite far before wrap-around effects arise.%
\footnote
    {%
    As discussed below,
    we use a significantly finer spectral grid, in both radial and
    spatial directions, for computing the colliding shock initial data.
    }
For 2D turbulence, we have used Fourier grids with several hundred points
(in each direction).
Here, the challenge is to make the spatial domain large enough to
contain flows whose Reynolds number is in the turbulent regime.


For the following examples, and our prior work
\cite{CY:isotropize,CY:boostinvar,CY:shocks},
we implemented the above-described approach and performed
calculations using \href{http://www.mathworks.com/products/matlab/}{MATLAB}\@.
It is possible that somewhat improved performance could be obtained by
carefully programming in a lower level language.
However, particularly for problems
(such as our examples in secs.~\ref{sec:shocks} and \ref{sec:2dfluid})
where symmetries at most reduce the problem to 2+1D or 3+1D PDEs,
the bulk of the computational time is spent in linear algebra routines
which are already highly optimized in MATLAB.
Consequently, we expect that any potential gain from coding in a lower
level language is quite modest.
(Far more important, for problems with non-trivial spatial dependence,
is that one implements the approach in a manner which allows easy
parallelization and hence benefits from multi-core processors.)

Our calculations have been performed with only desktop or laptop
scale computing resources.%
\footnote
    {%
    Our recent work has used a single six core
    Intel i7-3960x processor overclocked to 4.25\,GHz and a
    four core MacBook Pro with Intel i7 processor running at 2.5\,Ghz.
    }
Using these relatively limited computing resources,
calculating the homogeneous isotropization example discussed below
(where symmetries reduce the problem to 1+1D PDEs) is quick,
taking a few seconds.
Evolving the geometry in the colliding shock example
(where symmetries reduce the problem to 2+1D PDEs), for the
more demanding case of narrow shocks, required approximately 12 hours on a laptop computer.
Performing the numerical evolution of the geometry in our
turbulent fluid example (where one is dealing with 3+1D PDEs)
required approximately three weeks of time.

A different aspect of performance concerns achievable accuracy.
At a crude but important level, a key indicator of accuracy
is the 
absence of obvious
numerical instabilities which prevent continuing evolution
to arbitrarily late times.
Achieving stable evolution requires sensible choices for
the spectral grids and time step.
The use of UV filtering to control spectral blocking
(as described in section \ref{sec:filtering}) is important
for many problems.  

Once stable numerical evolution is achieved,
a more refined, physically important, measure of accuracy 
involves comparison of numerical results with analytically
derived late-time asymptotic forms.
This is discussed below in the context of our specific examples.
A further check of numerical accuracy can be obtained by monitoring
the validity of constraint equations.
As the constraint equations were used in deriving the horizon
stationarity condition (\ref{eq:apphoreq}), which determines the value of $A$
on the apparent horizon, correct numerical evolution 
of the gauge parameter $\lambda$ via
eqs.~(\ref{eq:apphoreq}) and (\ref{eq:lambdaevo}) is intimately
connected to the numerical validity of the constraint equations.
Hence, one simple test of the constraint equations comes from monitoring
how well the gauge parameter obtained from evolving eq.~(\ref{eq:lambdaevo}) 
agrees with the value obtained by directly solving the 
horizon fixing condition (\ref{eq:Sigdothor}).
If the gauge parameter evolved from eq.~(\ref{eq:lambdaevo})
drifts too far from the value which correctly
solves the horizon condition,
it may be necessary to make small periodic readjustments in $\lambda$
to ensure that the horizon remains at $r= 1$.

The bottom line is that with appropriate care
(in adjusting grids, filtering, etc.),
the achievable accuracy is, in our view,
remarkably good.


\addtocontents	{toc}{\tocsqueeze}
\section	{Examples}
\label		{sec:examples}

\subsection	{Homogeneous isotropization}
\label		{sec:iso}

\subsubsection	{Motivation}

Relativistic heavy ion collisions may be regarded as proceeding
through a sequence of stages.
Initially, the collision of the (overlapping portions of the)
highly Lorentz contracted nuclei may be viewed as liberating
a very large phase space density of partons from the colliding
nucleons.
Within the central rapidity region of the event, the initial
distribution of partons is highly anisotropic, with typical
transverse momenta much larger than longitudinal momenta.
These partons subsequently interact and scatter.
After a ``thermalization time'' (which, more properly,
should be called an isotropization time), the gas of interacting
partons may be modeled as a relativistic fluid --- a quark-gluon
plasma ---  whose stress tensor, in a local fluid rest frame,
is nearly isotropic.
This plasma expands, cools, and eventually reaches a temperature
where hadrons reform, fly outward, and ultimately reach the detector.%
\footnote
    {%
    For a more substantial introduction to heavy ion collisions see,
    for example, ref.~\cite{Heinz:2004qz}.
    }

Hydrodynamic modeling of the results of heavy ion collisions 
strongly suggests that the isotropization time of the dense parton
gas is remarkably short, less than 1 fm/$c$ \cite{Heinz:2004pj},
and that the resulting plasma behaves as a nearly ideal fluid.
Understanding the dynamics responsible for such rapid isotropization
in a far-from-equilibrium non-Abelian plasma is a challenge.
The nearly ideal (i.e., low viscosity) behavior of the produced plasma
is an indication that experimentally accessible quark-gluon plasma
is strongly coupled \cite{Shuryak:2008eq}.%
\footnote
    {%
    For systems with a quasiparticle interpretation,
    viscosity scales as energy density times the mean free time of excitations.
    Weakly coupled systems have excitations with long mean free times,
    and hence large viscosity relative to entropy density.
    Low viscosity, relative to entropy density, implies short
    mean free times or rapid scattering, and hence strong coupling.
    }

Due to the difficulty of studying real time quantum dynamics in
QCD at strong coupling, it is useful to examine far-from-equilibrium
behavior in an instructive toy model, namely $\Nfour$ SYM, whose
equilibrium behavior at non-zero temperature mimics many features
of real QCD plasma.
This was the motivation for our earlier study \cite{CY:isotropize} of
isotropization in spatially homogeneous but highly anisotropic states of
strongly coupled $\Nfour$ SYM, using the dual gravitational description.
In that work, we considered initial states which could be produced
by the action of time-dependent (but spatially homogeneous) background fields.
The background field which naturally couples to the stress-energy tensor
of the field theory is the metric of the four-dimensional geometry in
which the QFT is formulated.
A time-dependent 4D metric in the QFT description corresponds,
under the holographic mapping, to a time-dependent boundary
geometry in the dual gravitational description.
Our earlier work \cite{CY:isotropize} solved the resulting
gravitational dynamics (numerically), using the approach
presented in section \ref{sec:strategy},
with the simplifying assumption of spatial homogeneity but with
the complication of a time-dependent boundary geometry.

In the present paper we focus, for simplicity, on problems involving 
a flat Minkowski boundary geometry.
To illustrate the application of our methods, we will present results
on far-from-equilibrium isotropization in which the operational
driving via a time-dependent boundary geometry of ref.~\cite{CY:isotropize}
is replaced by a simple (and arbitrary) choice of initial data for our
characteristic formulation.

\subsubsection	{Setup}

The boundary dimension $D = 4$.
With the imposition of spatial homogeneity,
spatial parity invariance,
and $O(2)$ rotation invariance,
the only non-zero functions in the metric ansatz (\ref{eq:ansatz})
are $A$, $\Sigma$, and the diagonal elements of $\hat g_{ij}$
which we write in terms of a single ``anisotropy'' function $B$,
\begin{equation}
    \|\hat g_{ij}\| = \mathop{\rm diag}(e^B,\, e^B,\, e^{-2B}) \,.
\label{eq:gijiso}
\end{equation}
The unknown functions $A$, $B$, and $\Sigma$ depend only on $t$ and $r$.
Eqs.~(\ref{eq:Sigeqn}), (\ref{eq:Aeqn}), and (\ref{eq:Sigdoteqn})
for $\Sigma$, $A$, and $d_+\Sigma$, respectively, become
\begin{subequations}
\begin{align}
    &\Sigma'' + \half (B')^2 \, \Sigma = 0 \,,
\\
    &A'' = 6 (\Sigma'/ \Sigma^{2}) \, d_+\Sigma - \tfrac 32 B' \, d_+ B - 2 \,,
\\
    &(d_+\Sigma)' + 2 (\Sigma'/\Sigma) \, d_+\Sigma = 2 \Sigma \,,
\end{align}\label{eq:isoeqns}%
\end{subequations}%
while eq.~(\ref{eq:gdoteqn}) for $d_+\hat g$ reduces to
\begin{equation}
    (d_+B)' + \tfrac 32 (\Sigma'/\Sigma)\, d_+B
    = -\tfrac 32 B' \, (d_+\Sigma)/\Sigma \,.
\label{eq:Bdot}
\end{equation}
We replace the field redefinitions (\ref{eq:redefs})
involving the spatial metric
with the redefinitions
\begin{align}
    b \equiv u^{-3} \, B\,,\quad
    \dot b \equiv u^{-3} \, d_+ B\,.
\end{align}
for the anisotropy function and its (modified) time derivative.
The asymptotic behavior (\ref{eq:asymp}) implies that
$b$ vanishes at the AdS boundary while
$\dot b$ approaches a finite limit of $-2b^{(4)}$.
The latter boundary condition is imposed
when solving eq.~(\ref{eq:Bdot}) for $\dot b$.

The apparent horizon condition (\ref{eq:Sigdothor}) is just
\begin{equation}
    d_+\Sigma \big|_{\rh} = 0 \,.
\end{equation}
Since there is no spatial dependence, the horizon stationarity
equation (\ref{eq:apphoreq}) 
becomes a simple algebraic condition for the value of $A$ on the
apparent horizon,
\begin{equation}
    A_h = -\fourth \, (d_+B)^2 \,.
\end{equation}

Initial data consists of a choice of the
anisotropy function on the initial time slice, $B(t_0,r)$,
plus a value for the single asymptotic coefficient $a^{(4)}(t_0)$
which sets the initial energy density.
We make the simple but arbitrary choice:
\begin{equation}
    b(t_0,u) = \beta \, u \, e^{-(u-u_0)^2/w^2} \,,\qquad
    a^{(4)} = -\tfrac{1}{2} \, \alpha \,,
\label{eq:isoinitialdata}
\end{equation}
with $\beta = 5$, $u_0 = 0.25$, $w = 0.15$, and $\alpha = 1$.
Using the result (\ref{eq:TMN}) for the boundary stress-energy tensor and
inserting the holographic relation $G_N = \frac{\pi}{2} \, L^3/\Nc^2$
appropriate for $\Nfour$ SYM, the corresponding
energy density $T^{00} = \frac 38 \, \Nc^2 \, \alpha/\pi^2$.
The energy density of an equilibrium, strongly coupled $\Nfour$ SYM plasma
at temperature $T$ is given by
$T^{00}_{\rm eq} = \frac 38 \, \Nc^2 \, \pi^2 T^4$.
Hence, our chosen value of $\alpha$ corresponds to an equilibrium
temperature $T \equiv \alpha^{1/4}/\pi = 1/\pi$.

To evolve the geometry, we use a spectral grid in the radial
direction with 25 points, and employ
explicit fourth-order Runge-Kutta for the time-integrator
with a time step $\Delta t = 0.01$.  For this simple $1+1$ dimensional problem,
where all dynamics takes place in the radial direction only,
we do not employ any filtering.
Indeed, because radial lines are
infalling null geodesics, any high frequency numerical noise generated by the 
numerical evolution tends to get absorbed effectively instantaneously by the horizon.

\subsubsection	{Results}

\begin{figure}
    \begin{center}
    \tracingmacros=1
    \hspace*{-20pt}\suck[scale = 0.37]{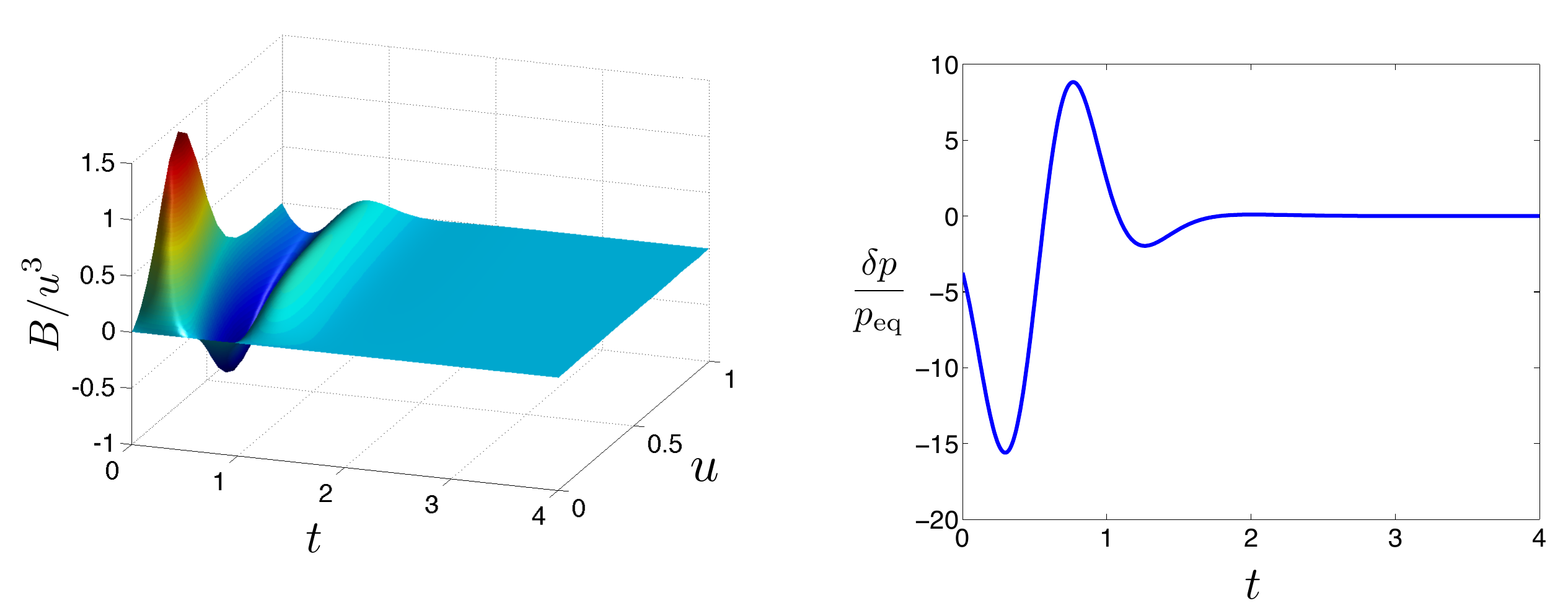}\hspace*{-20pt}
    \tracingmacros=0
    \vspace*{-15pt}
    \end{center}
\caption
    {%
    Homogeneous isotropization results.
    Left panel: Anisotropy function $B(t,u)/u^3$.
    The anisotropy function rapidly attenuates, with
    exponentially damped oscillations.
    Right panel: Pressure anisotropy $\delta p = T_{zz} - \half (T_{xx}+T_{yy})$,
    relative to the equilibrium pressure $p_{\rm eq} = \frac 18 \Nc^2 (\pi T)^4$,
    as a function of time.  At early times the pressures anisotropy is very large.  
    However, 
    just as the anisotropy function vanishes exponentially fast, so does 
    the pressure anisotropy.  
    }
\label{fig:isotropization}
\end{figure}

The resulting evolution of the anisotropy function $B$ is shown
in the left panel of fig.~\ref{fig:isotropization}.
The right panel displays a plot of the pressure anisotropy
$\delta p \equiv T_{zz} - \half(T_{xx} + T_{yy})$,
relative to the equilibrium pressure
$p_{\rm eq} = \frac 18 \Nc^2 (\pi T)^4$, as a function of time.
Inserting the diagonal form (\ref{eq:gijiso}) of the spatial
metric into the general result (\ref{eq:TMN}) for the
stress-energy tensor, one sees that the pressure anisotropy
is simply proportional to the coefficient $b^{(4)}$ of the
leading near-boundary behavior of the anisotropy function,
$B(t,u) \sim b^{(4)}(t)\, u^4 + O(u^5)$.

Examining fig.~\ref{fig:isotropization} one sees,
first and foremost, that the geometry evolves toward
an isotropic equilibrium geometry, which is just the
static Schwarzschild black-brane solution.
This is a basic test of the numerics;
no problems with numerical instabilities, potentially preventing
evolution to arbitrarily late times, are seen.
The approach to equilibrium shows exponentially damped oscillations.
With no spatial gradients, there is no excitation of hydrodynamic 
degrees of freedom, and hence no hydrodynamic regime in the response.

\begin{figure}
    \begin{center}
    \suck[scale = 0.45]{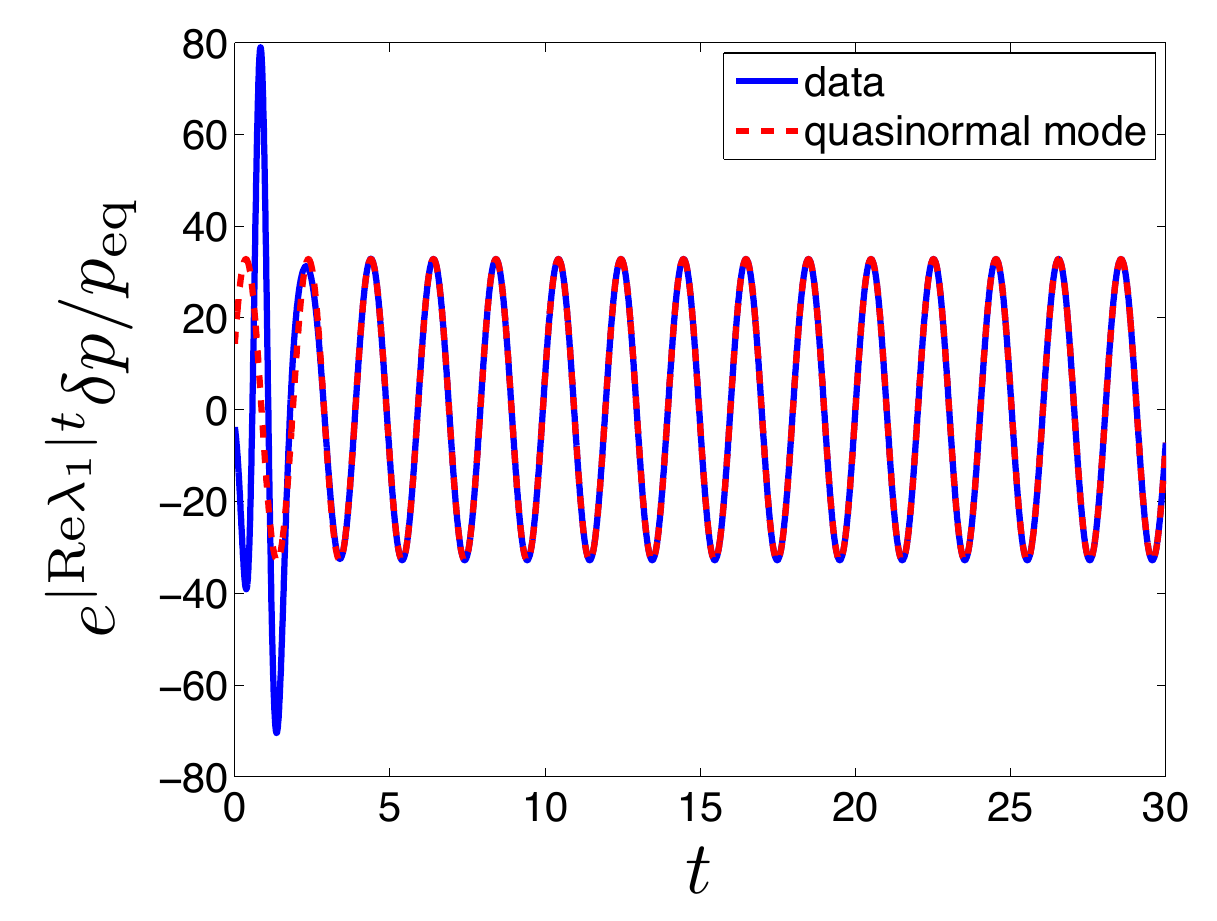}
    \vspace*{-20pt}
    \end{center}
\caption
    {A plot of $e^{|{\rm Re\, \lambda_1}| t}\delta p/ p_{\rm eq}$ as well as the lowest quasinormal mode (also multiplied by a factor of $e^{|{\rm Re \lambda_1}| t}$).
    The fit to the lowest quasinormal mode agrees with the numerics at the 1 part in $10^4$ level or better after time $t = 10.$
     }
\label{fig:isotropizationQM}
\end{figure}

At sufficiently late times,
the damped oscillations of the pressure anisotropy reflect the
discrete spectrum of complex quasinormal mode frequencies
characterizing infinitesimal departures from equilibrium
\cite{Starinets:2002br,Kovtun:2005ev},
specifically those of $\ell = 2$ metric perturbations whose
linearized dynamics around the AdS-Schwarzschild black brane solution
coincides with fluctuations of a minimally coupled scalar field.
The late time asymptotic response has the form
\begin{equation}
    \delta p(t) \sim \mathop{\rm Re} \sum_n c_n \, e^{-\lambda_n t} \,,
\label{eq:qnmresponse}
\end{equation}
where the first few quasinormal mode frequencies, at zero wavevector, are
given by
\cite{Starinets:2002br}:
\begin{equation}
    \frac{\lambda_1}{\pi T} = 2.746676 + 3.119452\, i \,,\quad
    \frac{\lambda_2}{\pi T} = 4.763570 + 5.169521\, i \,,\quad
    \frac{\lambda_3}{\pi T} = 6.769565 + 7.187931\, i \,.
\label{eq:qnmiso}
\end{equation}
As a check on the accuracy of the numerics,
in fig.~\ref{fig:isotropizationQM} we plot 
$e^{|{\rm Re\, \lambda_1}| t}\, \delta p/ p_{\rm eq}$,
as well as a fit to the lowest quasinormal mode.
As is evident from the figure,
the rescaled amplitude of $e^{|{\rm Re \lambda_1}| t}\,\delta p/ p_{\rm eq}$
is constant at late times.
Indeed, our fit to the lowest quasinormal mode agrees with the numerics
at the level of a part in $10^4$, or better, after time $t = 10.$

In terms of physics, perhaps the most significant result one sees
from fig.~\ref{fig:isotropization} (and from the results of ref.~\cite{CY:isotropize})
is that the characteristic relaxation time is comparable or shorter than $1/T$,
even when the system is initially quite far from equilibrium with
$\delta p/p_{\rm eq}$ of $O(10)$.
The gravitational infall time in the AdS-Schwarzschild
geometry is also order $1/T$.
This naturally suggests that, even far from equilibrium, one should regard
the gravitational infall time as characterizing the relaxation time of
non-hydrodynamic degrees of freedom.


\subsection	{Colliding planar shocks}
\label		{sec:shocks}


\subsubsection{Motivation}

Collisions of infinitely extended planar shock waves in $\Nfour$ SYM
may be viewed as instructive caricatures of collisions of
large, highly Lorentz-contracted nuclei.
In the dual description of strongly coupled (and large $\Nc$) SYM,
this becomes a problem of colliding gravitational shock waves
in asymptotically AdS$_5$ spacetime.
In this section, we discuss the setup, preparation of initial data,
and results for such planar shock collisions.

Numerical construction of a complete colliding planar shock geometry
was first performed in ref.~\cite{CY:shocks}.
More recently, the authors of ref.~\cite{Casalderrey-Solana:2013aba}
examined the sensitivity of the post-collision energy density and
pressure distributions to the width of the initial shocks.
In both of these previous works, a small background energy density
was added to the initial data to help control 
numerical instabilities deep in the bulk.

Using the filtering approach discussed 
in sec.~\ref{sec:filtering} to suppress spectral blocking,
it is possible to compute, accurately, colliding shock geometries,
even for very thin shocks, 
without adding any background energy density.
In other words, it is possible to study collisions of shocks which
are truly excitations of the vacuum state.
Even with a vanishing background energy density (or temperature),
we find no problems associated with caustics or non-planar 
horizon topology.

\subsubsection{Initial data}

The boundary dimension $D = 4$.
With the imposition of spatial homogeneity in transverse directions,
plus 2D rotation and reflection invariance in the transverse plane,
the only non-zero functions in the metric ansatz (\ref{eq:ansatz})
are $A$, $\Sigma$,
the longitudinal component $F_z$ of the spatial vector $F$,
and the diagonal elements of the rescaled spatial metric
$\hat g_{ij}$.
The latter we write in terms of a single anisotropy
function $B$ which distinguishes the transverse and longitudinal
directions,
\begin{equation}
    \|\hat g_{ij}\| \equiv \mathop{\rm diag}(e^B,\, e^B,\, e^{-2B}) \,.
\end{equation}
The unknown functions $A$, $B$, $F_z$, and $\Sigma$ all depend on $t$,
$u \equiv 1/r$, and $z \equiv x^3$.

From the series expansions (\ref{eq:asymp}) one sees that near the boundary
\begin{equation}
\label{eq:BshockAsymtotic}
B \sim u^4 \, b^{(4)} + O(u^5)\,.
\end{equation}
We replace the field redefinitions (\ref{eq:redefs})
involving the spatial metric
with the following field redefinitions for the anisotropy function
and its (modified) time derivative,
\begin{align}
    b \equiv u^{-3} \, B\,,\quad
    \dot b \equiv u^{-3} \, d_+ B\,.
\end{align}
The asymptotic behavior (\ref{eq:asymp}) implies that
$b$ vanishes at the AdS boundary while
$\dot b$ approaches a finite limit of $-2b^{(4)}$.
The latter boundary condition is imposed
when solving eq.~(\ref{eq:gdoteqn}) for $\dot b$.

We choose initial conditions corresponding to two well separated,
smooth, non-singular planar gravitational waves.
In Fefferman-Graham coordinates
[denoted $(\tilde t, \tilde {\bm x}_{\perp},\tilde z, \tilde \rho )$]
the pre-collision metric reads
\begin{equation}
\label{eq:initialshockgeometry}
    ds^2 = \tilde \rho^{-2}
    \left [-d\tilde x_{+} d \tilde x_{-} + d \tilde {\bm x}_{\perp}^2
    + d \tilde \rho^2 \right ]
    +\tilde \rho^2
    \left [h(\tilde x_-)\, d \tilde x_-^2 + h(\tilde x_+)\, d \tilde x_+^2 \right ],
\end{equation}
where $\tilde x_{\pm} \equiv \tilde t \pm \tilde z$ and $h(z)$ is an
arbitrary function characterizing the longitudinal profile of the shocks.
In what follows we choose a simple Gaussian profile of adjustable width
and amplitude, parameterized as
\begin{equation}
h(z) \equiv \mu^3 \, (2 \pi w^2)^{-1/2} \, e^{-\frac{1}{2} z^2/w^2}.
\label{eq:profile}
\end{equation}

In the distant past, the geometry both between and far away from the shocks
is just AdS$_5$, up to negligible, exponentially small corrections.
Via eq.~(\ref{eq:FGstress}), the initial boundary energy density
and longitudinal stress are given by
\begin{equation}
    \widehat T^{00}(\tilde t,\tilde z) =
    \widehat T^{zz}(\tilde t,\tilde z) =
    h(\tilde t-\tilde z) + h(\tilde t+\tilde z) \,,
\end{equation}
while the momentum density
\begin{equation}
    \widehat T^{0z}(\tilde t,\tilde z) =
    h(\tilde t-\tilde z) - h(\tilde t+\tilde z) \,.
\end{equation}
Therefore, the metric (\ref{eq:initialshockgeometry}),
with shock profile (\ref{eq:profile}), describes two localized
planar lumps of energy of width $w$ moving toward each
other at the speed of light and colliding at time $\tilde t = 0$.  
Restoring the overall factor of $\kappa = L^3/(4\pi G_N)$
[c.f. eq.~(\ref{eq:Thatdef})]
and inserting the holographic relation $G_N = \frac \pi2 \, L^3/\Nc^2$,
appropriate for $\Nfour$ SYM, shows that the energy per unit area
of each incoming shock is $\mu^3 (\Nc^2/2\pi^2)$.

Without loss of generality we may set $\mu = 1$ and measure all quantities
in units of $\mu$ to the appropriate power.
We will present results for the collisions of ``wide'' shocks with $w = 0.375$,
and ``narrow'' shocks with $w = 0.075$.
(For comparison, ref.~\cite{CY:shocks} used $w = 0.75$ and
ref.~\cite{Casalderrey-Solana:2013aba} investigated widths ranging
from 1.9 down to $0.05$.)

In the distant past, when the two functions $h(\tilde t \pm \tilde z)$ have
negligible overlap,
the metric (\ref{eq:initialshockgeometry}) is arbitrarily close
to an exact solution to Einstein's equations (\ref{eq:Einstein}).%
\footnote
    {%
    If we had chosen profile functions with compact support,
    then the metric (\ref{eq:initialshockgeometry}) would be
    an exact solution in the region of spacetime outside the
    causal future of the collision (i.e., not in the causal future
    of any event where
    $h(\tilde t {+} \tilde z)h(\tilde t{-}\tilde z)$ is non-zero).
    The fact that our Gaussian profile functions do not have compact support
    is irrelevant for all practical purposes.
    }
But near the collision time $\tilde t = 0$,
when the functions $h(\tilde t \pm \tilde z)$ begin to overlap significantly,
the metric (\ref{eq:initialshockgeometry}) ceases to be a (near) solution to
Einstein's equations,
and one must compute the future evolution numerically.
To do so we employ our characteristic formulation.

To obtain initial data suitable for our formulation,
the initial metric (\ref{eq:initialshockgeometry}) must be transformed 
from Fefferman-Graham coordinates to infalling Eddington-Finkelstein
coordinates, in which the metric takes the form (\ref{eq:ansatz}).
To do so
we compute, numerically, the needed coordinate transform for a single shock
moving in the $+z$ direction and thereby determine the set of functions
$\{b_+(t{-}z,u),\, a^{(4)}_+(t{-}z),\, f_{z+}^{(4)}(t{-}z),\, \lambda_+(t{-}z)\}$
characterizing a right-moving shock.
The substitution $z \to -z$ produces
the corresponding functions
$\{b_-(t{+}z,u)$, $a^{(4)}_-(t{+}z)$, $f_{z-}^{(4)}(t{+}z)$, $\lambda_-(t{+}z)\}$
for a left-moving shock.
As we discuss in greater detail below,
we then superimpose the pre-collision functions
\begin{equation}
    b(t,z,u) = b_+(t{-}z,u) + b_-(t{+}z,u) \,,\quad
    a^{(4)}(t,z) = a^{(4)}_+(t{-}z) + a^{(4)}_-(t{+}z) \,,
\label{eq:shocksuperpose}
\end{equation}
and likewise for $f_z^{(4)}$ and $\lambda$,
and then evolve  $\{b,\, a^{(4)},\, f_{z}^{(4)},\, \lambda\}$
forward in time by numerically solving Einstein's equations.

The metric of a single shock moving in the $+z$ direction is given by 
\cite{Janik:2005zt} 
\begin{equation}
\label{eq:initialshockgeometryrightmover}
    ds^2 = \tilde \rho^{\, -2}
    \left [-d\tilde x_{+} d \tilde x_{-} + d \tilde {\bm x}_{\perp}^2
	+ d \tilde \rho^2 \right ] 
    + \tilde \rho^2 \, h(\tilde x_-) \, d \tilde x_-^2 \,.
\end{equation}
The coordinate transformation taking this metric to the 
Eddington-Finkelstein form (\ref{eq:ansatz}) (with $u \equiv 1/r$)
can be expressed as
\begin{align}
    \label{eq:shockcoordtrans}
    \tilde t = t + u + \alpha(t{-}z,u), \quad
    \tilde {\bm x}_{\perp} = {\bm x}_{\perp}, \quad
    \tilde z= z - \gamma(t{-}z,u), \quad
    \tilde \rho = u + \beta(t{-}z,u)\,,
\end{align}
for suitable functions $\alpha$, $\beta$, and $\gamma$
whose determination will be described momentarily.

The functions
$\{b_+,\, a^{(4)}_+,\, f_{z+}^{(4)},\, \lambda_+\}$
providing the required initial data for
our characteristic formulation
can be expressed in terms of
the profile function $h$ and the transformation functions
$\alpha$, $\beta$, and $\gamma$.
A short exercise shows 
\begin{subequations}
\begin{equation}
\label{eq:a4f2coordtransshock}
    a^{(4)}_+ = -\tfrac{2}{3}\, h\,,\quad
     f_{z+}^{(4)} = h\,,\quad
     \lambda_+ = -\tfrac{1}{2}\, \partial_u^2 \beta\big|_{u = 0}\,,
\end{equation}
and
\begin{equation}
    \label{eq:Bcoordtransshock}
    b_+ = -\tfrac{1}{3} u^{-3}
    \log \left [-(\partial_t \alpha )^2
    +(\partial_t \beta )^2
    +\left(1+\partial_t \gamma\right)^2
    + (u{+}\beta)^4 \left(1+\partial_t \alpha +\partial_t \gamma \right)^2 h \right].
\end{equation}
\end{subequations}

The equations determining the coordinate transformation functions
(which follow from solving for infalling radial null geodesics
in the metric (\ref{eq:initialshockgeometryrightmover}), or equivalently
demanding that the transformed metric have the desired form
(\ref{eq:ansatz}))
are simplified by redefining
\begin{align}
\label{eq:coordtransredef}
\beta \equiv - \frac{u^2 \xi}{1 + u \, \xi }\,, \quad
\alpha \equiv - \gamma + \beta  + \delta\,.
\end{align}
In terms of $\xi$, $\delta$ and $\gamma$,
the equations of the coordinate transformation
reduce to a system of coupled radial ODEs for $\xi$ and $\delta$,
\begin{align}
\label{eq:coupledcoordtrans}
\frac{1}{u^2} \frac{\partial}{\partial u} \left ( u^2 \frac{\partial \xi}{\partial u} \right )+ \frac{2 u H}{(1 + u \xi)^5}  = 0\,,
\qquad
\frac{\partial \delta}{\partial u} - \frac{u^2 }{(1 + u \xi)^2} \frac{\partial \xi}{\partial u} = 0\,,
\end{align}
with $H \equiv h\left (t {-} z + u + \delta - u^2 \xi/(1 + u \xi) \right)$.  
The function $\gamma$ satisfies the first order radial ODE
\begin{align}
\label{eq:auxcoordtrans}
\frac{\partial \gamma}{\partial u} 
- \frac{u^2}{(1 + u \xi )^2} \frac{\partial \xi}{\partial u} 
+ \frac{u^4}{2 (1 + u \xi )^2}  \left (\frac{\partial \xi}{\partial u}  \right )^2 
+ \frac{u^4 H}{2 ( 1 + u \xi )^6} = 0\,.
\end{align}
As is clear from inspecting eq.~(\ref{eq:coupledcoordtrans}),
on each slice with fixed $t {-} z$, the functions $\xi$ and $\delta$ 
can be determined by integrating the coupled ODEs (\ref{eq:coupledcoordtrans})
from $u = 0$ to $u = 1$.
With $\xi$ and $\delta$ known,
one can then integrate the radial ODE (\ref{eq:auxcoordtrans})
on the same $t {-} z=$ const.\ slice to determine $\gamma$.
In other words, determination of the coordinate transformation functions
is local in $t{-}z$; one need only integrate radial ODEs.

The desired solutions to eqs.~(\ref{eq:coupledcoordtrans}) and
(\ref{eq:auxcoordtrans}) are specified by boundary
conditions at $u = 0$ and $u = 1$.
The conditions $t = \tilde t$, $z = \tilde z$, and $u = \tilde \rho$
near the AdS boundary imply that the fields $\xi$, $\delta$, and $\gamma$
have the asymptotic forms
\begin{align}
\xi = \xi_0 +O(u^3) \,,\quad
\delta = O(u^5) \,,\quad
\gamma = O(u^5)\,.
\end{align}
Defining further rescaled fields $\Delta \equiv \delta/u^4$ and
$\Gamma \equiv \gamma/u^4$, we therefore impose at the AdS boundary
the conditions
\begin{align}
    \partial_u \xi\big|_{u=0} = 0\,, \ \ \
    \Delta\big|_{u=0} = 0\,, \ \ \
    \Gamma\big|_{u=0} = 0\,,
\end{align}
and integrate eq.~(\ref{eq:coupledcoordtrans}) to find $\xi$ and $\Delta$, 
and eq.~(\ref{eq:auxcoordtrans}) to find $\Gamma$.
One additional boundary condition is needed to fully specify a solution
to eq.~(\ref{eq:coupledcoordtrans}).
At $u = 1$ we impose the condition
\begin{equation}
    \xi\big|_{u=1} = -1 + \frac{1}{\tilde \rho_{\rm max}}\,,
\label{eq:utildemax}
\end{equation}
for some choice of the function $\tilde \rho_{\rm max}(t {-} z)$.
This boundary condition determines how deep into the bulk
one determines the transformation of the initial geometry.
Via eqs.~(\ref{eq:shockcoordtrans}) and (\ref{eq:coordtransredef}),
one sees that at $u = 1$ the Fefferman-Graham coordinate
$\tilde \rho$ coincides with $\tilde \rho_{\rm max}$.
The boundary condition (\ref{eq:utildemax})
also largely determines the gauge parameter $\lambda_+$ since,
away from the shock where $h$ is negligible, one has
\begin{equation}
\label{eq:AsymGaugeFunction}
    \lambda_+ \to -1 + \frac{1}{\tilde \rho_{\rm max}}\,.
\end{equation}

Controlling how deep into the bulk one solves for
the initial geometry (in Eddington-Finkelstein coordinates) is essential.
If one integrates too far into the bulk, the metric functions become very
large, causing problems with loss of numerical precision.
This can already be seen in the single shock Fefferman-Graham metric
(\ref{eq:initialshockgeometryrightmover}),
where metric functions grow like $\tilde \rho^2$ for large $\tilde \rho$.
However, the apparent horizon of the colliding shock geometry exists
prior to the collision at $t = 0$ \cite{CY:shocks}.
The mapping of the initial geometry into Eddington-Finkelstein coordinates
must go sufficiently deep into the bulk so that the
apparent horizon lies within the chosen computational domain $u \in [0,1]$.
Selecting an appropriate value for $\tilde \rho_{\rm max}$ 
so that the apparent horizon lies in this interval,
and the bulk geometry is reasonably well behaved,
can require some trial and error.
We set 
\begin{equation}
\label{eq:singleshockgaugechoice}
\tilde \rho_{\rm max} = 8,
\end{equation}
independent of $t{-}z$, and comment below on more refined
choices of $\tilde \rho_{\rm max}(t{-}z)$.

We employ domain decomposition
in both the radial and longitudinal ($t{-}z$) directions
when solving eqs.~(\ref{eq:coupledcoordtrans}) and (\ref{eq:auxcoordtrans}).
We use $20$ Chebyshev polynomials in each subdomain in both directions.  
We employ 350 subdomains in the $t{-}z$ direction and $35$ subdomains
in the $u$ direction, and solve the equations in the interval
$-18 \leq t {-} z \leq 18$.
Using domain decomposition
in each direction is advantageous for several reasons.
First, as mentioned above, the coordinate transformation 
can become badly behaved deep in the bulk.
As discussed in sec.~\ref{sec:domaindecomp},
if the convergence of the spectral series very deep in the bulk becomes poor,
the use of domain decomposition serves to reduce the influence of such
poor convergence on fields closer to the boundary.
Second, the use of domain decomposition 
--- with relatively few points in each subdomain --- allows the function
$b_+$ [defined in eq.~(\ref{eq:Bcoordtransshock})],
and its near-boundary asymptotics,
to be determined numerically with very good and controllable accuracy.
In particular, the use of domain decomposition allows finely spaced
grid points to be used for rapidly varying functions, thereby
enabling good spectral convergence, while simultaneously avoiding the
significant round-off error that can occur when employing a single global domain
with a very large number of grid points.

\begin{figure}
\suck[scale = 0.4]{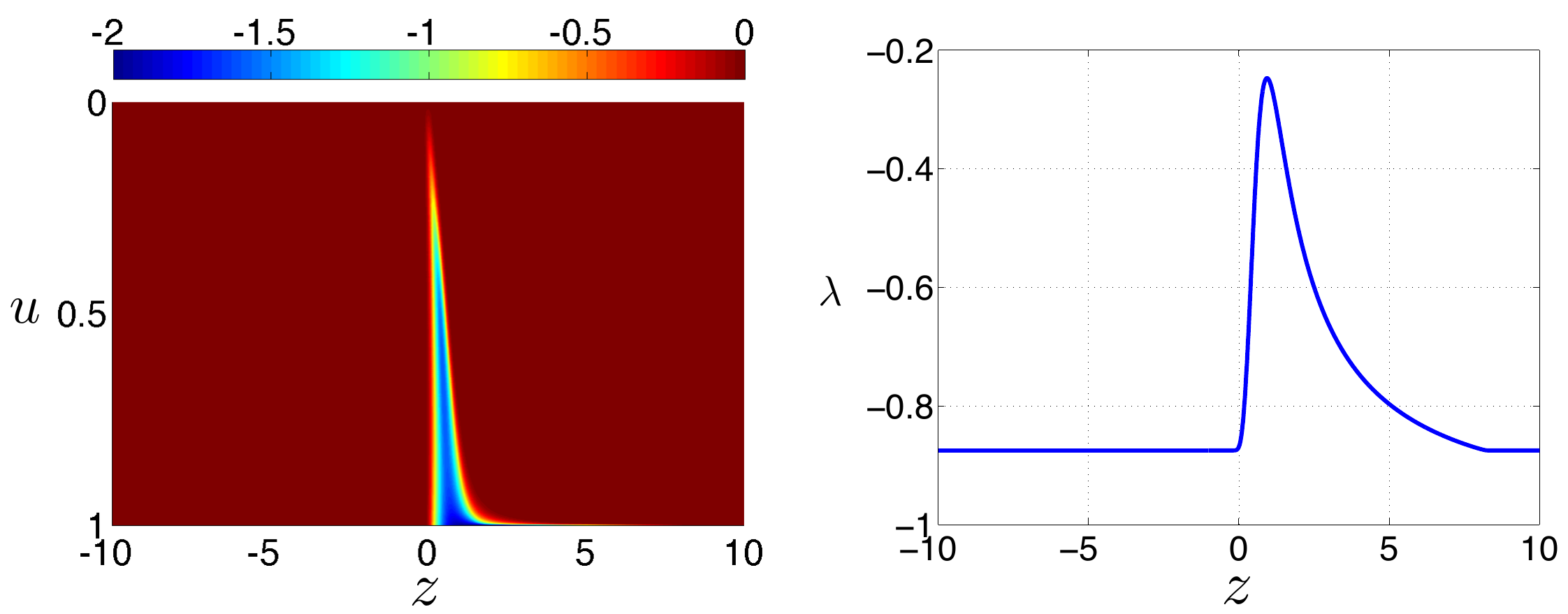}
\vskip -0.05in
\caption{Plots of $b_+$ (left) and $\lambda_+$ (right)
for a single narrow shock of width $w = 0.075$ moving in the $+z$ direction.
The choice of gauge parameter $\lambda_+$ is such that $u = 1$ corresponds with
Fefferman-Graham coordinate $\tilde \rho = 8$.
On the boundary, $u = 0$, the shock is centered at $z = 0$ at the time shown, $t = 0$.
However, in Eddington-Finkelstein coordinates
the shock increasingly extends into the $+z$ direction as one goes deeper into the bulk.
This also manifests itself in the gauge parameter $\lambda$, which differs significantly
from its background value in front of the shock.
In regions where $b_+ = 0$ the geometry is that of AdS$_5$.
\label{fig:singleshockinitialdata}
} 
\end{figure}

Fig.~\ref{fig:singleshockinitialdata} plots the resulting functions
$b_+$ and $\lambda_+$, at time $t = 0$,
for a  single narrow shock moving in the $+z$ direction.
One sees that $b_+$ is non-zero for positive values of $z$ (well beyond the
width of the shock) deep in the bulk.
Likewise, the gauge function $\lambda_+$ differs significantly from
its background value far in front of the shock.
This behavior is an unavoidable consequence of our use of infalling
Eddington-Finkelstein coordinates, combined with the fact that,
in Fefferman-Graham coordinates, the perturbation to the geometry due
to the shock extends arbitrarily deep into the bulk at any fixed value
of $\tilde t{-}\tilde z$ lying within the shock profile.
Any radially infalling null geodesic which begins at the boundary at
$t = 0$ and some $z \gg w$ eventually intersects the shock which
is moving in the $+z$ direction with unit speed.
Since all events along such a geodesic have common values of the
Eddington-Finkelstein coordinates $t$ and $z$ this shows that,
for any $z > 0$,
sufficiently deep in the bulk,
metric functions
are influenced by the shock.

In the neighborhood of slices with fixed $t{-}z \gg w$,
on which $a_+^{(4)}$ and $f^{(4)}_{z+}$ are negligible,
Einstein's equations imply that the local geometry is AdS$_5$
as long as $ b_+ $ is also negligible.
The geometry only ceases to be AdS$_5$ deep in the bulk where $b_+$ becomes non-negligible.
In other words,
the geometry corresponding the dark red ``background'' region in the left panel
of fig.~\ref{fig:singleshockinitialdata} is simply that of AdS$_5$.

The fact that the functions $b_+$ and $\lambda_+$ are non-zero in front of the shock may appear to constitute
a problem for computing the initial geometry of two colliding shocks
in Eddington-Finkelstein coordinates.
Incoming shocks which, near the boundary, have arbitrarily large separation at some initial time
are, in Eddington-Finkelstein coordinates, already colliding sufficiently deep in the bulk.
In other words, even for shocks which are widely separated on the boundary,
there will always be some region deep in the bulk where the simple superposition
(\ref{eq:shocksuperpose})
of the functions $\{b_\pm, a^{(4)}_\pm, f_{z\pm}^{(4)},\lambda_\pm\}$
is not correct.
However, as we demonstrate below, when the shocks are well separated on the boundary,
the region where the functions $b_{\pm}$ overlap significantly,
and hence where the shocks are already colliding deep in the bulk,
lies inside the apparent horizon and thus is causally disconnected from
the above-horizon geometry.  
Just as seen in the Fefferman-Graham metric (\ref{eq:initialshockgeometry}),
the initial above-horizon geometry 
both between and far away from the shocks is simply AdS$_5$.

Using $b_-(t{+}z,u) = b_+(t{-}z,u)$, we superimpose $b_{\pm}(t,z)$ and define the initial anisotropy function to be
\begin{equation}
b(t_0,z,u) \equiv b_+(t_0{+}z,u) + b_+(t_0{-}z,u),
\end{equation}
for some initial time $t_0$.  We choose $t_0 = -1$ for narrow shocks and $t_0 = -2$ for wide shocks.
Similarly, we have $a^{(4)}_-(t{+}z) =  a^{(4)}_+(t{-}z)$,
$f^{(4)}_{z-}(t{+}z) =  -f^{(4)}_{z+}(t{-}z)$, and 
$\lambda_-(t{+}z) =  \lambda_+(t{-}z)$.
We define
\begin{equation}
a^{(4)}(t_0,z) \equiv a^{(4)}_+(t_0{+}z) + a^{(4)}_+(t_0{-}z), \ \ \ 
f^{(4)}_z(t_0,z) \equiv f^{(4)}_{z+}(t_0{+}z) - f^{(4)}_{z+}(t_0{-}z).
\end{equation}
However, one should not simply superimpose the gauge functions $\lambda_\pm$
since these functions, as defined above, asymptote to non-zero background values away 
from the shocks;
if one simply superimposes
$\lambda_{\pm}$ then the value of the total gauge function $\lambda$ will differ 
significantly from the desired $\lambda_{\pm}$ near the individual shocks.  However, because
the initial above-horizon geometry between the shocks is just that of AdS$_5$,
one can freely adjust the gauge function $\lambda$ between the shocks
and not alter the geometry between them.
We therefore choose to superimpose the functions $\lambda_{\pm}$ via
\begin{equation}
\lambda_{\rm tot}(t_0,z) \equiv \theta(-z)\, \lambda_+(t_0{+}z) + \theta(z)\, \lambda_+(t_0{-}z),
\end{equation}
where $\theta(z)$ is a regularized step function,
\begin{equation}
\theta(z) = \frac{1}{2} \left [ 1 - {\rm erf} \left (- \frac{z}{\sqrt{2} w} \right ) \right ].
\end{equation}
With this choice,
provided $|t_0| \gg w$,
the function $\lambda$ differs negligibly from $\lambda_{\pm}$ in the vicinity of
each shock.

After determining $\{b,a^{(4)},f_z^{(4)},\lambda\}$ on the domain decomposition grid
used to find the transformation functions,
we then interpolate the functions to the spectral grid used to evolve Einstein's equations.
The interpolation is performed using the spectral representations of
the functions in each subdomain, and hence entails no lose of spectral accuracy.  
For the evolution of the geometry,
we choose to use a Fourier grid in the $z$ direction with $N_z$ points,
with periodicity enforced at $z = \pm z_{\rm max}$ with $z_{\rm max} \equiv 10$.
For narrow shock collisions we use $N_z = 801$ and for wide shock collisions we use
$N_z = 401$.
We use domain decomposition in the radial direction with 4 domains, each having
$20$ Chebyshev points.
After computing the functions $\{b,a^{(4)},f_z^{(4)},\lambda\}$ on the new grid,
we then apply a radial gauge transform to reposition the
apparent horizon to radial coordinate $u = 1$.

\begin{figure}
\suck[scale = 0.4]{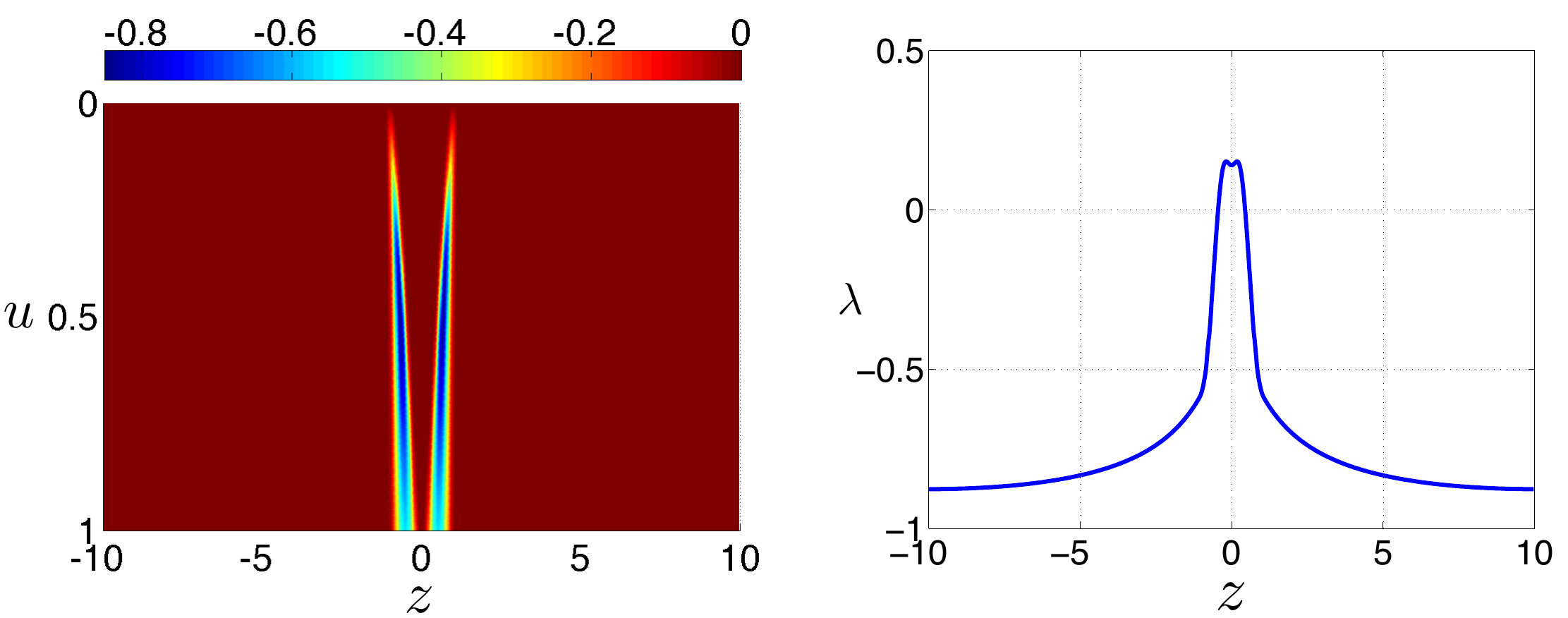}
\vskip -0.05in
\caption{Plots of the anisotropy function $b$ (left) and
gauge parameter $\lambda$ (right) for two incoming narrow shocks
of width $w = 0.075$.
The choice of gauge parameter $\lambda$ is such that the apparent horizon lies at $u = 1$.
In the dark red region where $b$ differs negligibly from zero, the geometry is that of AdS$_5$.  
\label{fig:InitialConditionsShocks}
} 
\end{figure}

Before proceeding, we address two more technical points.
First, in the infinite volume ($z_{\max} \to \infty$) limit,
the apparent horizon asymptotes to the Poincar\'e horizon at Fefferman-Graham coordinate
$\tilde \rho \to \infty$. In this limit, our choice of constant $\tilde \rho_{\rm max}$
 in eq.~(\ref{eq:singleshockgaugechoice}) 
will not yield the the entire above-horizon geometry in the computational domain $0 \leq u \leq 1$.  
This can present a problem since, for any finite choice of $z_{\rm max}$,
one cannot compute the location of the apparent horizon
and thereby know how big $\tilde \rho_{\rm max}$ should be
until the functions $\{b,a^{(4)}, f^{(4)}_z,\lambda\}$ are computed
(which requires a choice of $\tilde \rho_{\rm max}$).
However, the above-horizon pre-collision geometry at large $|z|$ is simply AdS$_5$.
Because of this, one may freely adjust $\lambda(t_0,z)$ at large $|z|$ without changing 
the initial geometry.  In other words, one may make the redefinition
\begin{equation}
\label{eq:Wtrans}
\lambda(t_0,z) \to W(z)\, \lambda(t_0,z),
\end{equation}
with $W(z) = 1$ in the vicinity of the shocks and $W(z)$ arbitrary at large $|z|$.
This freedom allows one to compute and superpose the single shock profiles using
$\tilde \rho_{\rm max} = {\rm const.}$,
and then apply the transformation (\ref{eq:Wtrans}) with
$W(z)$ chosen such that the apparent horizon lies in the computational domain
for any choice of $z_{\rm max}$.  
With this technique, $\tilde \rho_{\rm max}$ need only be chosen large enough such
that the horizon lies in the computational domain $u \leq 1$ near $z = 0$.
We employ this technique and parameterize $W(z)$ via
\begin{equation}
W(z) = \frac{1}{(K - 1)^2} \left [ K+ {\rm erf}\left( -\frac{z + z_0}{\sqrt{2} s} \right ) \right ]
\left [ K+ {\rm erf}\left(\frac{z - z_0}{\sqrt{2} s} \right ) \right ],
\end{equation}
with $K$, $z_0$, and $s$ adjustable parameters.
We choose $K = 21$ and $s = 0.25$.
For narrow shocks we use $z_0 = 3$, and for wide shocks we use $z_0 = 6$.

Second, after computing $\{b,a^{(4)},f_z^{(4)},\lambda\}$ on the grid used to solve
Einstein's equations, but before gauge transforming
to reposition the apparent horizon at $u = 1$, 
we have found it advantageous to filter high momentum modes.  
This helps eliminate numerical noise generated in the numerical calculation of $b$ and $\lambda$.  
We perform the filtering by Fourier transforming $b$ and $\lambda$ in $z$ and then setting the coefficients of 
modes with momentum $|k|  \leq k_{\rm max}/2$ to vanish.  

Fig.~\ref{fig:InitialConditionsShocks} shows the resulting gauge transformed 
initial anisotropy function $b$ and gauge parameter $\lambda$
for incoming narrow shocks with width 0.075.  The apparent horizon is at $u = 1$.
In between the shocks, the functions
$b$, $a^{(4)}$, and $f^{(4)}$ differ negligibly from zero.
As mentioned above, in the neighborhood of $z = {\rm const.}$ slices 
on which $a_+^{(4)} = f^{(4)}_{z+} = b_+ = 0$, Einstein's
equations imply that the local geometry is AdS$_5$;
only deep in the bulk where $b$ becomes significant does the geometry deviate from AdS$_5$.
Therefore, the geometry corresponding the background dark red region in the figure (everywhere
except in the vicinity of the shocks) is simply that of AdS$_5$.
Exactly the same description holds for the wide shock initial data.

\subsubsection{Results}

\begin{figure}
\vskip -0.10in
\begin{center}
\suck[scale = 0.59]{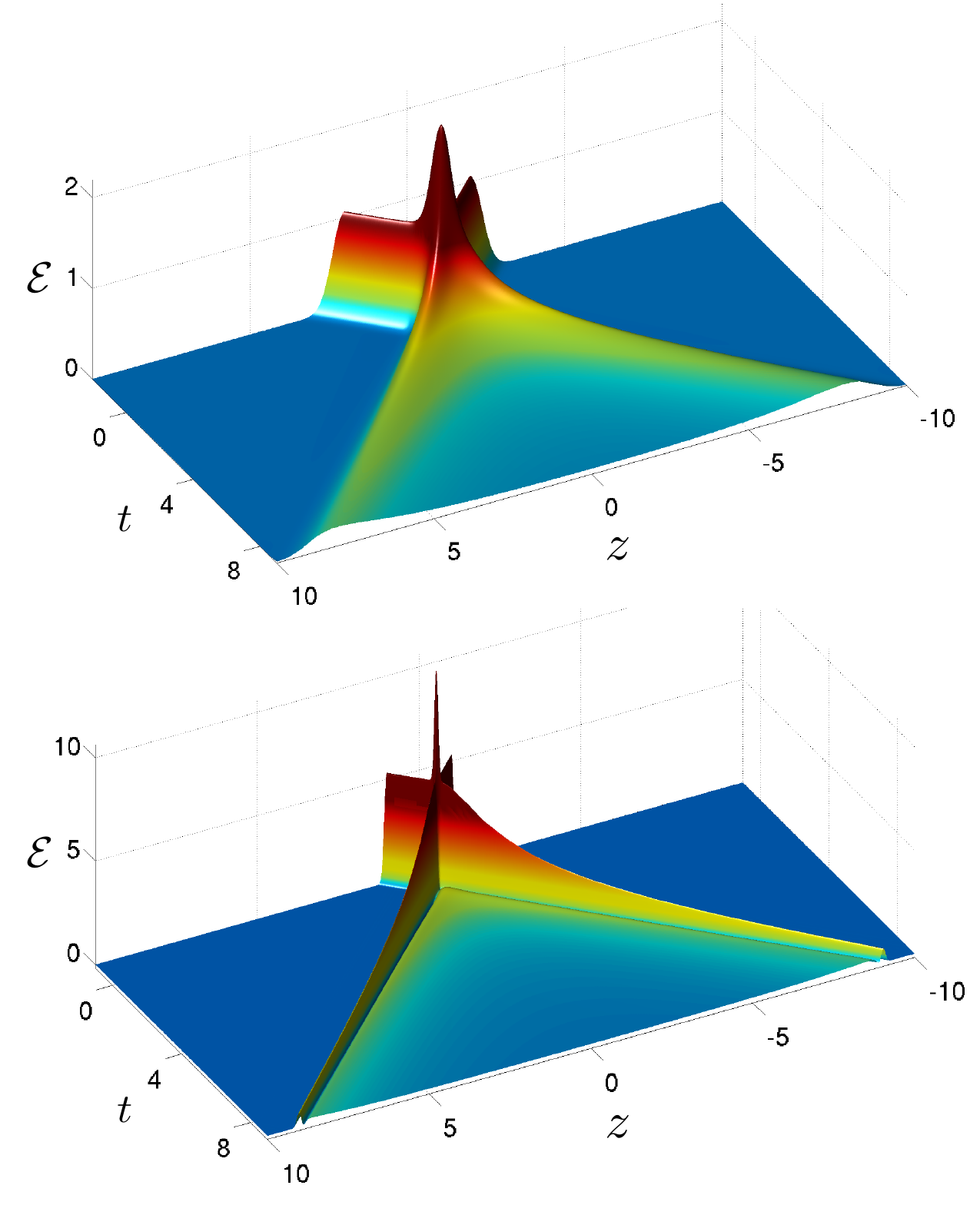}
\end{center}
\vskip -0.35in
\caption{%
Energy density (rescaled by $\kappa = \Nc^2/(2\pi^2)$)
in planar shock collisions,
as a function of time $t$ and longitudinal position $z$.
Top figure: wide shocks with $w= 0.375$.
Bottom figure: narrow shocks with $w = 0.075$.
In both plots, the shocks approach each other along the $z$ axis
and collide at $z = 0$ at time $t = 0$.  The collisions
produce debris that fills the forward light cone.
In the case of narrow shock collisions, the amplitude of
the visible remnants of the shocks on the forward light cone
falls like  $t^{-p}$ with $p \approx 0.9$.%
\label{fig:shocksenergy}
} 
\end{figure}

\begin{figure}
\hspace*{-15pt}%
\suck[scale = 0.35]{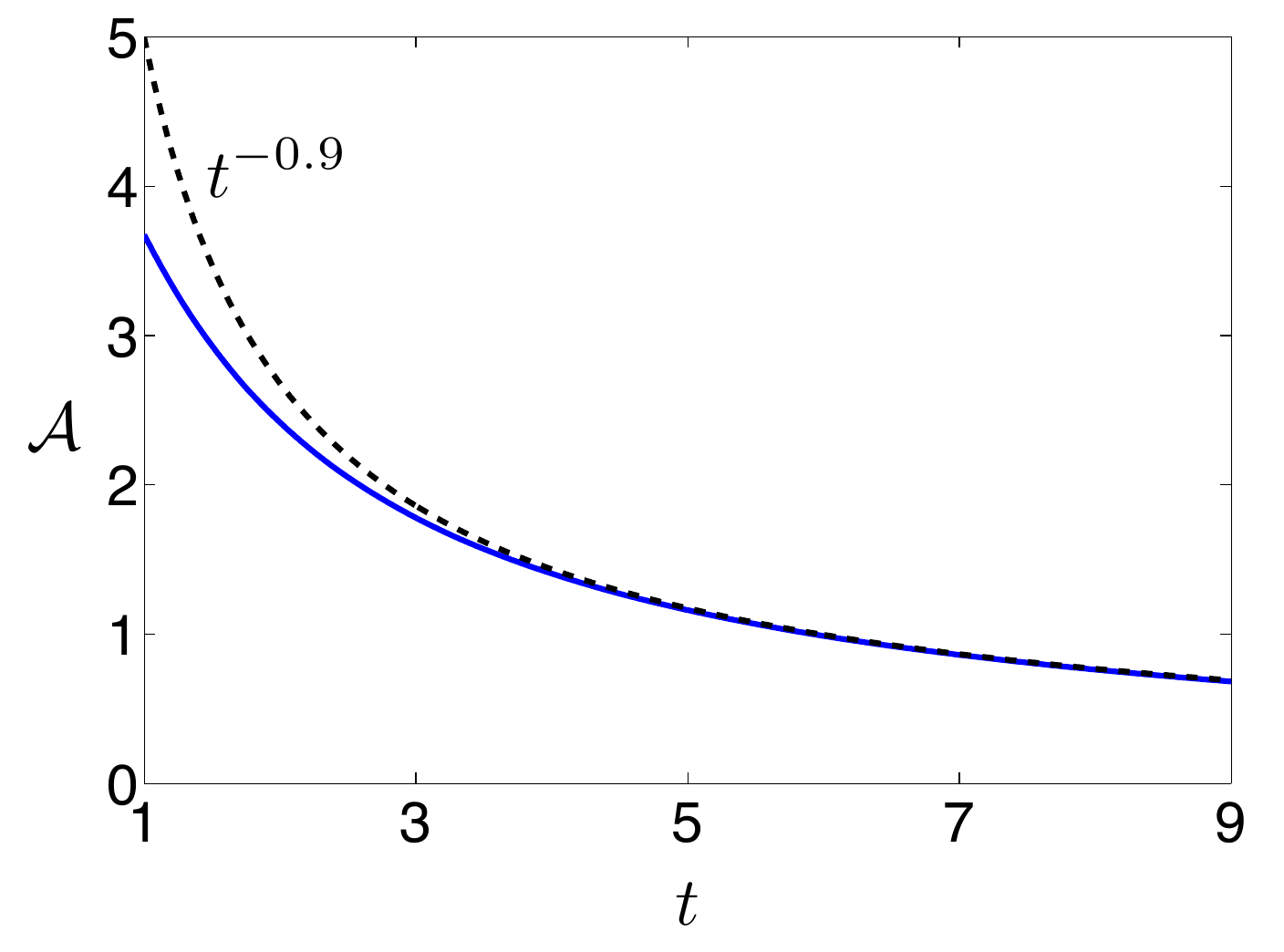}%
\suck[scale = 0.35]{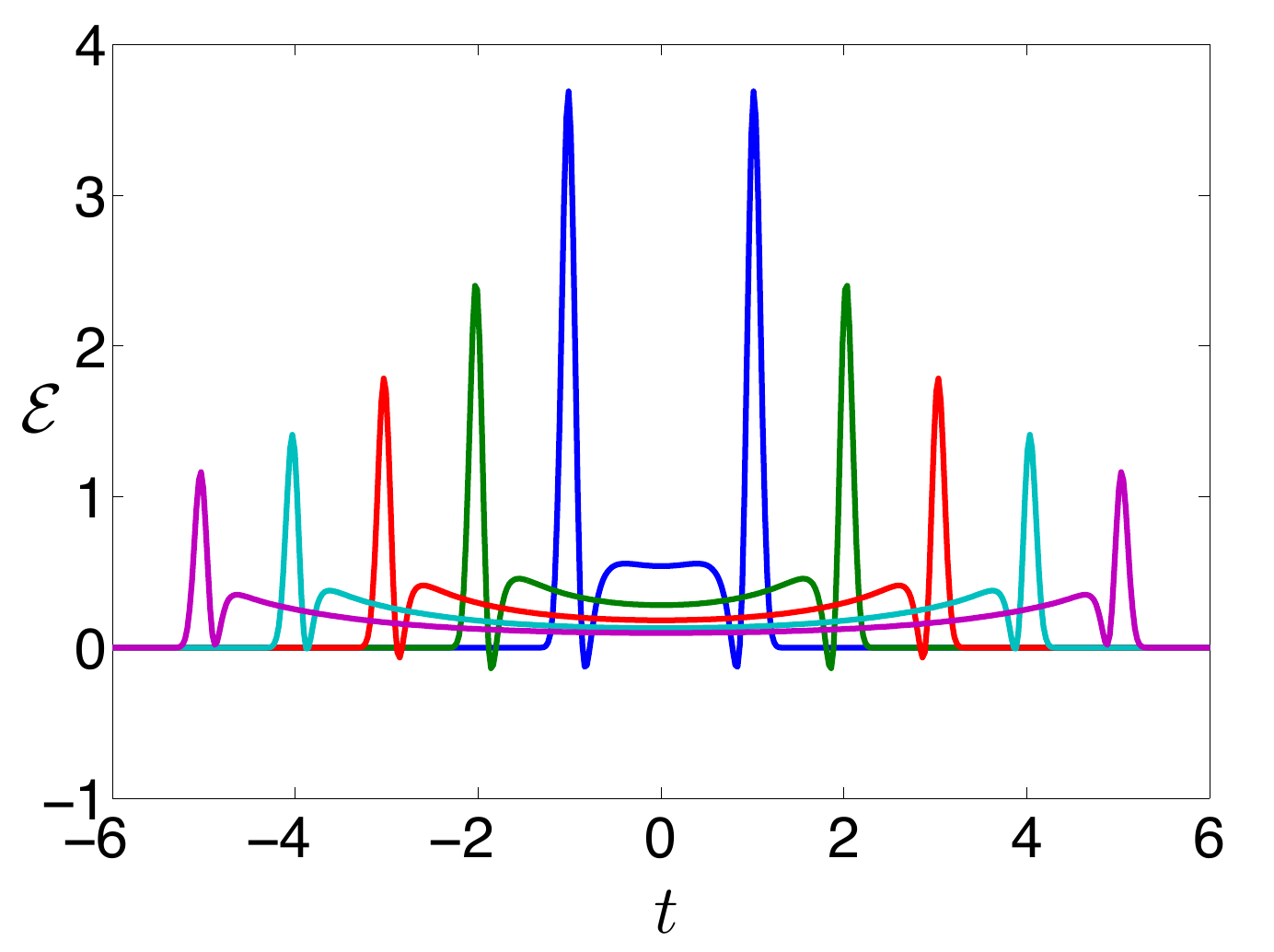}%
\hspace*{-15pt}
\vskip -0.15in
\caption{Left: Plot of the amplitude $\mathcal A$ of the outgoing decaying
null maxima in the energy density, as a function of time,
for the narrow shock collisions.
At late times our results are consistent with
$\mathcal A \sim t^{-p}$ with $p \approx 0.9$.
\label{fig:negenergy}%
Right: Plot of the energy density for the narrow shock collision
at successive times $t = 1,\ 2, \ 3, \ 4, \ 5$.  Small regions 
behind the decaying null maxima with negative energy density are
visible at $t = 1$, 2 and 3.
By time $t = 4$, and thereafter, the energy density is everywhere positive.%
\label{fig:maxofenergy}%
} 
\end{figure}

Fig~\ref{fig:shocksenergy} displays
$\mathcal E \equiv \langle \widehat T^{00} \rangle$,
the energy density rescaled by a factor of $\kappa = \Nc^2/(2\pi^2)$,
for both wide (top) and narrow (bottom) shock collisions.
The shocks approach each other at the speed of light in the $\pm z$ 
direction and collide at $z = 0$ at time $t = 0$.
For both cases, the debris leaving the collision event
appears dramatically different than the initial incoming shocks.
Prior to the collision, all the shock energy 
lies near the lightcone (smeared only by the width of the shock),
while long after the collision nearly all the energy lies
inside the lightcone.
Inspecting fig.~\ref{fig:shocksenergy},
one sees qualitative differences between narrow and wide shock collisions.
For wide shocks,
there is no sign of any distinct remnant of the shock remaining
on the forward light cone;
the energy density of the post-collision debris is smoothly distributed
in the interior of the forward light cone \cite{CY:shocks}.  
In contrast, for the narrow shock collisions
there are clear remnants of the initial shocks 
propagating outward
on the forward light cone \cite{Casalderrey-Solana:2013aba}.  
But, as can easily be seen in fig.~\ref{fig:shocksenergy},
immediately after the collision energy density is
transported inside the lightcone and the portion remaining
very near the lightcone steadily attenuates.
On the left side of fig.~\ref{fig:maxofenergy}
we plot the amplitude $\mathcal A$ of the energy density on the lightcone as a 
function of time for the narrow shock collisions.
At late times
our results are consistent with the power-law decay $\mathcal A \sim t^{-0.9}$.
By time $t = 9$,
the amplitude of the null maxima has decreased to 13\% its pre-collision value.
Evidently, for both wide and narrow shocks the collision event results in
the subsequent annihilation of the shocks
with essentially all energy lying well inside the
forward light cone at late times.  

\begin{figure}
\begin{center}
\suck[scale = 0.35]{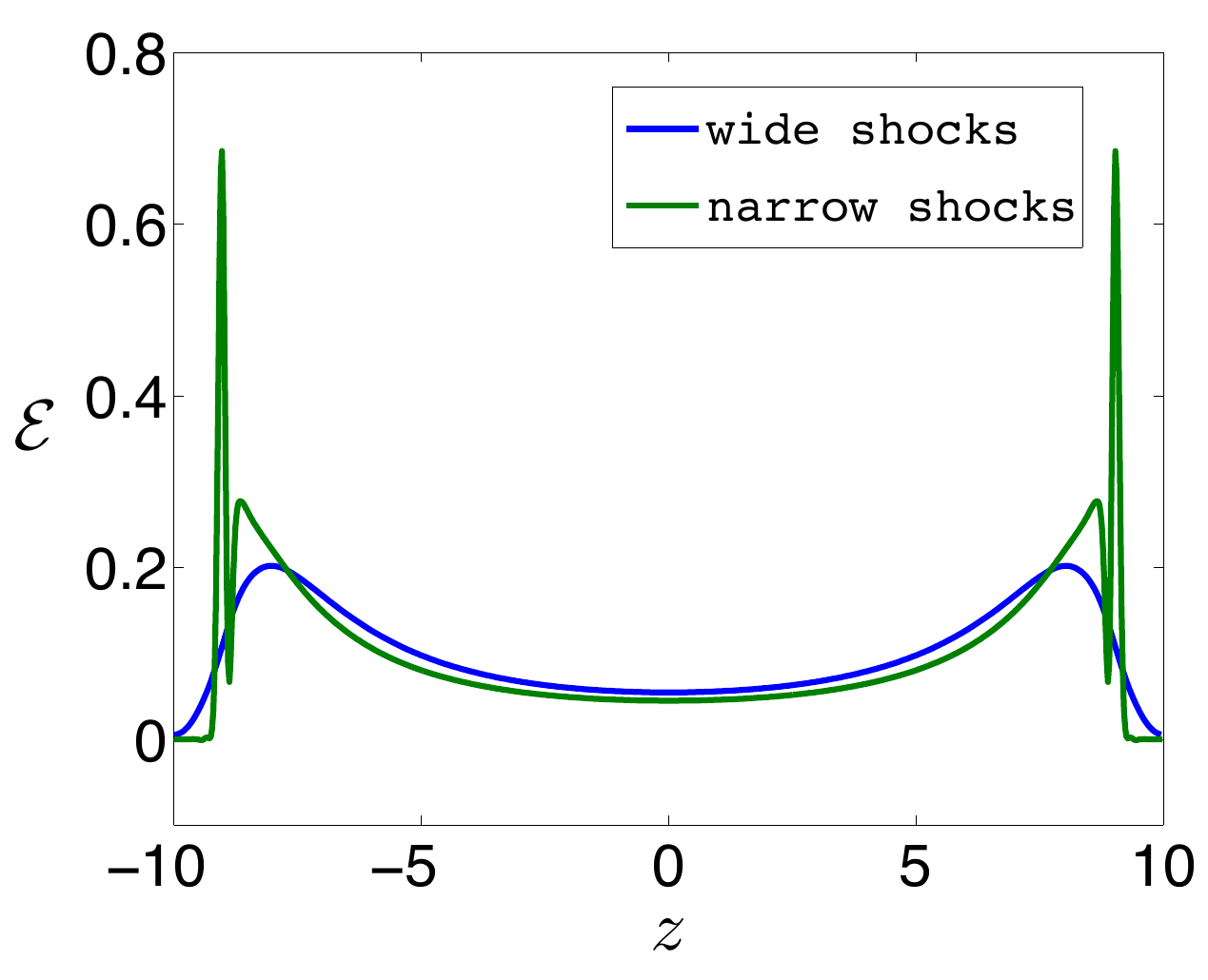}
\end{center}
\vskip -0.30in
\caption{Comparison of the longitudinal distribution of energy density for
wide and narrow shocks at time $ t = 9$.%
\label{fig:energycomparison}%
} 
\end{figure}

Aside from the decay of the null peaks in the energy density, there is
another qualitative difference between collisions of narrow and wide shocks.
On the right side of
fig.~\ref{fig:negenergy} we plot the energy density for the narrow shock
collision at successive times $t = 1, \ 2, \ 3, \ 4, \ 5$.
As is evident from the figure,
there is a brief period of time after the collision when the energy density
just behind the receding null peaks is locally negative
\cite{Casalderrey-Solana:2013aba}.  However, by time $t = 4$ 
the energy density is everywhere positive,
just as it always is for wide shock collisions.
Evidently, 
the presence of negative energy density is a transient effect.
Indeed, as shown in fig.~\ref{fig:energycomparison}, aside from the decaying 
null maxima on the light cone, at late times the distribution of
energy density produced by
both wide and narrow shock collisions looks quite similar.

It is instructive to compare our results with predictions from the
fluid/gravity correspondence~\cite{Hubeny:2011hd}.
In the limit of asymptotically slowly varying fields (compared to the dissipative scale set by the local temperature $T$ of the system)
Einstein's equations (\ref{eq:Einstein}) can be solved perturbatively with
a gradient expansion 
\begin{equation}
g_{MN}(x,r) \sim \sum_{n = 0}^{\infty} \> g_{MN}^{(n)}(x,r)\,, 
\end{equation}
where $g_{MN}^{(n)}$
is of order $\left (\partial/\partial x^\mu \right )^n$ in 
boundary spacetime derivatives \cite{Bhattacharyya:2008jc}.
Via eq.~(\ref{eq:Tmunu}), this implies that 
the boundary stress tensor also admits a gradient expansion.
In $\dim = D{-}1$ spatial dimensions, the resulting gradient expansion of the
boundary stress begins
\begin{equation}
\label{eq:hydroconst}
    T^{\mu \nu}_{\rm hydro} = \frac{\kappa\varepsilon}{\dim}
    \left [\eta^{\mu \nu} + (\dim{+}1) \, u^\mu u^\nu \right ]
    - 2 \eta \, \sigma_{\mu \nu} + O(\partial^2)\,,
\end{equation}
where $\varepsilon$ is the (rescaled) proper energy density,
$u$ the fluid velocity, $\eta$ the shear viscosity, and 
\begin{equation}
\label{eq:shear}
    \sigma_{\mu \nu } \equiv
    \half \left[\, \partial_\mu u_\nu
    + \partial_\nu u_\mu
    + u^\rho \partial_\rho (u_\mu u_\nu) \right]
    - \tfrac{1}{\dim}\, (\partial_\alpha u^\alpha)
	\left [ \eta_{\mu \nu} {+}u_\mu u_\nu \right].
\end{equation}
is the relativistic shear tensor
(which is symmetric, traceless, and orthogonal to the flow velocity $u$).
The fluid velocity and proper energy density
satisfy $T^{\mu \nu}_{\rm hydro} \, u_\nu = - \kappa\varepsilon \, u^\mu$.
Moreover, the fluid/gravity gradient expansion 
yields expressions for all transport coefficients as functions of the
proper energy density. 
For $D = 4$, the shear viscosity $\eta = \frac{1}{4} \kappa (\pi T)^3$,
where the local temperature $T$ is defined by
$\varepsilon = \frac{3}{4} \kappa (\pi T)^4$ \cite{Policastro:2001yc}.
Eq.~(\ref{eq:hydroconst}) is precisely the constitutive relation of
first order relativistic conformal hydrodynamics.  

To compare our numerical results with the asymptotic predictions of the
fluid/gravity correspondence,
we first extract the fluid velocity $u$ and rescaled
proper energy density $\varepsilon$
from the numerically computed stress-energy tensor
(by finding the timelike eigenvector and associated eigenvalue
of $\langle \widehat T^{\mu}_{\ \nu} \rangle$,
as discussed in section \ref{sec:setup}).
With $u$ and $\varepsilon$ obtained via eq.~(\ref{eq:flowfield}),
we then use eq.~(\ref{eq:hydroconst}) to 
construct the hydrodynamic approximation to the spatial
stress tensor, $T^{ij}_{\rm hydro}$.
Rotational symmetry in the transverse plane implies that all
off-diagonal elements of the spatial stress tensor vanish,
and that $\langle T^{xx}\rangle = \langle T^{yy}\rangle$.
Therefore, we define a simple dimensionless residual function,
\begin{equation}
\mathcal R \equiv  \frac{1}{p_{\rm ave}}
\left[{ \left (\langle T^{xx}\rangle  - T^{xx}_{\rm hydro} \right )^2 + \left (\langle T^{zz} \rangle - T^{zz}_{\rm hydro} \right )^2 }\right]^{1/2},
\end{equation}
where the average pressure
$
    p_{\rm avg} \equiv
    \frac 23 \langle T^{xx}\rangle + \frac 13 \langle T^{zz}\rangle
$.
The residual $\mathcal R$ gives a measure of the relative deviation of
the spatial stress from the prediction of the hydrodynamic constitutive
relation (\ref{eq:hydroconst}).
\begin{figure}
\vskip -0.20in
\begin{center}
\suck[scale = 0.55]{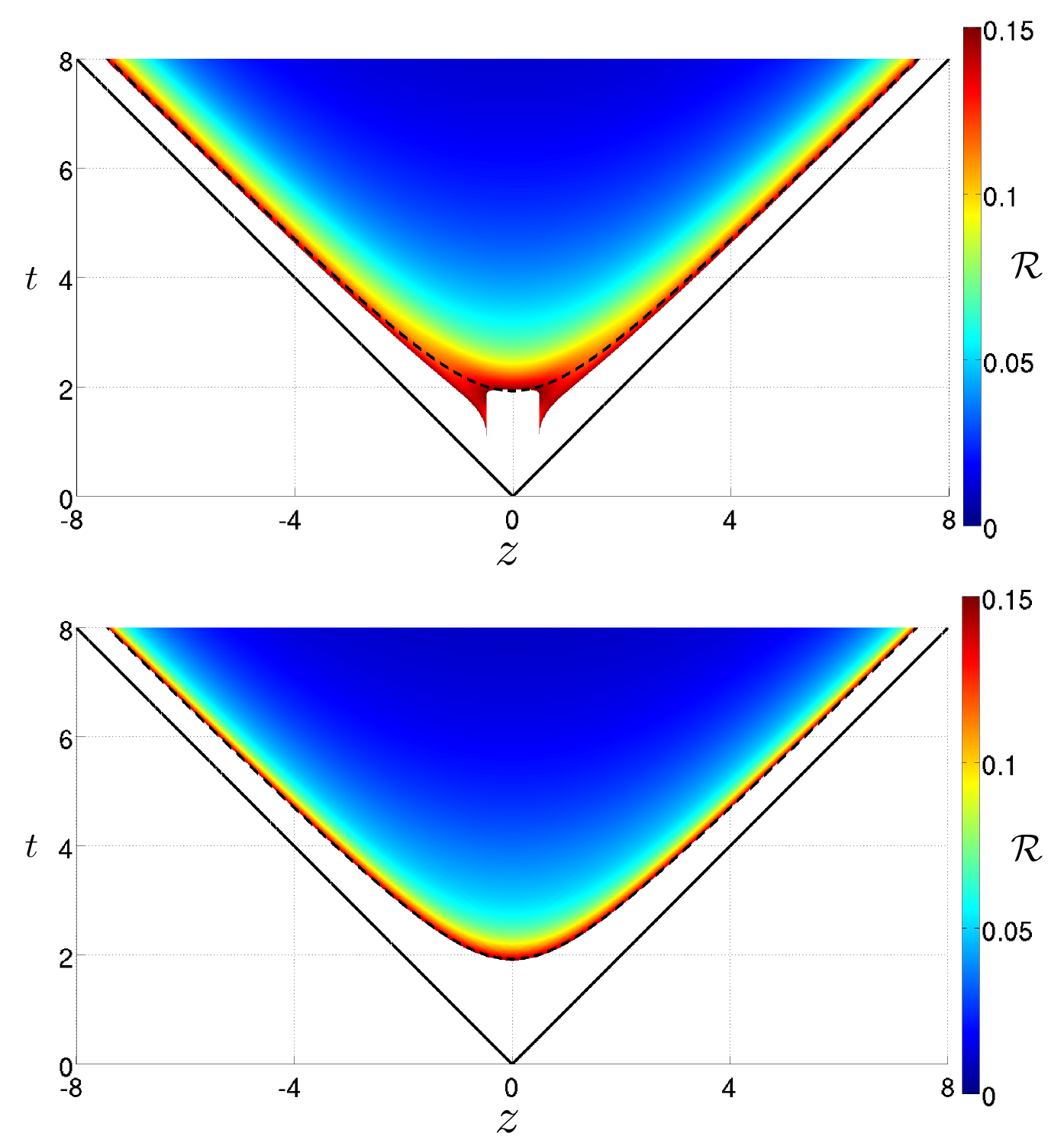}
\end{center}
\vskip -0.30in
\caption{%
The relative deviation $\mathcal R$ of the
spatial stress tensor from prediction of first order viscous hydrodynamics
for the case of wide shocks (top) and narrow shocks (bottom).
As detailed in the text, we only display the region
$\mathcal H = \{(t,z)\!: \mathcal R(t,z)  \leq 0.15\}$
where the residual is no more than 0.15.
The dashed curve, discussed in the text,
is defined by eq.~(\ref{eq:hydrodynamizationtime}).
For both cases, viscous hydrodynamics becomes a good description 
near mid rapidity when $t \gtrsim 2$.
\label{fig:hydroresidual}
} 
\end{figure}
Fig.~\ref{fig:hydroresidual} plots $\mathcal R$ for collisions of
both wide shocks (top) and narrow shocks (bottom).
In each plot we exclude the region where $\mathcal R > 0.15$.  Specifically, 
for every value of $z$,
we define $t_*(z)$ as the last time for which $\mathcal R(t,z) > 0.15$
and exclude from the plot all points $(t,z)$ for which $t \leq t_*(z)$.  
We will denote by $\mathcal H$ the region where viscous hydrodynamics
works at  the 15\% level or better (as measured by $\mathcal R$).
The dashed line in each plot is the curve
\begin{equation}
\label{eq:hydrodynamizationtime}
\tau_{\rm hydro}^2 = (t - \Delta t)^2 - z^2,
\end{equation}
with $ \Delta t = 0.43$ and $\tau_{\rm hydro} = 1.5$
which, as seen in the figure, nicely approximates the boundary of
region $\mathcal H$.
Fig.~\ref{fig:hydroresidual} clearly shows that our planar shock collisions
result in the formation of an expanding volume of fluid which is
well described by hydrodynamics everywhere except near the light cone,
where non-hydrodynamic effects become important.
At mid-rapidity, viscous hydrodynamics becomes a good description
when $t \gtrsim 2$ \cite{CY:shocks}.

\begin{figure}
\begin{center}
\suck[scale = 0.35]{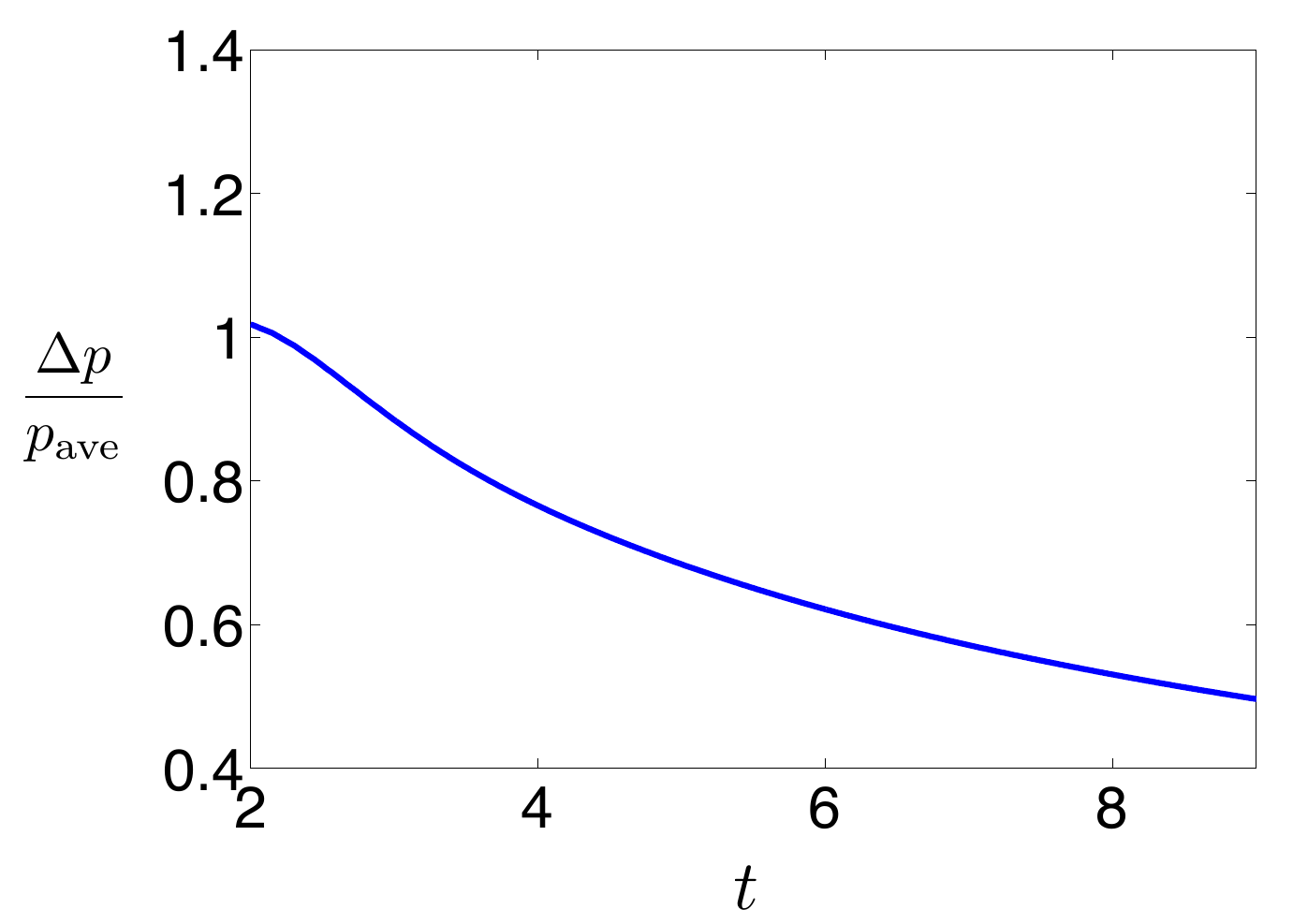}
\end{center}
\vskip -0.20in
\caption{%
The normalized anisotropy in the spatial stress,
$\Delta p/p_{\rm avg}$, at $z = 0 $ for the narrow shocks.
The anisotropy is $O(1)$ indicating that viscous effects are not small
compared to the ideal fluid terms.%
\label{fig:anisotropy}
} 
\end{figure}

As was noted in refs.~\cite{CY:shocks,CY:boostinvar},
even in the region $\mathcal H$ where viscous
hydrodynamics works at the  15\% level or better,
the first order viscous corrections are not small.
The viscous stress tensor $-2 \eta \sigma_{\mu \nu}$ in eq.~(\ref{eq:hydroconst}) can be just as large as the zeroth order ideal fluid term.
One manifestation of this is that in the local rest frame of the fluid
(where $u^\mu = \delta^\mu_0$),
the spatial stress $\langle T_{ij}^{\rm local} \rangle$ 
can be highly anisotropic with very different eigenvalues
(\textit{i.e.} pressures) in each direction.
In the local fluid rest frame, this anisotropy is solely due to the
gradient corrections in eq.~(\ref{eq:hydroconst}).
To illustrate this point,
fig.~\ref{fig:anisotropy} plots, for narrow shocks, the difference
$\Delta p = \langle T_{xx} \rangle  - \langle T_{zz} \rangle$ in the
eigenvalues of the spatial stress at $z = 0$
(where by $z \to -z$ symmetry the fluid is at rest),
normalized by the average pressure $p_{\rm avg}$.
As just asserted, $\Delta p/p_{\rm avg}$ is $O(1)$.
Given the size of the first order gradient corrections,
it is quite remarkable that the hydrodynamic constitutive
relation works so well.

It is also illuminating to examine how well
boost invariant flow approximates our numerical results.
As the name suggests, boost invariant flow is defined by the 
condition that the system be invariant under arbitrary boosts
in the longitudinal direction.
Our initial conditions corresponding to two colliding shocks with
non-zero widths are not boost invariant, and hence neither is the debris
produced by the collision.
Nevertheless, in a qualified sense which we make precise below,
the produced debris does display some characteristics of nearly
boost invariant flow.
In what follows we focus on the case of narrow shock collisions,
and on the dynamics in the region $\mathcal H$,
shown in fig.~\ref{fig:hydroresidual},
where viscous hydrodynamics is applicable at the 15\% level.

From the fluid/gravity correspondence,
the fluid velocity and proper energy density (rescaled by $\kappa$)
for boost invariant flow,
up to second order in gradients, are given by
\cite{Heller:2008mb}
\begin{subequations}
\begin{align}
& u_\mu \, dx^\mu = d\tau \equiv \cosh y \, dt + \sinh y \, dz \,,
\\\label{eq:BIflow}
&\varepsilon =
\tfrac 34 \frac{(\pi \Lambda)^4}{(\Lambda \tau)^{4/3}}
\left [ 1  - \frac{C_1}{(\Lambda \tau)^{2/3}}
    + \frac{C_2}{(\Lambda \tau)^{4/3}}
    + O\Big( \frac 1{(\Lambda \tau)^2}\Big)
\right ],
\end{align}
\end{subequations}
where $\tau \equiv \sqrt{t^2 - z^2}$ is proper time,
$y \equiv \tanh^{-1} \frac{z}{t}$ is rapidity, and 
\begin{subequations}
\begin{align}
C_1 &= \frac{2}{3 \pi} \approx 0.21\,,
\qquad
C_2 = \frac{1 + 2 \log 2}{18 \pi^2} \approx 0.013\,.
\end{align}
\end{subequations}
The energy scale $\Lambda$ is set by initial conditions and is otherwise arbitrary.
Each subsequent gradient correction to the proper energy density is suppressed by an additional power of $(\Lambda \tau)^{-2/3}$;
for boost invariant flow, the fluid/gravity gradient expansion
is precisely a late time expansion in inverse powers of proper time.

\begin{figure}
\begin{center}
\suck[scale = 0.38]{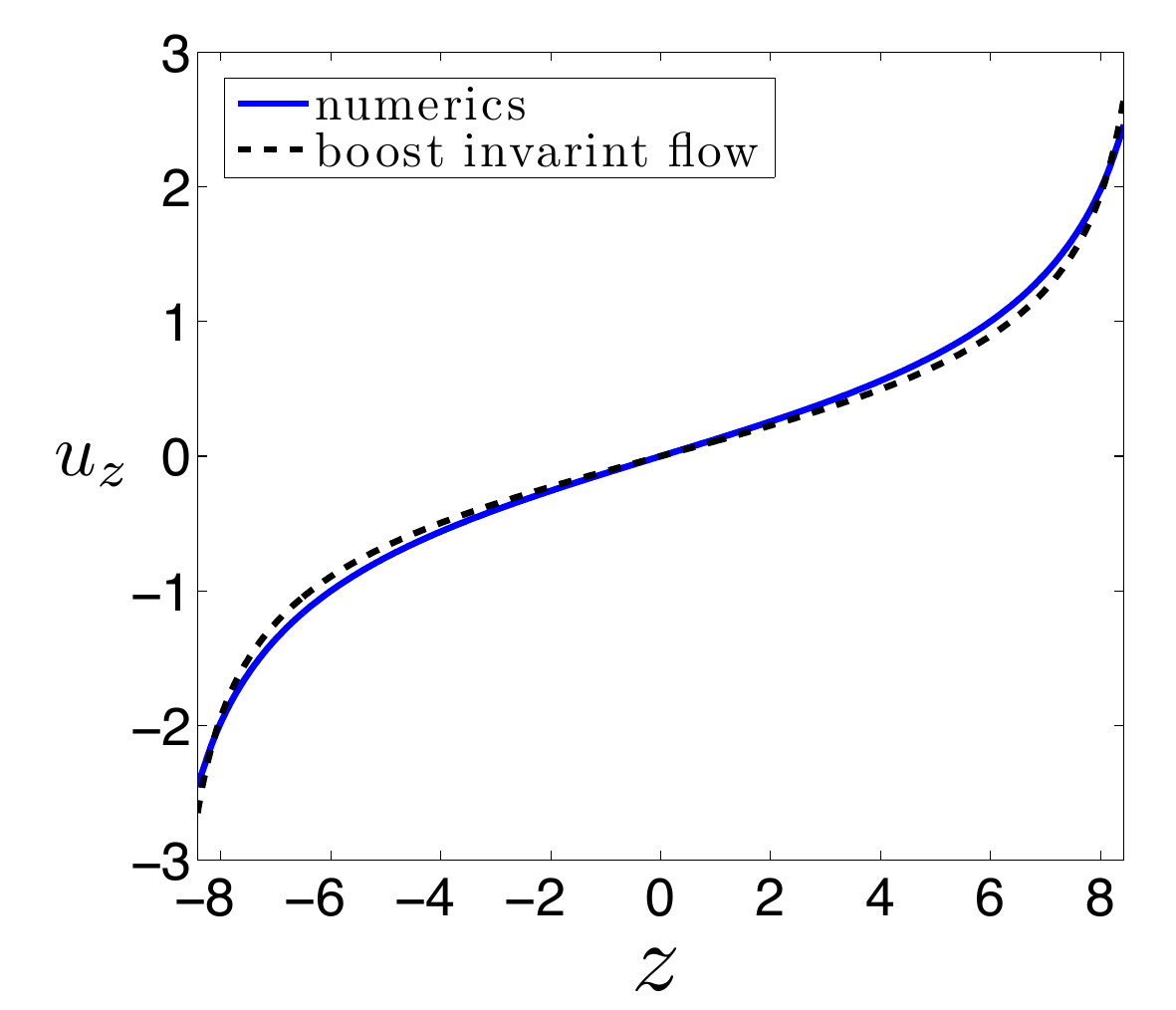}
\end{center}
\vskip -0.20in
\caption{%
The longitudinal fluid velocity $u_z$ for narrow shock collisions,
at time $t = 9$, in the region $\mathcal H$ where viscous hydrodynamics 
works at the 15\% level or better.
The boost invariant flow result, $u_z = {z}/\tau$,
fits the numerical result quite well.
\label{fig:vel2BI}
} 
\end{figure}

Our first comparison to boost invariant flow is shown in fig.~\ref{fig:vel2BI},
where we plot the longitudinal component $u_z$ of the fluid velocity
at time $t = 9$ for the narrow shock collision.
Also shown in the plot is the 
boost invariant flow result $u_z = \sinh y = {z}/{\tau}$.
Again, we display $u_z$ only in the region $\mathcal H$ 
where viscous hydrodynamics works at the 15\% level or better.
As is evident from the figure, the numerical result agrees quite nicely
with this prediction of boost invariant flow.

\begin{figure}
\begin{center}
\suck[scale = 0.35]{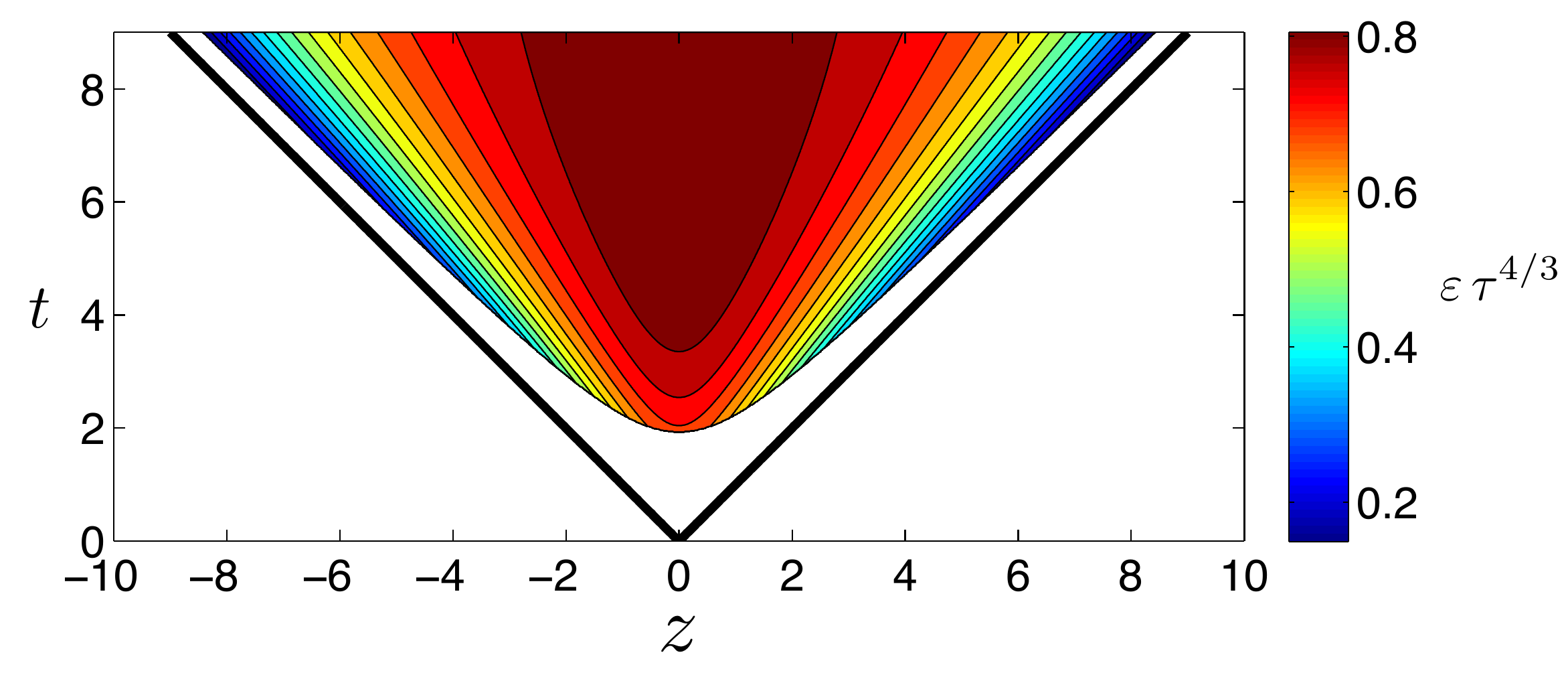}
\end{center}
\vskip -0.20in
\caption{%
The normalized proper energy density $\varepsilon \, \tau^{4/3}$
in the region $\mathcal H$ for the narrow shock collision.
At late times,
lines of constant $\varepsilon \tau^{4/3}$ are approximately
straight lines from the origin, $t \approx z \coth y$.%
\label{fig:properenergycontours}%
} 
\end{figure}

Fig.~\ref{fig:properenergycontours} shows a contour plot of the 
proper energy density $\varepsilon$
extracted from our numerical results and
multiplied by a factor of $\tau^{4/3}$.
Lines through the origin corresponds to events with fixed rapidity,
$t = z \coth y$.
Inspecting eq.~(\ref{eq:BIflow}), it is evident 
that if the flow was truly boost invariant then
$\varepsilon \tau^{4/3}$ would asymptote to a constant,
independent of rapidity, in the $\tau \to \infty$ limit.
Fig.~\ref{fig:properenergycontours} shows that this is not at all the case;
the flow is not \textit{globally} boost invariant
(as was also found in ref.~\cite{Casalderrey-Solana:2013aba}).
However, one striking feature of fig.~\ref{fig:properenergycontours} is
that contours of $\varepsilon \tau^{4/3}$, at late times,
are approximately straight lines through the origin,
$t \approx  z \coth(y)$.
This observation suggests that on each slice of constant rapidity $y$,
the proper energy density is 
approximately given by eq.~(\ref{eq:BIflow}) but with a 
rapidity dependent scale parameter, $\Lambda = \Lambda(y)$.
\begin{figure}
\begin{center}
\hspace*{-40pt}
\suck[scale = 0.32]{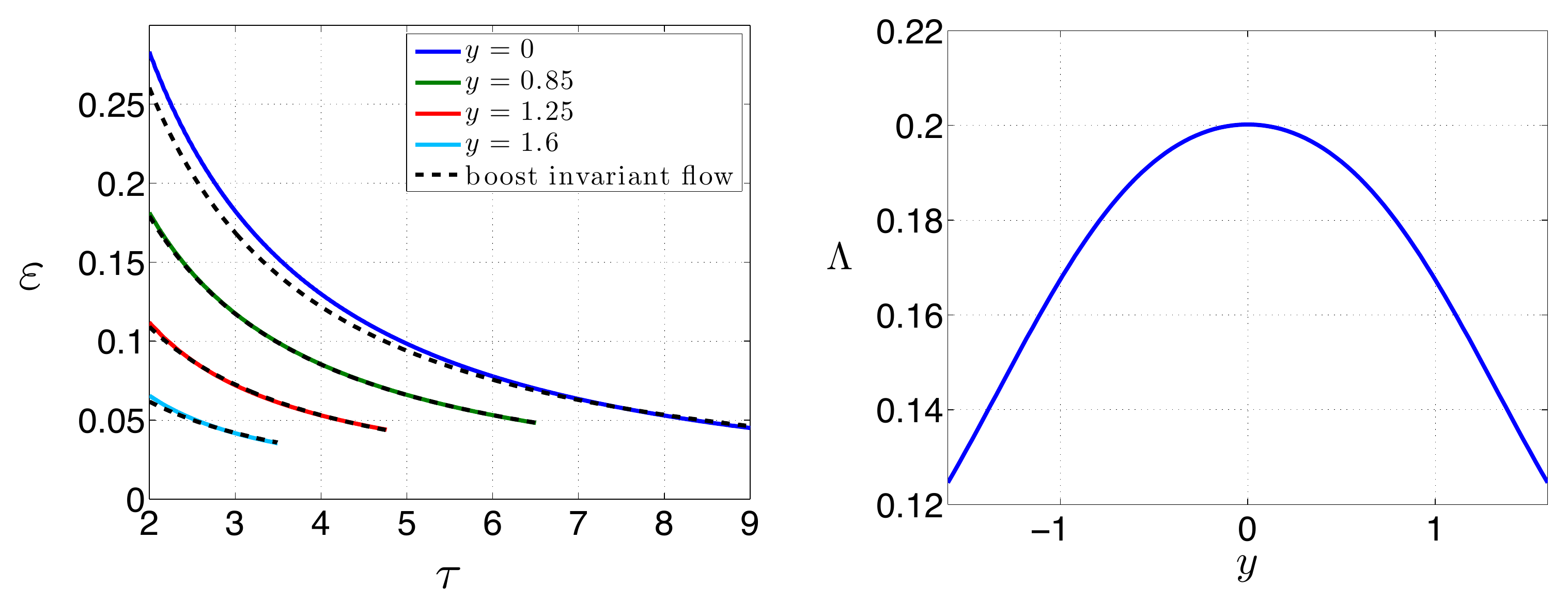}
\hspace*{-20pt}
\end{center}
\vskip -0.15in
\caption{%
Left: the proper energy density $\varepsilon$ on several slices of constant rapidity $y$, as a function of proper time $\tau$.
The dashed curves are fits to the boost invariant flow result
(\ref{eq:BIflow}) with a rapidity dependent scale parameter $\Lambda(y)$.
Right: the resulting scale parameter $\Lambda(y)$ as a function of rapidity.%
\label{fig:energy2bI}%
} 
\end{figure}
To test this hypothesis, on each slice of constant $t/z = \coth y$ we fit
the proper energy density $\varepsilon$ to the boost invariant expression
(\ref{eq:BIflow}) allowing $\Lambda$ to depend on $y$.
In the left panel of fig.~\ref{fig:energy2bI} we plot $\varepsilon$ at
$ y = 0$, 0.85, 1.25, and 1.6,
and the corresponding fit to eq.~(\ref{eq:BIflow}).
The agreement with eq.~(\ref{eq:BIflow}) is remarkable.
In the right panel of fig.~\ref{fig:energy2bI} we plot the resulting
scale parameter $\Lambda(y)$ emerging from this fit to local
(in rapidity) boost invariant flow.

It would be interesting to study more carefully the dependence of
$\Lambda(y)$ on the width of the incoming shocks,
and to evolve longer in time in order to examine the asymptotic behavior
of $\Lambda(y)$ at large rapidity.

\subsection	{Two-dimensional turbulence}
\label		{sec:2dfluid}

\subsubsection{Motivation}

Turbulent flows in relativistic boundary
conformal field theories with $\nu$ spatial dimensions should be dual,
via holography, to dynamical black hole solutions in  
asymptotically AdS$_{\nu+2}$ spacetime.
This connection raises many interesting questions in gravitational physics.
For example, what distinguishes turbulent black holes from 
non-turbulent ones?
And what is the gravitational origin of the Kolmogorov scaling and
energy cascades observed in turbulent fluid flows? 

Gravitational dynamics may also provide insight into turbulence,
in particular for problems where microscopic physics plays a crucial
role in turbulent evolution.
For superfluids, whose turbulent evolution is not governed by ordinary
hydrodynamics, holography has already yielded insight into
two dimensional turbulent flows \cite{Adams:2012pj}.
In particular, ref.~\cite{Adams:2012pj}
found that turbulence in a two dimensional holographic superfluid exhibits a 
direct energy cascade into the UV.
This stands in stark contrast to turbulence in normal fluids 
in two spatial dimensions,
where enstrophy conservation gives rise to an inverse cascade to the IR.
A fully consistent microscopic description of normal turbulence in
three dimensions may also prove useful.
Turbulence in three spatial dimensions is characterized by a cascade of
energy from the IR to the UV,
with dissipation occurring at microscopic length scales which may
lie outside the hydrodynamic regime governed by the Navier-Stokes equation.
Via holography, black hole solutions to Einstein's equations
provide a laboratory in which one can
study the domain of validity and late-time regularity
of turbulent solutions to the Navier-Stokes equation.

In this section, we numerically construct black hole solutions
in asymptotically AdS$_4$ spacetime dual to $\nu \,{=}\, 2$ turbulent flows,
where energy flows from the UV to the IR in an inverse cascade.
The following discussion summarizes work first presented 
in ref.~\cite{Adams:2013vsa}.

\subsubsection{Setup}

The boundary dimension $D = 3$.
We choose an explicit parameterization of the rescaled spatial metric
$\hat g_{ij}$ that manifestly satisfies $\det \hat g = 1$,
\begin{equation}
    ||\hat g_{ij} || = \left[ \begin{array}{cc}
    e^{B} \cosh C & \sinh C \\
    \sinh C & e^{-B} \cosh C
    \end{array} \right].
\end{equation}
From the series expansions (\ref{eq:asymp}) we see that $B$ and $C$ have the near-boundary asymptotics
\begin{align}
    B(x,u) \sim u^3 \, B^{(3)}(x) + O(u^4)\,, \qquad
    C(x,u) \sim u^3 \, C^{(3)}(x) + O(u^4)\,.
\label{eq:2dasym}
\end{align}
We choose to replace the field redefinitions (\ref{eq:redefs})
involving the spatial metric
with the following field redefinitions for the spatial metric functions
\begin{align}
    b \equiv u^{-2} \, B\,,\quad
    \dot b \equiv u^{-1} \, d_+ B\,,\quad
    c \equiv u^{-2} \, C\,,\quad
    \dot c \equiv u^{-1}  \, d_+ C\,.
\end{align}
The asymptotic behavior (\ref{eq:2dasym}) implies that
$b = \dot b = c = \dot c = 0$
at the AdS boundary $u = 0$.
Therefore, when solving eq.~(\ref{eq:gdoteqn}) for $\dot b$ and 
$\dot c$, we impose the Dirichlet boundary conditions
$\dot b = \dot c = 0$ at $u = 0$.

We choose initial conditions corresponding to a locally boosted black brane.
With our metric ansatz,
\begin{equation}
    ds^2 =
    r^2  g_{\mu\nu}(x,r) \, dx^\mu \, dx^\nu + 2 \, dt \, dr
    \,,
\label{eq:ansatz2}
\end{equation}
a boosted black brane geometry
is described by
\begin{equation}
\label{eq:gradmetric}
    g_{\mu\nu}(x,r) =
    \left( \frac{\mathcal R(x,r)}{r} \right)^2
    \left [ \eta_{\mu \nu} +
	\left (\frac{\rh(x)}{\mathcal R(x,r)} \right )^3 u_\mu(x)\, u_\nu(x)
    \right ],
\end{equation}
where $u^\mu(x)$ is the boost velocity and $\rh(x) \equiv 4 \pi T(x)/3$,
with $T(x)$ the local temperature of the brane.
(We are using simple Cartesian boundary coordinates for the boundary geometry.)
The function $\mathcal R(x,r)$ satisfies 
\begin{equation}
    \frac{\partial \mathcal R(x,r)}{\partial r}
    =\Big[1 + \rh(x)^3 \, \mathcal R(x,r)^{-3}  \, \bm u(x)^2 \Big]^{1/2},
\end{equation}
where $\bm u^2 \equiv u^i u_i$.
For constant values of $u_\mu$ and $T$,
the metric (\ref{eq:gradmetric}) is an exact solution to Einstein's equations.

After applying the time-space split (\ref{eq:names}) to $g_{\mu\nu}$,
the initial data for integrating Einstein's equations consists of
the rescaled spatial metric with unit determinant,
\begin{subequations}
\label{eq:initialdataturb}
\begin{equation}
    \hat g_{ij}(\x,r)
    =
    \frac{ \delta_{ij}
	    + \rh(\x)^3 \, \mathcal R(\x,r)^{-3} \>
		u_i(\x) \, u_j(\x) }
    {\Big[
	1 + \rh(\x)^3 \, \mathcal R(\x,r)^{-3} \> \bm u(\x)^2
    \Big]^{1/2} }\,,
\end{equation}
together with the asymptotic coefficients describing the energy
and momentum density on the initial slice,
\begin{equation}
    a^{(3)}(\x) = - \half \, \rh(\x)^3
	\left [ -1+ 3 \, u_0(\x)^2 \right ],\qquad
    f_i^{(3)}(\x) = -\rh(\x)^3 \, u_0(\x) \, u_i(\x)\,,
\end{equation}
\end{subequations}
with all functions evaluated at the initial time $t_i \equiv 0$.


We apply the above setup to the specific case of a boost velocity
with sinusoidal variations plus
small random perturbations
(which serve to break the symmetry of the initial conditions),
\begin{equation}
    u_i(\x)
    =
    \cos (Q x^1) \, \delta_i^2 + \delta u_i(\x) \,.
\label{eq:initialvelocity}
\end{equation}
We study evolution in a periodic square spatial box of size $L_s$
and choose the wavevector $Q= 10 \pi/L_s$.
The small fluctuations $\delta u_i$ are chosen
to be a sum of the first four spatial Fourier modes with random coefficients,
with the overall amplitude of the fluctuation adjusted to make
$|\delta u_i(\x)|_{\rm max} = 1/5$.  These initial conditions are unstable and capable of producing subsequent turbulent evolution
if the Reynolds number $Re$ is sufficiently large. For our initial conditions $Re \sim L_s T $.  
We choose box size $L_s = 1500$ and the initial temperature $4 \pi T/3 = 1$.

A linear combination of the first $20$ Chebyshev polynomials
is used to represent the radial dependence of all functions,
while an expansion of 305 plane waves (in each direction)
is used to represent the spatial dependence.
The discretized geometry was evolved from the initial time $t_i \equiv 0$
to a final time $t_f \equiv 3001$,
using AB3 with timestep $\Delta t = 1/25$.
Computations were performed on a single six core Intel i7-3960x
processor overclocked to 4.25GHz.  With this relatively limited
computing resource, producing the following results required
approximately three weeks of running time.

\subsubsection{Results}

\begin{figure}
\vspace*{-0.4cm}
\hspace*{-2.8cm}\suck[scale = 0.35]{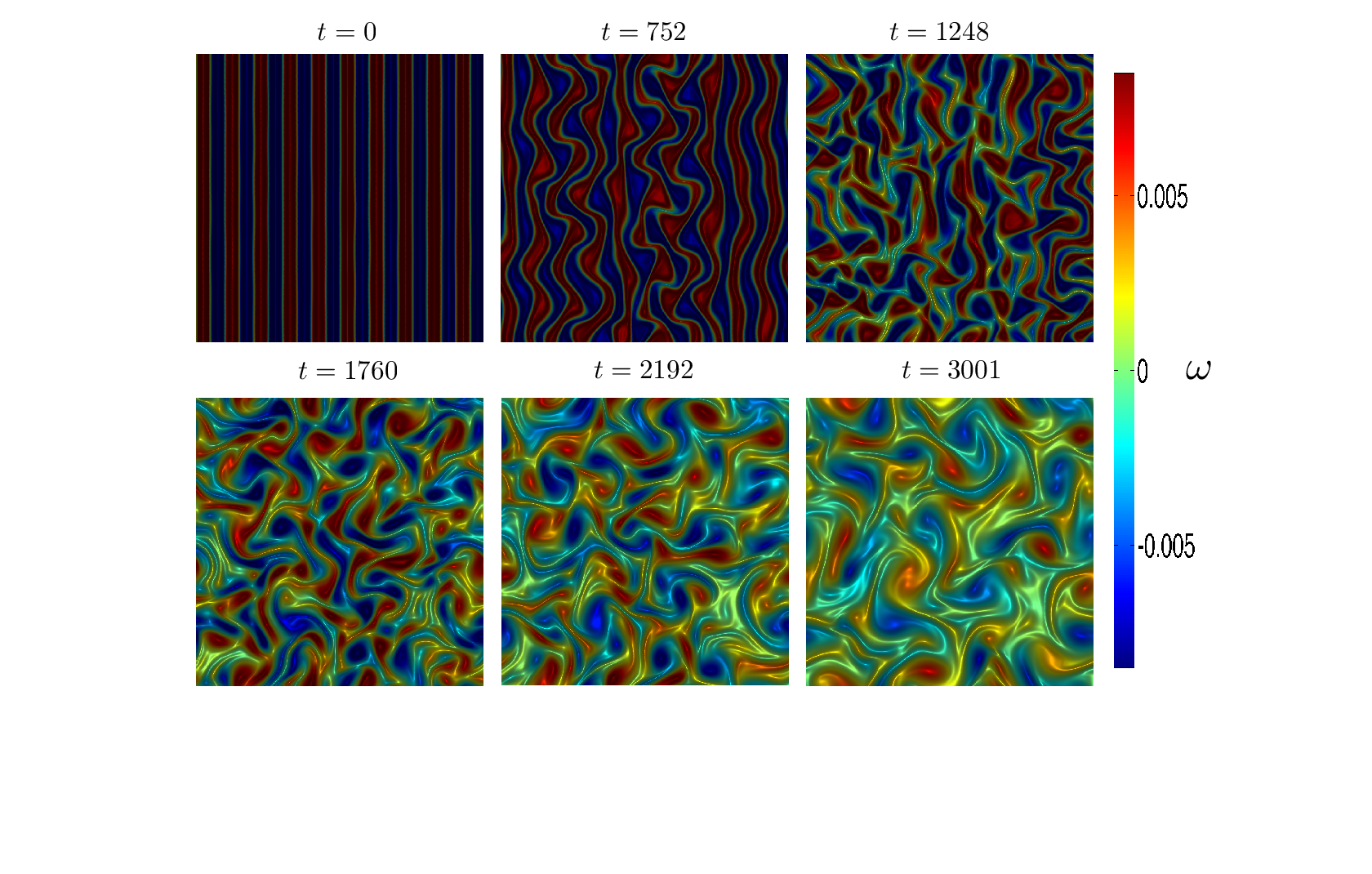}\hspace*{-1cm}
\vspace*{-2.8cm}
\caption{The boundary vorticity at six different times.
The initial conditions shown at time $t = 0$ give rise to an instability which 
produces many vortices as seen in the subsequent evolution at times
$t \geq 1248$.  Vortices colored red (blue) correspond to clockwise 
(counterclockwise) fluid rotation.
As time progresses, vortices of like rotation tend to combine to
produce larger and larger vortices.
\label{fig:vorticity}
} 
\bigskip
\begin{center}
\suck[scale = 0.44]{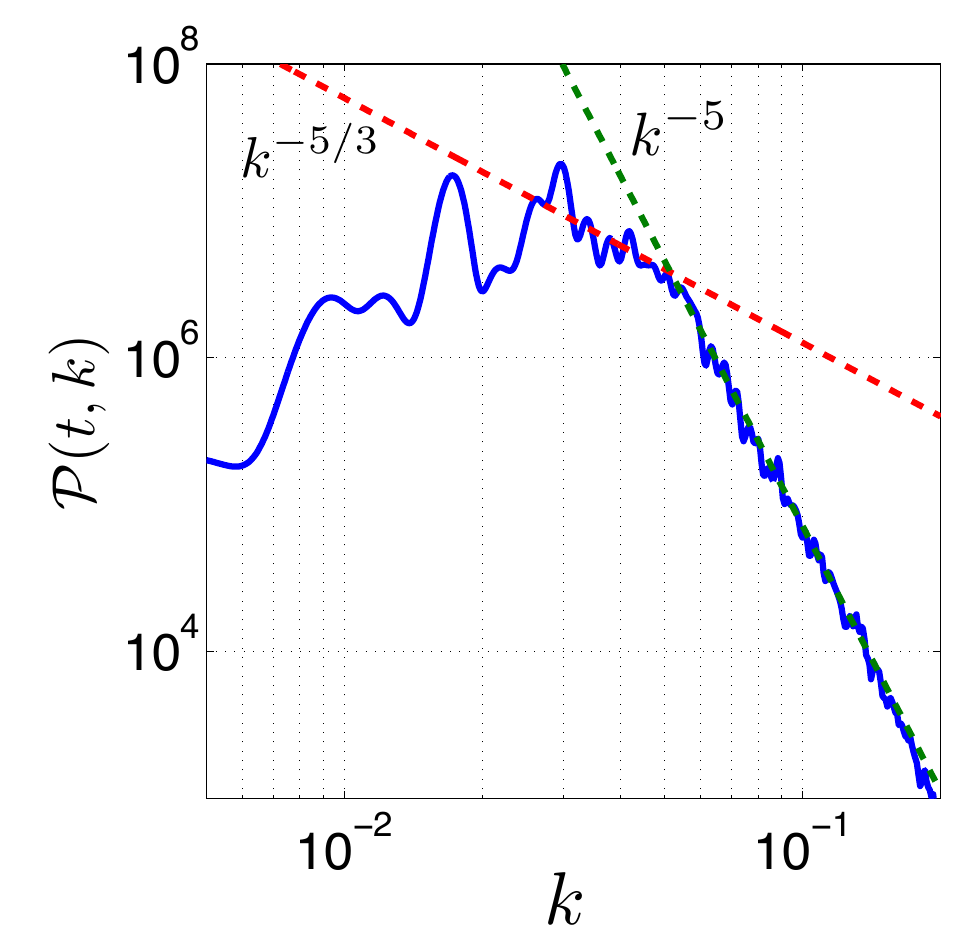}
\vspace*{-20pt}
\end{center}
\caption{The power spectrum $\mathcal P(t,k)$ of the fluid velocity at $t = 1008$.
Also shown as dashed lines are $k^{-5/3}$ and $k^{-5}$ power laws.%
\label{fig:spectrum}
} 
\vspace*{-25pt}
\end{figure}

To illustrate the turbulent flow which emerges from the
solution to Einstein's equations, we plot
in fig.~\ref{fig:vorticity} the boundary vorticity,
\begin{equation}
    \omega \equiv
    \epsilon^{\mu \nu \alpha} \, u_\mu \, \partial_\nu u_{\alpha}\,,
\end{equation}
at six different times.  We extract the fluid velocity $u^\mu$ from the
boundary stress tensor $\langle T^{\mu \nu} \rangle$
via eq.~(\ref{eq:flowfield}),
just as we did for the shock collisions in Section~\ref{sec:shocks}.

At time $t = 0$, when the fluid velocity is given by eq.~(\ref{eq:initialvelocity}), 
the vorticity is approximately sinusoidal in the $x^1$ direction and translationally invariant in the $x^2$ direction.  
By time $t = 752$, an instability is visible and the approximate symmetry
of the initial conditions is destroyed.
By time $t = 1248$, the instability has generated many small vortices
with fluid rotating clockwise (red) and 
counterclockwise (blue).
Subsequently,
vortices with the same rotation tend to merge together
producing larger and larger vortices, as seen in
the evolution snapshots at times $t = 1760$, 2192, and 3001.
As time progresses, the number of vortices decreases
while the typical vortex size grows.
This is a characteristic signature of an inverse cascade.

It is instructive to compare the gravitational evolution with
predictions from the Kolmogorov theory of turbulence.
A simple quantity to study is the power spectrum of the fluid velocity,
defined as
\begin{equation}
    \mathcal P(t,k)
    \equiv
    \frac{\partial}{\partial k} \int\limits_{|\bm k'| \leq k} 
    \frac{d^\nu k'}{(2 \pi)^\nu} \> | \tilde {\bm u}(t,\bm k')|^2,
\end{equation}
where 
\begin{equation}
    \tilde {\bm u}(t,\bm k) \equiv
    \int d^\nu x \> \bm u(t,\bm x)\, e^{-i \bm k \cdot \bm x}.
\end{equation} 
A celebrated result of Kolmogorov 
is that for driven steady-state turbulence the power spectrum $\mathcal P$
obeys the scaling
\begin{equation}
\label{eq:kol}
\mathcal P(t,k) \sim k^{-5/3},
\end{equation}  
within an \textit{inertial range} $k \in (\Lambda_-,\Lambda_+)$.
The lower limit $\Lambda_-$ is determined by the size of
the largest eddies in the system,
while the upper limit $\Lambda_+$ is set by the scale on which
viscous effects damps small eddies.

Despite the fact that 
our system is not driven or in a steady-state configuration,
 we do see hints of Kolmogorov scaling.
In fig.~\ref{fig:spectrum} we plot $\mathcal P$ at time $t = 1008$.  
Our numerical results are consistent with the scaling (\ref{eq:kol}) in the 
inertial range $k \in (0.025,0.055)$.   As we are not driving the system, 
evidence of the $k^{-5/3}$ scaling is transient and is destroyed
first in the UV, with the UV knee at $k = 0.055$ shifting to the IR as time progresses further.
Beyond the inertial range the spectrum decreases like $\mathcal P \sim k^{-p}$ 
with $p \sim 5$ until $k \sim 0.15$ beyond which $\mathcal P$ decreases
exponentially.

\begin{figure}
\vspace*{-0.4cm}
\hspace*{-2.1cm}\suck[scale = 0.35]{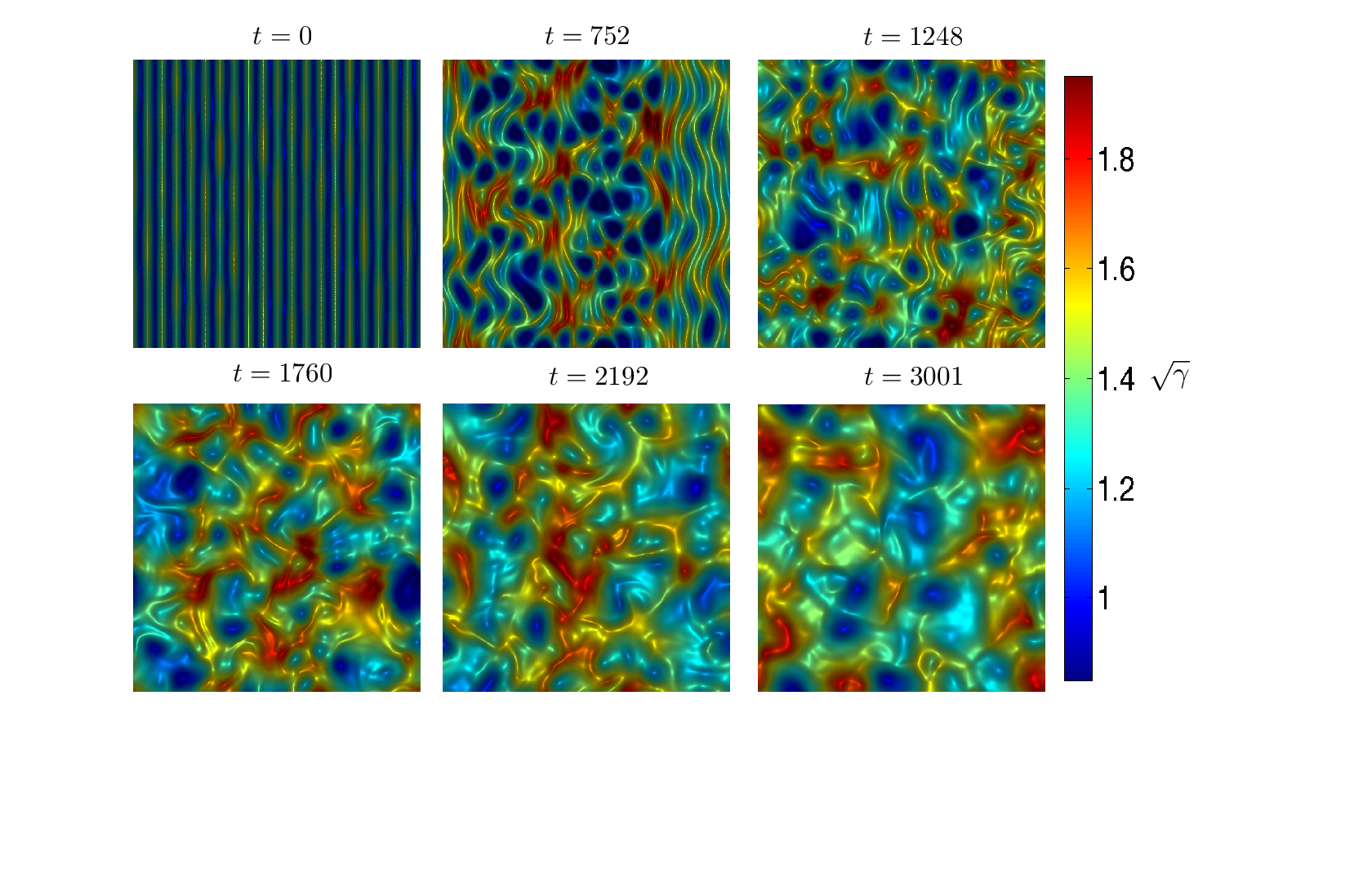}\hspace*{-1cm}
\vspace*{-2.4cm}
\caption{The horizon area element $\sqrt{\gamma}$ at six different times.  At time $t = 0$ 
$\sqrt{\gamma}$ varies sinusoidally in the $x^1$ direction and is approximately translationally invariant in the $x^2$ direction. 
By time $t = 752$, an instability is visible and the approximate symmetry
of the initial conditions is destroyed.  
By time $t = 1248$ $\sqrt{\gamma}$ exhibits structure over a large hierarchy of scales.
However, as time progresses $\sqrt{\gamma}$ becomes smoother and 
smoother just as the fluid vorticity $\omega$ does due to the inverse cascade.%
\label{fig:horizonareaturb}
} 
\end{figure}

The  inverse cascade also manifests itself in bulk gravitational quantities.  One interesting quantity to consider is the horizon
area element $\sqrt{\gamma}$.  In our coordinate system,
and in the limit of large Reynolds number $Re \gg 1$,
the event and apparent horizons approximately coincide at $r = 1$
and the horizon area element is $\sqrt{\gamma} \approx \sqrt{-g}\big|_{r = 1}$.%
\footnote
  {%
  In the fluid/gravity gradient expansion,
  the apparent and event horizons are identical up to second order in gradients.
  Hence their positions should coincide in the $Re \to \infty$ limit.
  }
In fig.~\ref{fig:horizonareaturb} we plot $\sqrt{\gamma}$ for the same
sequence of times displayed in fig.~\ref{fig:vorticity}.  
The evolution of $\sqrt{\gamma}$ closely mirrors the evolution of the vorticity on the boundary
shown in fig.~\ref{fig:vorticity}.  At time $t = 0$ when the fluid velocity is given by eq.~(\ref{eq:initialvelocity}), 
$\sqrt{\gamma}$ varies sinusoidally in the $x^1$ direction and is approximately translationally invariant in the $x^2$ direction. 
By time $t = 752$, an instability is visible and the approximate symmetry
of the initial conditions is destroyed.  By time $t = 1248$ the area element $\sqrt{\gamma}$ exhibits structure over a large hierarchy of scales
and is fractal-like in appearance.
During the subsequent evolution, $\sqrt{\gamma}$ becomes progressively
smoother, just like the fluid vorticity $\omega$,
reflecting the inverse cascade.

The velocity power spectrum $\mathcal P$ also imprints 
itself in bulk quantities.  One observable to consider is the 
extrinsic curvature $\Theta_{MN}$ of the event horizon.
The horizon curvature
$\Theta_{MN}$ can be constructed from the null normal $n_M$
to the horizon and an auxiliary null vector $\ell_{M}$ whose normalization is 
conveniently chosen to satisfy $\ell_M \, n^M = -1$.
The extrinsic curvature is then given by 
\begin{equation}
    \Theta_{MN} \equiv \Pi^P_{\ M} \, \Pi^{Q}_{\ N} \, \nabla_P \, n_Q \,,
\end{equation}
where the projection operator
$\Pi^M_{\ N} \equiv \delta^{M}_{\ N} + \ell^M n_N$.
Since the horizon is at $r \approx 1$ we choose $n_M \, dx^M = dr$ and $\ell_{M} \, dx^M= -dt$.   
In our coordinate system,
the horizon curvature satisfies
$\Theta^M_{\ N} \, \Theta^N_{\ M} = \Theta^i_{\ j} \, \Theta^j_{\ i}$,
where $i,j$ run only over the spatial coordinates.   
For later convenience we define the rescaled traceless horizon curvature
$\theta^{i}_{\ j} \equiv ({\gamma}/{\kappa^2})^{1/4}\, \Sigma^{i}_{ \ j}$,
where 
$
    \Sigma^{i}_{\ j} \equiv
    \Theta^i_{\ j} - \frac{1}{\dim} \, \Theta^k{}_k \, \delta^i{}_j
$
is the traceless 
part of the extrinsic curvature,
and $\kappa$ is the eigenvalue of the geodesic equation,
\begin{equation}
    n^M \nabla_M \, n_Q = \lambda \, n_Q \,.
\label{eq:nonaffinegeodesic}
\end{equation}

\begin{figure}
\begin{center}
\vspace*{-10pt}
\suck[scale = 0.44]{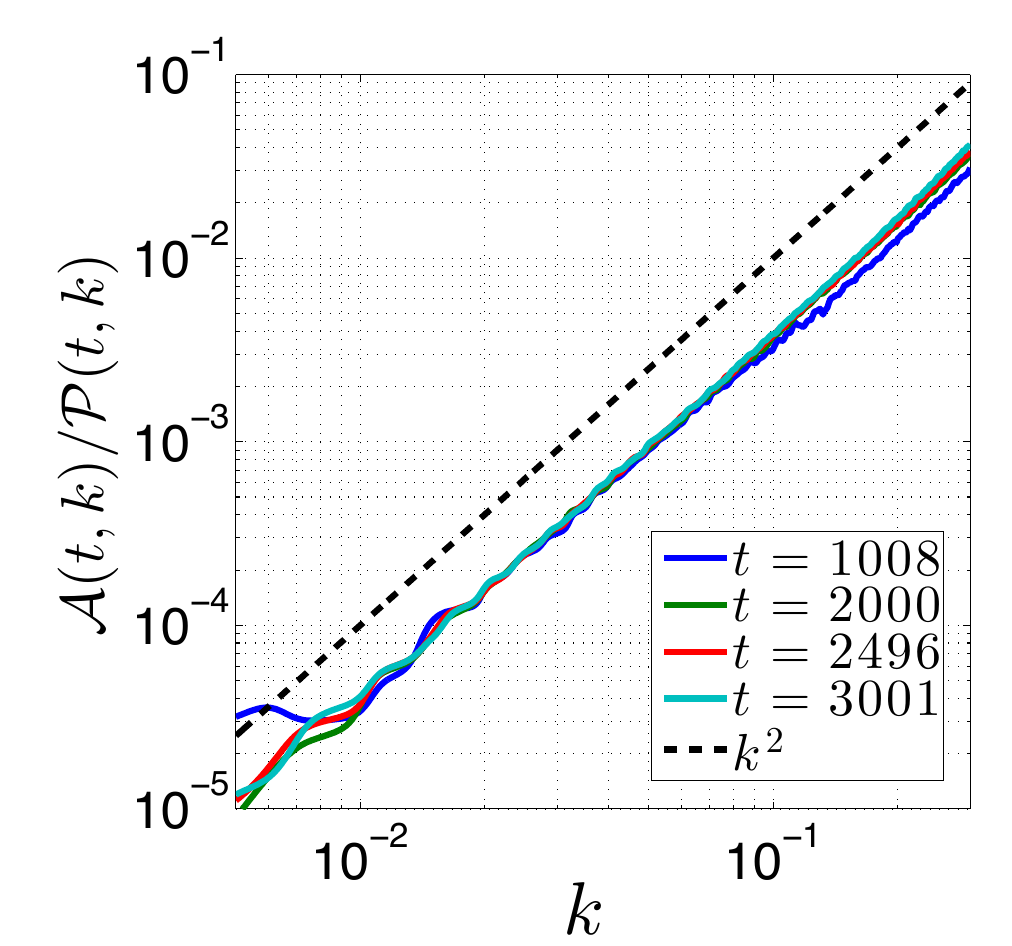}
\vspace*{-15pt}
\end{center}
\caption{The horizon curvature power spectrum $\mathcal A$ divided by the
boundary velocity power spectrum $\mathcal P(t,k)$.  
Different curves correspond to different times, as indicated in the legend.
For comparison, the dashed line plots $k^2$.%
\label{fig:horizonspectrum}
} 
\vspace*{-5pt}
\end{figure}

We define the horizon curvature power spectrum
\begin{equation}
\label{eq:Adef}
    \mathcal A(t,k) \equiv \frac{\partial}{\partial k}
    \int\limits_{|\bm k'| \leq k}  \frac{d^\dim k'}{(2 \pi)^d}  \>
    \widetilde {\theta}^i{}_j(t,\bm k')^* \>
    \widetilde {\theta}^{j}_{\ i}(t,\bm k') \,,
\end{equation}
with 
$
    \widetilde {\theta}^{i}_{\ j} \equiv
    \int d^\dim x \> \theta^i{}_j \, e^{-i \bm k \cdot \bm x}
$,
and plot the ratio $\mathcal A(t,k)/\mathcal P(t,k)$
in fig.~\ref{fig:horizonspectrum}.
As this figure makes clear,
our numerical results are consistent with
the simple scaling relation
\begin{equation}
\label{eq:AP}
\mathcal A(t,k) \sim k^2 \, \mathcal P(t,k)\,.
\end{equation}
Evidently, these horizon and boundary observables are highly correlated.
This follows directly from the applicability of the fluid/gravity
correspondence.

Both qualitative and quantitative features of our numerical results can be understood in terms of ideal conformal hydrodynamics and the 
fluid/gravity correspondence.   As discussed in sec.~\ref{sec:shocks}, in the limit of long wavelength spatial fluctuations (compared to $1/T$)
Einstein's equations can be solved perturbatively with a gradient expansion.  
At leading order the metric is precisely the locally boosted black
brane (\ref{eq:gradmetric}), with the evolution of the fluid velocity
$u$ and temperature $T$ governed by relativistic ideal 
conformal hydrodynamics \cite{Bhattacharyya:2008jc, VanRaamsdonk:2008fp}.
In other words, to leading order in the gradient expansion,
solutions to Einstein's equations can be generated merely by
solving the equations of relativistic ideal conformal hydrodynamics
and constructing the bulk metric from the resulting fluid velocity
and temperature via eq.~(\ref{eq:gradmetric}).

It was recently demonstrated that turbulent evolution in two dimensional ideal relativistic conformal hydrodynamics 
gives rise to an inverse cascade and Kolmogorov scaling (\ref{eq:kol}) \cite{Carrasco:2012nf}.  
Since the relativistic hydrodynamic equations reduce to the non-relativistic incompressible Navier-Stokes equation at low velocities
\cite{Bhattacharyya:2008kq}, this connects directly to classic results on non-relativistic  two dimensional  turbulence.  It is well known
that two dimensional  non-relativistic  incompressible turbulent flows exhibit 
Kolmogorov scaling and an inverse cascade, with the latter a consequence of  
conservation of enstrophy (the square of the vorticity).
As demonstrated in ref.~\cite{Carrasco:2012nf},
the equations of two dimensional ideal relativistic conformal hydrodynamics
conserve a relativistic generalization of enstrophy.

\begin{figure}
\begin{center}
\suck[scale = 0.45]{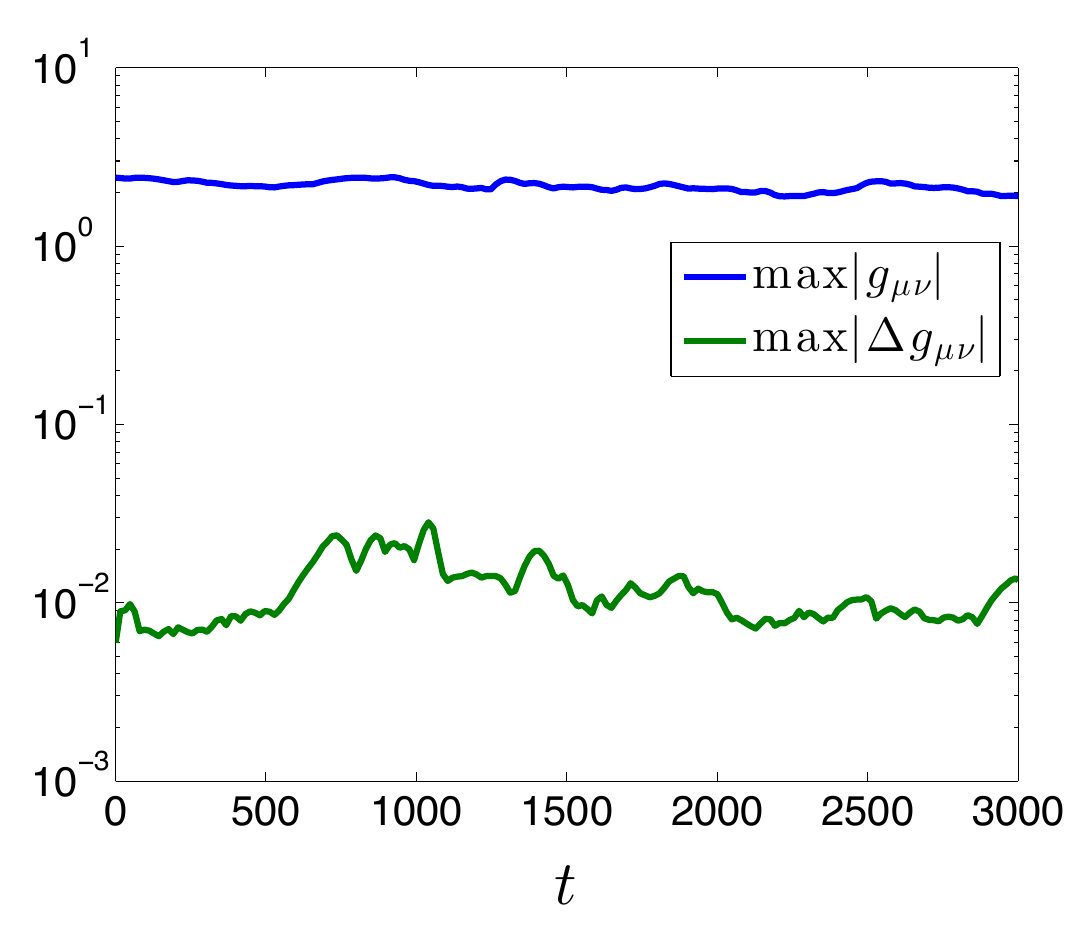}
\vspace*{-20pt}
\end{center}
\caption{Absolute deviation, as a function of time, between the
numerically computed spacetime metric and a boosted black brane metric,
with fluid flow and local temperature extracted from the numerical solution.
The upper (blue) curve shows the maximum size of metric components
on the timeslice $t$, and the much lower (green) curve shows the
maximum, on the given timeslice, of the difference between the
numerically computed metric and the boosted black brane ansatz.%
\label{fig:error}
} 
\end{figure}

We find that our numerical metric is surprisingly well approximated
by the boosted black brane metric (\ref{eq:gradmetric}).  
To perform the comparison,
we extract
the flow field $u$ and the proper energy density $\varepsilon$
from $\langle \widetilde T^{\mu \nu} \rangle$
via eq.~(\ref{eq:flowfield}).
The proper energy density is converted to a local temperature via
the (static AdS$_4$ black brane) relation
$T \equiv  \frac{3}{4 \pi}\left (\frac{3}{2} \varepsilon \right )^{1/3}$.
The flow field $u$ and local temperature $T$ are then used to construct
the boosted black brane metric (\ref{eq:gradmetric}).
Finally, we compute the difference $\Delta g_{\mu \nu}$ between the numerical
metric and the boosted black brane metric and define the error to be
${\rm max} \{ | \Delta g_{\mu \nu}|  \}$ on a given timeslice $t$.
As shown in fig.~\ref{fig:error}, the boosted black brane metric ansatz
(\ref{eq:gradmetric}) approximates the complete
geometry, even at early times, to better than 1\%!  

Although the accuracy with which the simple boosted black brane ansatz
approximates the numerical solution is remarkable,
it should not be too surprising that turbulent evolution in two spatial
dimensions gives rise to dual geometries which are reasonably
well approximated by the locally boosted black brane ansatz.
First of all, irrespective of the dimensionality,
turbulent flows require large Reynolds number,
$Re \gg 1$, which (in a strongly coupled fluid)
is equivalent to small gradients compared to the local temperature $T$.
This is precisely the regime where the fluid/gravity gradient expansion
should be well behaved.
Second, the inverse cascade of turbulence in two spatial dimensions
implies that  
gradients become smaller and smaller as energy cascades from the UV to the IR.
Therefore, the leading term (\ref{eq:gradmetric}) should become a better and better approximation to the metric as time progresses and the inverse cascade develops.

At least for $\nu = 2$,
the above observation has powerful consequences for studying turbulent
black holes.  
Instead of numerically solving the equations of general relativity,
one can simply study the equations
of hydrodynamics and construct the bulk geometry via the fluid/gravity gradient expansion.  
This is particularly illuminating in the limit of non-relativistic fluid
velocities $|\bm u| \ll 1$, where the bulk geometry 
and boundary stress are asymptotically close to equilibrium.
As shown in ref.~\cite{Bhattacharyya:2008kq}, under the rescalings  
    $t \to t/s^2$,
    $\bm x \to \bm x/s$,
    $\bm u \to s \, \bm u$, and
    $\delta T \to s^2 \, \delta T$
(with $\delta T$ the variation in the temperature away from equilibrium),
as $s \to 0$ the boundary evolution of $\delta T$ and $u$ implied by the
fluid/gravity correspondence
reduces to the non-relativistic incompressible Navier-Stokes equation.
Indeed, the above rescalings are 
symmetries of the Navier-Stokes equation.
Likewise, in the $s \to 0$ limit the geometry dual to the
Navier-Stokes equation can be computed analytically
\cite{Bhattacharyya:2008kq}.
At least for two spatial dimensions,
where is it known that solutions to the Navier-Stokes equation remain regular, 
it should be possible to (re)derive results from classic studies of turbulence,
such as Kolmogorov scaling (\ref{eq:kol}), directly from the dual
gravitational dynamics.
This is discussed in more detail in ref.~\cite{Adams:2013vsa}.



\addtocontents	{toc}{\tocsqueeze}
\section	{Conclusions}

We have presented a characteristic formulation of gravitational dynamics
which permits accurate and efficient study of a wide variety of gravitational
initial value problems in asymptotically anti-de Sitter spacetimes.
The requirement of the approach that geometries of interest have
an apparent horizon cloaking any caustics in the infalling null
congruence has, in practice, not been a limitation.
Problems with numerical stability are less severe than is
often the case with numerical relativity, due to helpful
attributes of our characteristic formulation, the presence of
an apparent horizon, and the asymptotic anti-de Sitter geometry.
With only modest computing resources, we have shown that problems 
whose symmetries reduce the dynamics to 1+1 dimensional partial
differential equations (homogeneous isotropization),
2+1 dimensional PDEs (colliding planar shocks),
or 3+1 dimensional PDEs (turbulence in two space dimensions),
are quite manageable.
An obvious question concerns the feasibility of solving 4+1 dimensional
gravitational dynamics with no simplifying symmetry restrictions.
We are optimistic that various problems in this category,
such as studying turbulent fluids in three spatial dimensions,
or off-center ``heavy ion'' collisions, will also be feasible.

\acknowledgments

The work of LY is supported by the U.S. Department
of Energy under Grant No.~DE-FG02-\-96ER\-40956.
The work of PC is supported by a Pappalardo Fellowship in Physics at MIT.
We are grateful to
Scott Hughes,
Andreas Karch,
Luis Lehner,
Frans Pretorius,
and Ruben Rosales
for helpful discussions.

\appendix

\addtocontents	{toc}{\tocsqueeze}
\section	{Bulk sources}\label{sec:bulkTMN}

For future convenience, we record here the explicit form of Einstein's equations
in our ansatz (\ref{eq:ansatz}), with the addition of a bulk matter
stress-energy tensor $T^{MN}$.
To make the radial gauge invariance manifest (in addition to spatial covariance)
it is convenient to decompose $T^{MN}$, or other tensors, using the
frame defined by the modified derivatives $d_+$ (\ref{eq:d+}),
$d_i$ (\ref{eq:di}), plus $\partial_r$, so that basis vectors are
\begin{equation}
    e'_0 \equiv d_+ = \partial_t + A \, \partial_r \,, \quad
    e'_i \equiv d_i = \partial_i + F_i \, \partial_r \,, \quad
    e'_r \equiv \partial_r \,.
\label{eq:frame}
\end{equation}
The corresponding coframe has basis one-forms
\begin{equation}
    \omega'^{\,0} \equiv dt \,,\quad
    \omega'^{\,i} \equiv dx^i \,,\quad
    \omega'^{\,r} \equiv dr - A \, dt - F_i \, dx^i \,.
\label{eq:coframe}
\end{equation}
(To make notation concise, we name the radial basis vector and dual one-form
$e'_r$ and $\omega'^{\,r}$ instead of $e'_{D+1}$ and $\omega'^{\,D+1}$.)
The metric, expressed in terms of these basis forms, is simply
\begin{equation}
    ds^2 = G_{ij} \, \omega'^{\,i} \, \omega'^{\,j}
	    + 2\, \omega'^{\,0} \, \omega'^{\,r} \,.
\label{eq:coframemetric}
\end{equation}
Using the coframe (\ref{eq:coframe}),
the stress-energy tensor decomposes into
\begin{equation}
    \langle T, \, \omega'^{\,0} \otimes \omega'^{\,0} \rangle
    = T^{00}, \quad
    \Pi^i \equiv
    \langle T, \, \omega'^{\,0} \otimes \omega'^{\,i} \rangle
    = T^{0i}, \quad
    s^{ij} \equiv
    \langle T, \, \omega'^{\,i} \otimes \omega'^{\,j} \rangle
    = T^{ij} \,,
\label{eq:Pidef}
\end{equation}
along with
\begin{align}
    \kappa &\equiv
    \langle T, \, \omega'^{\,0} \otimes \omega'^{\,r} \rangle
    = T^{0r} - A \, T^{00} - F_j \, T^{0j} \,,
\label{eq:kappa}
\\
    q^i &\equiv
    \langle T, \, \omega'^{\,i} \otimes \omega'^{\,r} \rangle
    = T^{ir} - A \, T^{i0} - F_j \, T^{ij} \,,
\label{eq:qi}
\\
    \tau &\equiv
    \langle T, \, \omega'^{\,r} \otimes \omega'^{\,r} \rangle
    = T^{rr} - 2A \, T^{0r} - 2F_i \, T^{ir}
    + A^2 T^{00} +2 A \, F_j \, T^{0j} + F_i \, F_j \, T^{ij} \,.
\label{eq:tau}
\end{align}
These combinations all transform as scalars with respect to
radial shifts (\ref{eq:radialshift}).

Einstein's equations, now in the presence of bulk sources,
decompose into the scalar equations:
\begin{align}
    0 &= \tr \!\left( G'' - \half G'^{\,2} \right) + 2 \Too ,
\label{eq:SigrrT}
\\[4pt]
    0 &= A''
    + \half \grad\cdot F'
    + \half F' \cdot F'
    + \half (\tr \, d_+G)'
    + \fourth\, \tr (G' \, d_+G)
    + \tfrac 2\dim \, \Lambda
    - \tfrac 1\dim \, \tr s
    + (1{-}\tfrac 2\dim) \, \kappa \,,
\label{eq:ArrT}
\\[4pt]
    0 &= \tr[d_+(d_+G)
    - A' \, (d_+G)
    - \half (d_+G)^2]
    + 2 \, \grad\cdot E
    + \half \tr (\Omega^2)
    + 2 \tau \,,
\label{eq:SigddotT}
\end{align}
two vector equations:
\begin{align}
    0 &= G_{ik} \big[G^{1/2} \, F'^{\,k} \big]' \, G^{-1/2}
    - G'^{\,k}{}_{i|k} + (\tr G')_{|i}
    + 2 \Pi_i \,,
\label{eq:FrrT}
\\[4pt]
    0 &=
    d_+F'_i
    + (d_+G)^k{}_{i|k}
    - (\tr d_+G)_{|i}
    + \half (\tr d_+G) F'_i
    - 2 A'_{|i}
    - G'_i{}^k E_k
    + \Omega^k{}_{i|k}
    + F'_k \, \Omega^k{}_i
    -2 q_i \,,
\label{eq:FdotT}
\end{align}
and the symmetric tensor equation:
\begin{align}
    0 &= \Bigl\{
      G_{ik}  \big[G^{1/4} (d_+G)^k_{\;j} \big]' \, G^{-1/4}
    + \fourth G'_{ij} \, \tr(d_+G) 
    - \widetilde R_{ij}
    + \tfrac 2\dim \, \Lambda \, G_{ij}
    + F'_{\,i|j}
    + \half F'_i F'_j
\nonumber\\ &\qquad {}
    + s_{ij}
    - \tfrac 1\dim \, G_{ij} \, (2\kappa + \tr s)
    \Bigr\}
    + (i \leftrightarrow j) \,.
\label{eq:GdotT}
\end{align}
Recall that $\Omega_{ij}$ (\ref{eq:Omegadef}) and $E_i$ (\ref{eq:Edef})
are the ``magnetic'' and ``electric''
parts of the radial shift field strength.
The trace of the last equation
separates from the traceless part, as before, and reads
\begin{align}
    0 &= 
      \big[G^{1/2} \, \tr (d_+G) \big]' \, G^{-1/2}
    - \widetilde R + 2\Lambda
    + \grad \cdot F'
    + \half F' \cdot F'
    -2 \kappa
    \,.
\label{eq:SigdotT}
\end{align}
The simple integration strategy described in
section \ref{sec:strategy} relies on the
nesting of the equations.
To remain applicable,
the $\Sigma$ equation (\ref{eq:SigrrT}) must only require
knowledge of $\hat g_{ij}$, 
the $F$ equation (\ref{eq:FrrT}) must only require
knowledge of $\hat g_{ij}$ and $\Sigma$,
the $d_+\Sigma$ equation (\ref{eq:SigdotT}) must only depend
on $\hat g_{ij}$, $\Sigma$ and $F$, and
the $d_+\hat g_{ij}$ equation (\ref{eq:GdotT})
must not depend on $A$.
The radial shift invariance of the linear combinations
(\ref{eq:kappa})--(\ref{eq:tau}) guarantees that the
explicit factors of $A$ and $F_i$ which appear in
these expressions cancel when combined with the
metric dependence inside $T^{MN}$.
For either an electromagnetic field, or a scalar field
$\phi$ with arbitrary potential $V(\phi)$,
one may easily confirm that the bulk source
terms do not upset the nesting of equations
which underlie the integration strategy.%
\footnote
    {%
    To see this, one must rewrite time and space
    derivatives of matter fields appearing in $T^{MN}$ in terms of
    $d_+$ and $d_i$ derivatives and, for electromagnetism,
    choose radial gauge.
    }

The apparent horizon condition (\ref{eq:Sigdothor}) is not affected by the
addition of a bulk stress-energy tensor, but the horizon stationarity
condition (\ref{eq:apphoreq}) receives modifications from
source terms (due to the use of Einstein's equations in
its derivation) and becomes
\begin{align}
    0 &= \nabla^2 A
	- \nabla A \cdot (F' - G' F)
    \nonumber\\ & \quad {}
	+ \half A \Bigl[
			- R^{(\nu)} + 2\Lambda
			+ \half (F' {-} G'F) \cdot (F' {-} G'F)
			- \nabla \cdot (F' - G'F)
			+ 2\Tor
			- \Too \, F\cdot F
		\Bigr]
    \nonumber\\ & \quad {}
	+ \half \, F \cdot F \Bigl[
			- \half \tr[(d_+G)']
			- (\nabla\cdot F)'
			- F_{i;j} G'^{\,ji}
			- \fourth (F \cdot F)' \tr G'
		    \Bigr]
    \nonumber\\[3pt] & \quad {}
	- \fourth \tr [(d_+G)^2]
	- (d_+G)^{ji} F_{i;j}
	+ F \cdot \nabla^2 F
	- \half (F'-G'F) \cdot \nabla(F\cdot F)
    \nonumber\\[3pt] & \quad {}
	- \fourth (F_{i;j}-F_{j;i})(F^{j;i}-F^{i;j})
	- \Trr
	+ \Tor \, F \cdot F
	\Bigr|_{r=\rh}
	\,.
\label{eq:apphoreqT}
\end{align}

\addtocontents	{toc}{\tocsqueeze}
\section {Riemann tensor components}\label{sec:Riemann}

For some purposes, such as evaluating curvature invariants,
it is desirable to have explicit expressions for the
Riemann tensor components generated by
our metric ansatz (\ref{eq:ansatz}).
Defining components with respect to the frame (\ref{eq:frame})
is convenient, as this makes the results
transform as scalars with respect to radial shifts.
(Moreover, the corresponding components of the
metric (\ref{eq:coframemetric}) are especially simple.)
One finds:
\begin{align}
    R_{t'r't'r'}
    &=
    A'' + \fourth F'\cdot F' ,
\\[5pt]
    R_{t'r'i'r'}
    &=
    \half F''_i - \fourth ( G'\cdot F' )_i \,,
\\[5pt]
    R_{r'i'j'r'}
    &=
    \half G''_{ij} - \fourth (G'\cdot G')_{ij} \,,
\\[5pt]
    R_{t'r'i'j'}
    &=
    \half F'_{i|j} + \fourth [ G' \cdot (d_+G + \Omega) ]_{ij}
	- (i \leftrightarrow j) \,,
\\[5pt]
    R_{t'r't'i'}
    &=
    A'_{|i}
    -\fourth [F' \cdot (d_+G + \Omega)]_i
    -\half (d_+ F')_i
    +\half (G' \cdot E)_i \,,
\\[5pt]
    R_{t'i'j'r'}
    &=
    \half (d_+G)'_{ij}
    -\fourth [ G' \cdot (d_+G + \Omega) ]_{ij}
    +\half F'_{j|i}
    +\fourth F'_i F'_j \,,
\\[5pt]
    R_{t'i'j't'}
    &=
    \half (d_+d_+G)_{ij}
    -\half A' \, (d_+G)_{ij}
    +\half (E_{i|j} + E_{j|i})
    -\fourth [ (d_+G -\Omega)\cdot(d_+G + \Omega) ]_{ij} \,,
\\[5pt]
    R_{r'i'j'k'}
    &=
    \half G'_{ij|k} + \fourth G'_{ij} F'_k 
    - (j \leftrightarrow k) \,,
\\[5pt]
    R_{t'i'j'k'}
    &=
    \half (d_+G + \Omega)_{ij|k}
    -\fourth (d_+G + \Omega)_{ij} F'_k
    -\fourth F'_i \, \Omega_{jk}
    - (j \leftrightarrow k) \,,
\\[5pt]
    R_{i'j'k'l'}
    &=
    \widehat R_{ijkl}
    -\fourth \big[
	  (d_+G)_{ik} \, G'_{jl}
	- (d_+G)_{jk} \, G'_{il}
	+ (d_+G)_{jl} \, G'_{ik}
	- (d_+G)_{il} \, G'_{jk} \big] \,,
\end{align}
with $\widehat R_{ijkl}$ defined in footnote \ref{fn:modR}.

\addtocontents	{toc}{\tocsqueeze}
\section	{Spatially covariant expressions}\label{sec:spatialcovariant}
\addtocontents	{toc}{\tocsqueeze}

Using our metric ansatz (\ref{eq:ansatz}),
explicit forms of Einstein's equations and Riemann
curvature components are most compact when written
using the modified spatial derivatives (\ref{eq:modderiv})
which are covariant under both spatial diffeomorphisms
and radial shifts,
as done in section \ref{sec:Einstein} and appendices
\ref{sec:bulkTMN} and \ref{sec:Riemann}.
Neverthess, there may be occasions where it is helpful
to have available equivalent expressions written using
ordinary spatial covariant derivatives.
These are recorded below, using the
decomposition (\ref{eq:Pidef})--(\ref{eq:tau}) of any
bulk stress-energy tensor.

Einstein's equations may be separated into
three scalar equations,
\begin{align}
    0 &= \tr \!\left( G'' - \half G'^{\,2} \right) + 2 \Too ,
\label{eq:uglySigrrT}
\\[6pt]
    0 &= A''
    + \half \nabla\cdot (F')
    + \half (\sqrt G \, F \cdot F')' / \sqrt G
    + \half (\tr d_+G)'
    + \fourth \tr (G' \, d_+G)
    + \tfrac 2\dim \, \Lambda
\nonumber\\ &\quad{}
    - \tfrac 1\dim \, \tr s
    + (1{-}\tfrac 2\dim) \, \kappa \,,
\label{eq:uglyArrT}
\\[6pt]
    0 &= \tr[d_+(d_+G)
    - A' (d_+G)
    - \half (d_+G)^2]
    + 2 \nabla\cdot (d_+F - \nabla A - A' F)
\nonumber\\[3pt] & \quad{}
    + 2F \cdot \big[\sqrt G \,
		(d_+F - \nabla A - A' F)
		\big]'/\sqrt G
    + \half \tr (\Omega^2)
    + 2 \tau 
    \,,
\label{eq:uglySigddotT}
\end{align}
two vector equations,
\begin{align}
      0 &= \big[G^{1/2} \, G_{ik} (F^k)' \big]' \, G^{-1/2}
    - G'^{\;k}{}_{i;k} + (\tr G')_{;k}
    - 2 \Too \, F_i
    + 2 \Pi_i
    \,,
\label{eq:uglyFrrT}
\\[6pt]
    0 &=
    G_{ij} \big[ ((d_+F)^j)' - ((d_+G)^j_k)' F^k \big]
    + \big( (d_+G)^{\;k}_i + (F \cdot F) \, G'^{\,k}_i \big)_{;k}
    + G'^{\;k}_i A_{;k}
    + (\nabla^2 F)_i
\nonumber\\[3pt] & \quad{}
    - (F'_k - G'_{kj} F^j)_{;i} F^k
    + (F'_i - G'_{ij} F^j)_{;k} F^k 
    + (F'_i - G'_{ij}F^j) \big[\half \tr (d_+G) -A' + \nabla\cdot F \big]
\nonumber\\[3pt] & \quad{}
    - \big[
	\tr (d_+G) + 2A' + F \cdot F'
	+ \half F \cdot F \, \tr (G') + \nabla\cdot F
    \big]_{;i}
    - \half (F \cdot F)\, \tr(G')_{;i} 
    + R^{(\nu)}_{ij} F^j
\nonumber\\[3pt] & \quad{}
    + 2(\kappa + F\cdot\Pi) F_i
    - 2 s_{ij} F^j
    - 2 Q_i
    \,,
\label{eq:uglyFdotT}
\end{align}
and the symmetric tensor equation,
\begin{align}
    0 &= \Bigl\{
      G_{ik}  \big(G^{1/4} (d_+G)^k_{\;j} \big)' \, G^{-1/4} \,
    + \half G_{ik}
	\big(\sqrt G \, (F \cdot F) \, G'^k{}_{\!j} \big)' / \sqrt G
    - R^{(\nu)}_{ij}
    + \tfrac 2\dim \, \Lambda \, G_{ij}
\nonumber\\ & \quad{}
    - (G'_{ik;j} - G'_{ij;k}) F^k
    - F_{i;k} \, G'^k{}_{\!j}
    + F'_{\,i;j}
    + \half F_{i;j} \, (\tr G')
    + \fourth G'_{ij} \left[ \tr(d_+G) + 2\nabla\cdot F \right]
\nonumber\\[2pt] & \quad{}
    + \half (F'_i-G'_{ik}F^k) (F'_j-G'_{jl}F^l)
    + \Too F_i \, F_j 
    - 2 F_i \, \Pi_j 
    + \stress_{ij}
    - \tfrac 1\dim \, G_{ij} \, ( 2\kappa + \tr s) 
    \Bigr\}
\nonumber\\ & \qquad{}
    + (i \leftrightarrow j) \,.
\label{eq:uglyGdotT}
\end{align}
The trace of this last equation separates from the
traceless part and reads%
\footnote
    {%
    Note that
    $(\nabla\cdot F)'=\nabla\cdot (F'{-}G'F) + \half F \cdot \nabla (\tr G')$.
    }
\begin{align}
    0 &= 
      \big[\sqrt G
	\big(
	    \tr (d_+G)
	    + \nabla\cdot F
	    + \half \tr (G') F \cdot F
	    \big)
	\big]' /\sqrt G
    + \half \nabla \cdot (\tr(G') \, F)
    - R^{(\nu)} + 2\Lambda
\nonumber\\[6pt]& \quad{}
    + \half (F'{-}G'F) \cdot (F'{-}G'F)
    + \Too \, F\cdot F
    - 2F\cdot\Pi
    - 2\kappa
    \,.
\label{eq:uglySigdotT}
\end{align}
In the above,
$R^{(\dim)}_{ij}$ and $R^{(\dim)}$
denote the spatial Ricci tensor
and Ricci scalar, respectively, and
$
    \Omega_{ij} \equiv
    F_{j,i} -F_{i,j} + F_i F'_j - F_j F'_i
$.

Components of the Riemann tensor are given by:
\begin{align}
    R_{trtr}
    &=
    A'' + \fourth (F'\cdot F') \,,
\\[6pt]
    R_{trir}
    &=
    \half F''_i - \fourth (G'\cdot F')_i \,,
\\[6pt]
    R_{rijr}
    &=
    \half G''_{ij} + \fourth (G'\cdot G')_{ij} \,,
\\[6pt]
    R_{trij}
    &=
    \half F'_{i;j}
    - \fourth
	\big[ (d_+G)_{ik} + F_{i;k} - F_{k;i} + F'_{i} \, F_k \big]
	G'^k{}_{\!j}
    - (i {\leftrightarrow} j) \,,
\\[6pt]
    R_{trti}
    &=
    A'_{;i}
    -\fourth (d_+G)_i{}^k F'_k
    + \half G'_i{}^k \big[ (d_+F)_k - A_{;k} \big]
    + \half A' \, \big[ F'_i - (G'\cdot F)_i \big]
\nonumber \\[3pt] &{}\quad
    + \half A \, \big[ F''_i - \half (G'\cdot F')_i \big]
    - \half (d_+F)'_i
    - \fourth (F'\cdot F) F'_i
    - \fourth (F_{i;k}-F_{k;i}) F'^{\,k} \,,
\\[6pt]
    R_{tijr}
    &=
    \half (d_+G)'_{ij}
    - \fourth G'_i{}^k (d_+G)_{kj}
    + \half F'_{j;i}
    + \fourth \big[ F'_i-(G'\cdot F)_i \big]  F'_j
    + \fourth G'_{ij} \, (F \cdot F')
\nonumber \\[3pt] &{}\quad
    + \fourth G'_i{}^k (F_{k;j} - F_{j;k})
    - \half A \, \big[ G''_{ij} - \half (G'\cdot G')_{ij} \big] \,,
\\[6pt]
    R_{tijt}
    &=
	\half (d_+d_+G)_{ij}
	- A \, (d_+G)'_{ij}
	- \half A' \, (d_+G)_{ij} 
	+ \half A^2 \, G''_{ij}
\nonumber \\[5pt] &{}\quad
    - \fourth
	\big[
	    (d_+G)_{ik} - A G'_{ik}
	    + F_{i;k} {-} F_{k;i} + F'_i F_k
	\big]
	\big[
	    (d_+G)^k{}_{\!j} - A G'^k{}_{\!j}
	    + F_j{}^{;k} {-} F^k{}_{\!;j} + F'_j F^k
	\big]
\nonumber \\[5pt] &{}\quad
    + \half \big[ (d_+F)_{i;j} {+} (d_+F)_{j;i} \big]
    - \half A \, ( F'_{i;j} {+} F'_{j;i} + F'_i F'_j )
    - \half A' \, ( F_{i;j} {+} F_{j;i} )
    - (A_{;ij} {+} A_{;ji})
\nonumber \\[5pt] &{}\quad
    + \half G'_{ij} \, F \cdot \big[ d_+F -(AF)' - \nabla A \big] \,,
\\[6pt]
    R_{rijk}
    &=
    \half G'_{ij;k} + \fourth G'_{ij}\big[ F'_k - (G'\cdot F)_k \big]
    - (j {\leftrightarrow} k) \,,
\\[6pt]
    R_{tijk}
    &=
    \half (d_+G)_{ij;k}
    - \half A \, G'_{ij;k}
    - \fourth (d_+G)_{ij} F'_k
    - \fourth G'_{ij} (d_+G)_k{}^l F_l
    + \half F_{j;ki}
    - \fourth (F_{i;j}+F_{j;i}) F'_k 
\nonumber \\[3pt] &{}\quad
    - \fourth G'_{ij}
	\big[
	    F'_k \,(F\cdot F + A)
	    + (F_{k;l} {-} F_{l;k})F^l
	    - A \, (G'\cdot F)_k 
	\big]
    - (j {\leftrightarrow} k) \,,
\\[6pt]
    R_{ijkl}
    &=
    \fourth R^{(\nu)}_{ijkl}
    -\fourth (d_+G)_{ik} \, G'_{jl}
    -\fourth (F_{i;k}+F_{k;i}) \, G'_{jl}
    -\tfrac 18 G'_{ik} \, G'_{jl} \, (F \cdot F)
\nonumber \\[3pt] &{}\quad
    - (i {\leftrightarrow} j)
    - (k {\leftrightarrow} l)
    + (ij {\leftrightarrow} kl) \,.
\end{align}

\end		{document}